\documentclass[superscriptaddress,floatfix,twocolumn,showpacs,preprintnumbers,amsmath,amssymb]{revtex4}

\usepackage{graphicx}
\usepackage{dcolumn}
\usepackage{bm}
\allowdisplaybreaks

\begin{document}

\preprint{\large{\textit{Submitted to Physical Review C}}}

\title{Measurements of the neutron electric to magnetic form factor
       ratio $\bm{G_{En}/G_{Mn}}$ via the
       $\bm{^{2}}$H$\bm{(}\vec{\bm{e}}\bm{,e'}\vec{\bm{n}}\bm{)^{1}}$H
       reaction to $\bm{Q^{2}=1.45}$ (GeV/$\bm{c}$)$\bm{^{2}}$}

\author{B.~Plaster}
\email{plaster@caltech.edu}
\affiliation{Massachusetts Institute of Technology, Cambridge, Massachusetts 02139, USA}
\affiliation{California Institute of Technology, Pasadena, California 91125, USA}
\author{A.~Yu.~Semenov}
\email{semenov@jlab.org}
\affiliation{Kent State University, Kent, Ohio 44242, USA}
\affiliation{Joint Institute for Nuclear Research, Dubna 141980, Russia}
\author{A.~Aghalaryan}
\affiliation{Yerevan Physics Institute, Yerevan 375036, Armenia}
\author{E.~Crouse}
\affiliation{The College of William and Mary, Williamsburg, Virginia 23187, USA}\author{G.~MacLachlan}
\affiliation{Ohio University, Athens, Ohio 45701, USA}
\author{S.~Tajima}
\affiliation{Duke University and TUNL, Durham, North Carolina 27708, USA}
\author{W.~Tireman}
\affiliation{Kent State University, Kent, Ohio 44242, USA}
\affiliation{Northern Michigan University, Marquette, Michigan 49855, USA}
\author{A.~Ahmidouch}
\affiliation{North Carolina A\&T State University, Greensboro, North Carolina 27411, USA}
\author{B.~D.~Anderson}
\affiliation{Kent State University, Kent, Ohio 44242, USA}
\author{H.~Arenh\"{o}vel}
\affiliation{Johannes Gutenberg-Universit\"{a}t, D-55099 Mainz, Germany}
\author{R.~Asaturyan}
\affiliation{Yerevan Physics Institute, Yerevan 375036, Armenia}
\author{O.~K.~Baker}
\affiliation{Hampton University, Hampton, Virginia 23668, USA}
\author{A.~R.~Baldwin}
\affiliation{Kent State University, Kent, Ohio 44242, USA}
\author{D.~Barkhuff}
\altaffiliation{Now at Renaissance Technologies, East Setauket, New York 11733, USA.}
\affiliation{Massachusetts Institute of Technology, Cambridge, Massachusetts 02139, USA}
\author{H.~Breuer}
\affiliation{University of Maryland, College Park, Maryland 20742, USA}
\author{R.~Carlini}
\affiliation{Thomas Jefferson National Accelerator Facility, Newport News,
Virginia 23606, USA}
\author{E.~Christy}
\affiliation{Hampton University, Hampton, Virginia 23668, USA}
\author{S.~Churchwell}
\altaffiliation{Now at University of Canterbury, Christchurch 8020, New Zealand.}
\affiliation{Duke University and TUNL, Durham, North Carolina 27708, USA}
\author{L.~Cole}
\affiliation{Hampton University, Hampton, Virginia 23668, USA}
\author{S.~Danagoulian}
\affiliation{North Carolina A\&T State University, Greensboro, North Carolina 27411, USA}
\affiliation{Thomas Jefferson National Accelerator Facility, Newport News,
Virginia 23606, USA}
\author{D.~Day}
\affiliation{University of Virginia, Charlottesville, Virginia 22904, USA}
\author{T.~Eden}
\altaffiliation{Now at National Center for Atmospheric Research, Boulder, Colorado 80307, USA.}
\affiliation{Kent State University, Kent, Ohio 44242, USA}
\affiliation{Hampton University, Hampton, Virginia 23668, USA}
\author{M.~Elaasar}
\affiliation{Southern University at New Orleans, New Orleans, Louisiana 70126, USA}
\author{R.~Ent}
\affiliation{Thomas Jefferson National Accelerator Facility, Newport News,
Virginia 23606, USA}
\author{M.~Farkhondeh}
\affiliation{Massachusetts Institute of Technology, Cambridge, Massachusetts 02139, USA}
\author{H.~Fenker}
\affiliation{Thomas Jefferson National Accelerator Facility, Newport News,
Virginia 23606, USA}
\author{J.~M.~Finn}
\affiliation{The College of William and Mary, Williamsburg, Virginia 23187, USA}
\author{L.~Gan}
\affiliation{Hampton University, Hampton, Virginia 23668, USA}
\author{A.~Gasparian}
\affiliation{North Carolina A\&T State University, Greensboro, North Carolina 27411, USA}
\affiliation{Hampton University, Hampton, Virginia 23668, USA}
\author{K.~Garrow}
\affiliation{Thomas Jefferson National Accelerator Facility, Newport News,
Virginia 23606, USA}
\author{P.~Gueye}
\affiliation{Hampton University, Hampton, Virginia 23668, USA}
\author{C.~R.~Howell}
\affiliation{Duke University and TUNL, Durham, North Carolina 27708, USA}
\author{B.~Hu}
\affiliation{Hampton University, Hampton, Virginia 23668, USA}
\author{M.~K.~Jones}
\affiliation{Thomas Jefferson National Accelerator Facility, Newport News,
Virginia 23606, USA}
\author{J.~J.~Kelly}
\affiliation{University of Maryland, College Park, Maryland 20742, USA}
\author{C.~Keppel}
\affiliation{Hampton University, Hampton, Virginia 23668, USA}
\author{M.~Khandaker}
\affiliation{Norfolk State University, Norfolk, Virginia 23504, USA}
\author{W.-Y.~Kim}
\affiliation{Kyungpook National University, Taegu 702-701, Korea}
\author{S.~Kowalski}
\affiliation{Massachusetts Institute of Technology, Cambridge, Massachusetts 02139, USA}
\author{A.~Lung}
\affiliation{Thomas Jefferson National Accelerator Facility, Newport News,
Virginia 23606, USA}
\author{D.~Mack}
\affiliation{Thomas Jefferson National Accelerator Facility, Newport News,
Virginia 23606, USA}
\author{R.~Madey}
\affiliation{Kent State University, Kent, Ohio 44242, USA}
\affiliation{The College of William and Mary, Williamsburg, Virginia 23187, USA}
\affiliation{Thomas Jefferson National Accelerator Facility, Newport News,
Virginia 23606, USA}
\author{D.~M.~Manley}
\affiliation{Kent State University, Kent, Ohio 44242, USA}
\author{P.~Markowitz}
\affiliation{Florida International University, Miami, Florida 33199, USA}
\author{J.~Mitchell}
\affiliation{Thomas Jefferson National Accelerator Facility, Newport News,
Virginia 23606, USA}
\author{H.~Mkrtchyan}
\affiliation{Yerevan Physics Institute, Yerevan 375036, Armenia}
\author{A.~K.~Opper}
\affiliation{Ohio University, Athens, Ohio 45701, USA}
\author{C.~Perdrisat}
\affiliation{The College of William and Mary, Williamsburg, Virginia 23187, USA}
\author{V.~Punjabi}
\affiliation{Norfolk State University, Norfolk, Virginia 23504, USA}
\author{B.~Raue}
\affiliation{Florida International University, Miami, Florida 33199, USA}
\author{T.~Reichelt}
\affiliation{Rheinische Friedrich-Wilhelms-Universit\"{a}t, D-53115 Bonn,
Germany}
\author{J.~Reinhold}
\affiliation{Florida International University, Miami, Florida 33199, USA}
\author{J.~Roche}
\affiliation{The College of William and Mary, Williamsburg, Virginia 23187, USA}
\author{Y.~Sato}
\affiliation{Hampton University, Hampton, Virginia 23668, USA}
\author{N.~Savvinov}
\affiliation{University of Maryland, College Park, Maryland 20742, USA}
\author{I.~A.~Semenova}
\affiliation{Kent State University, Kent, Ohio 44242, USA}
\affiliation{Joint Institute for Nuclear Research, Dubna 141980, Russia}
\author{W.~Seo}
\affiliation{Kyungpook National University, Taegu 702-701, Korea}
\author{N.~Simicevic}
\affiliation{Louisiana Tech University, Ruston, Louisiana 71272, USA}
\author{G.~Smith}
\affiliation{Thomas Jefferson National Accelerator Facility, Newport News,
Virginia 23606, USA}
\author{S.~Stepanyan}
\affiliation{Yerevan Physics Institute, Yerevan 375036, Armenia}
\author{V.~Tadevosyan}
\affiliation{Yerevan Physics Institute, Yerevan 375036, Armenia}
\author{L.~Tang}
\affiliation{Hampton University, Hampton, Virginia 23668, USA}
\author{S.~Taylor}
\affiliation{Massachusetts Institute of Technology, Cambridge, Massachusetts 02139, USA}
\author{P.~E.~Ulmer}
\affiliation{Old Dominion University, Norfolk, Virginia 23508, USA}
\author{W.~Vulcan}
\affiliation{Thomas Jefferson National Accelerator Facility, Newport News,
Virginia 23606, USA}
\author{J.~W.~Watson}
\affiliation{Kent State University, Kent, Ohio 44242, USA}
\author{S.~Wells}
\affiliation{Louisiana Tech University, Ruston, Louisiana 71272, USA}
\author{F.~Wesselmann}
\affiliation{University of Virginia, Charlottesville, Virginia 22904, USA}
\author{S.~Wood}
\affiliation{Thomas Jefferson National Accelerator Facility, Newport News,
Virginia 23606, USA}
\author{Chen Yan}
\affiliation{Thomas Jefferson National Accelerator Facility, Newport News,
Virginia 23606, USA}
\author{Chenyu Yan}
\affiliation{Kent State University, Kent, Ohio 44242, USA}
\author{S.~Yang}
\affiliation{Kyungpook National University, Taegu 702-701, Korea}
\author{L.~Yuan}
\affiliation{Hampton University, Hampton, Virginia 23668, USA}
\author{W.-M.~Zhang}
\affiliation{Kent State University, Kent, Ohio 44242, USA}
\author{H.~Zhu}
\affiliation{University of Virginia, Charlottesville, Virginia 22904, USA}
\author{X.~Zhu}
\affiliation{Hampton University, Hampton, Virginia 23668, USA}

\collaboration{The Jefferson Laboratory E93-038 Collaboration}
\noaffiliation

\date{\today}

\begin{abstract}
We report values for the neutron electric to magnetic form factor
ratio, $G_{En}/G_{Mn}$, deduced from measurements of the neutron's
recoil polarization in the quasielastic
$^{2}$H$(\vec{e},e'\vec{n})^{1}$H reaction, at three $Q^{2}$ values of
0.45, 1.13, and 1.45 (GeV/$c$)$^{2}$.  The data at $Q^{2}=1.13$ and
1.45 (GeV/$c$)$^{2}$ are the first direct experimental measurements of
$G_{En}$ employing polarization degrees of freedom in the $Q^{2}>1$
(GeV/$c$)$^{2}$ region and stand as the most precise determinations of
$G_{En}$ for all values of $Q^{2}$.
\end{abstract}


\maketitle

\section{Introduction}
\label{sec:introduction}

The nucleon elastic electromagnetic form factors are fundamental
quantities needed for an understanding of the nucleon's
electromagnetic structure.  The Sachs electric, $G_{E}$, and magnetic,
$G_{M}$, form factors \cite{sachs62}, defined in terms of linear
combinations of the Dirac and Pauli form factors, are of particular
physical interest, as their evolution with $Q^{2}$, the square of the
four-momentum transfer, is related to the spatial distribution of
charge and current within the nucleon.  As such, precise measurements
of these form factors over a wide range of $Q^{2}$ are needed for a
quantitative understanding of the electromagnetic structure not only
of the nucleon, but also of nuclei (e.g.,
\cite{frullani84,drechsel89,sick01}).  Further, in the low-energy
regime of the nucleon ground state, the underlying theory of the
strong interaction, Quantum Chromodynamics (QCD), cannot be solved
perturbatively, and a proper description of even the static properties
of the nucleon, the lowest stable mass excitation of the QCD vacuum,
in terms of the QCD quark and gluon degrees of freedom still stands as
one of the outstanding challenges of hadronic physics.  Indeed, one of
the most stringent tests to which non-perturbative QCD (as formulated
on the lattice or in a model of confinement) can be subjected is the
requirement that the theory reproduce experimental data on the nucleon
form factors (e.g., \cite{thomas01,gao03,hyde-wright04}).

Because of the lack of a free neutron target, the neutron form factors
are known with less precision than are the proton form factors, and
measurements have been restricted to smaller ranges of $Q^{2}$.  A
precise measurement of the neutron electric form factor, $G_{En}$, has
proven to be especially elusive as the neutron's integral charge is
zero.  Prior to the realization of experimental techniques utilizing
polarization degrees of freedom, values for $G_{En}$ were extracted
from measurements of the unpolarized quasielastic
$^{2}$H$(e,e'n)^{1}$H cross section and the deuteron elastic structure
function $A(Q^{2})$.  Those results for $G_{En}$ deduced from
measurements of the quasielastic $^{2}$H$(e,e'n)^{1}$H cross section
provided little information on $G_{En}$, as all results were
consistent with zero over all ranges of $Q^{2}$ accessed, $0 < Q^{2} <
4$ (GeV/$c$)$^{2}$ (e.g., \cite{lung93}).  Similarly, results for
$G_{En}$ deduced from measurements of $A(Q^{2})$, although
establishing $G_{En} > 0$ for $0 < Q^{2} < 0.7$ (GeV/$c$)$^{2}$, were
plagued with large theoretical uncertainties ($\sim\pm 40$\%) related
to the choice of an appropriate $NN$-potential for the deuteron
wavefunction (e.g., \cite{platchkov90}).

With the advent of high duty-factor polarized electron beam facilities
and state-of-the-art polarized nuclear targets and recoil nucleon
polarimeters, experimental efforts over the past 15 years have now
yielded the first precise determinations of $G_{En}$.  Our experiment
\cite{madey99} was designed to extract the neutron electric to
magnetic form factor ratio, $G_{En}/G_{Mn}$, from measurements of the
neutron's recoil polarization in quasielastic
$^{2}$H$(\vec{e},e'\vec{n})^{1}$H kinematics at three $Q^{2}$ values
of 0.45, 1.13, and 1.45 (GeV/$c$)$^{2}$.  These results were published
rapidly by Madey \textit{et al}.\ \cite{madey03}; here we provide a
more detailed report of the experiment and analysis procedures.

The remainder of this paper is organized as follows.  We begin, in
Section \ref{sec:neutron-form-factors}, with a brief overview of the
experimental techniques utilizing polarization degrees of freedom that
have been employed for measurements of the neutron form factors.  We
continue with an overview of our experiment in Section
\ref{sec:experiment}, and then discuss our neutron polarimeter in
Section \ref{sec:neutron-polarimeter}.  Details of the analysis
procedure are discussed in Section \ref{sec:data-analysis}.  Our final
results are then presented in Section \ref{sec:final-results} and
compared with selected theoretical model calculations of the nucleon
form factors.  Finally, we conclude with a brief summary in Section
\ref{sec:summary-conclusions}.  A more detailed account of the
discussion that follows may be found in \cite{plaster03}.

\section{Neutron Form Factors}
\label{sec:neutron-form-factors}

\subsection{Electron kinematics}
\label{sec:neutron-form-factors-notation}

We will use the following notation for the electron kinematics:
$(E_{e},{\bf{p}}_{e})$ will denote the four-momentum of the initial
electron, $(E_{e'},{\bf{p}}_{e'})$ will denote the four-momentum of
the scattered electron, $\theta_{e'}$ will denote the electron
scattering angle, $\omega = E_{e} - E_{e'}$ will denote the energy
transfer, ${\bf{q}} = {\bf{p}}_{e} - {\bf{p}}_{e'}$ will denote the
three-momentum transfer, and $Q^{2} = {\bf{q}}^{2} - \omega^{2} =
4E_{e}E_{e'}\sin^{2}(\theta_{e'}/2)$ will denote the square of the
spacelike four-momentum transfer in the high-energy limit of massless
electrons.  The electron scattering plane is defined by ${\bf{p}}_{e}$
and ${\bf{p}}_{e'}$.

\subsection{Measurements via polarized electron beams and recoil
            nucleon polarimetry}
\label{sec:neutron-form-factors-recoilpolarimetry}

\subsubsection{Elastic $N(\vec{e},e'\vec{N})$ scattering}
\label{sec:neutron-form-factors-recoilpolarimetry-elastic}

The polarization of the recoil nucleon, $\bf{P}$, in elastic
polarized-electron, unpolarized-nucleon scattering is well-known to be
of the form \cite{akhiezer58,dombey69,akhiezer74,arnold81}
\begin{equation}
\frac{\mathrm{d}\sigma}{\mathrm{d}\Omega_{e'}}{\bf{P}} =
\sigma_{0}\left({\bf{P}}^{(0)} + h{\bf{P}}^{(h)}\right),
\end{equation}
where $\sigma_{0}$ denotes the unpolarized cross section,
${\bf{P}}^{(0)}$ denotes the helicity-independent recoil polarization,
${\bf{P}}^{(h)}$ denotes the helicity-dependent recoil polarization,
and $h = \pm 1$ denotes the electron helicity.  The polarization is
customarily projected onto a $(\hat{t},\hat{n},\hat{\ell})$ unit
vector basis, with the longitudinal component, $\hat{\ell}$, along the
recoil nucleon's momentum; the normal component, $\hat{n}$,
perpendicular to the electron scattering plane; and the transverse
component, $\hat{t}$, perpendicular to the $\hat{\ell}$-component in
the scattering plane.  In the one-photon exchange approximation,
${\bf{P}}^{(0)} = {\bf{0}}$, and ${\bf{P}}^{(h)}$ is confined to the
scattering plane (i.e., $P^{(h)}_{n} = 0$).  The transverse,
$P^{(h)}_{t}$, and longitudinal, $P^{(h)}_{\ell}$, components are
expressed in terms of kinematics and nucleon form factors as
\cite{akhiezer58,dombey69,akhiezer74,arnold81}
\begin{subequations}
\begin{eqnarray}
P^{(h)}_{t} &=& P_{e}\frac{-2G_{E}G_{M}\sqrt{\tau(1+\tau)}
\tan\displaystyle{\frac{\theta_{e'}}{2}}}
{G_{E}^{2}+\left[\tau+2\tau(1+\tau)
\tan^{2}\displaystyle{\frac{\theta_{e'}}{2}}\right]G_{M}^{2}}, \\
P^{(h)}_{\ell} &=& P_{e}\frac{2G_{M}^{2}\tau\sqrt{(1+\tau) +
(1+\tau)^{2}\tan^{2}\displaystyle{\frac{\theta_{e'}}{2}}}
\tan\displaystyle{\frac{\theta_{e'}}{2}}}
{G_{E}^{2}+\left[\tau+2\tau(1+\tau)
\tan^{2}\displaystyle{\frac{\theta_{e'}}{2}}\right]G_{M}^{2}},
\nonumber \\
\end{eqnarray}
\end{subequations}
where $P_{e}$ denotes the electron beam polarization, $\tau =
Q^{2}/4m^{2}$, and $m$ denotes the nucleon mass.

Access to both $P^{(h)}_{t} \propto G_{E}G_{M}$ and $P^{(h)}_{\ell}
\propto G_{M}^{2}$ via a secondary analyzing reaction in a polarimeter
is highly advantageous, as the analyzing power of the polarimeter,
denoted $A_{y}$, and $P_{e}$ cancel in the
$P^{(h)}_{t}/P^{(h)}_{\ell}$ ratio, yielding a measurement of
$G_{E}/G_{M}$ that is relatively insensitive to systematic
uncertainties associated with these quantities.  For the case of the
neutron form factor ratio, as suggested by Arnold, Carlson, and Gross
\cite{arnold81} and first implemented experimentally by Ostrick
\textit{et al}.\ \cite{ostrick99}, a vertical dipole field located
ahead of a polarimeter configured to measure an up-down scattering
asymmetry sensitive to the projection of the recoil polarization on
the $\hat{t}$-axis permits access to both $P^{(h)}_{t}$ and
$P^{(h)}_{\ell}$.  During transport through the magnetic field, the
recoil polarization vector will precess through some spin precession
angle $\chi$ in the $\hat{t}$-$\hat{\ell}$ plane, leading to a
scattering asymmetry, $\xi(\chi)$, which is sensitive to a mixing of
$P^{(h)}_{t}$ and $P^{(h)}_{\ell}$,
\begin{eqnarray}
\xi(\chi) &=& A_{y}\big(P^{(h)}_{t} \cos \chi +
P^{(h)}_{\ell} \sin \chi \big) \nonumber \\
&=& A_{y}\big|{\bf{P}}^{(h)}\big| \sin \left(\chi + \delta \right).
\label{eq:xi-chi}
\end{eqnarray}
In the above, $\big|{\bf{P}}^{(h)}\big| =
\big[\big(P^{(h)}_{t}\big)^{2} +
\big(P^{(h)}_{\ell}\big)^{2}\big]^{1/2}$, and we define the
phase-shift parameter $\delta$ according to
\begin{equation}
\tan\delta = \frac{P^{(h)}_{t}}{P^{(h)}_{\ell}} =
-\frac{G_{E}}{G_{M}} \frac{\cos \displaystyle{\frac{{\theta}_{e'}}{2}}}
{\sqrt{\tau + \tau^{2}\sin^{2}\displaystyle{\frac{{\theta}_{e'}}{2}}}}
\label{eq:phase-shift-delta}
\end{equation}

\subsubsection{Quasielastic $^{2}H(\vec{e},e'\vec{n})^{1}H$ scattering}
\label{sec:neutron-form-factors-recoilpolarimetry-quasi}

The above formalism is directly applicable to an extraction of the
proton form factor ratio, $G_{Ep}/G_{Mp}$, from measurements of the
proton's recoil polarization in elastic $^{1}$H$(\vec{e},e'\vec{p})$
scattering.  An extraction of the neutron form factor ratio,
$G_{En}/G_{Mn}$, from measurements of the neutron's recoil
polarization in quasielastic $^{2}$H$(\vec{e},e'\vec{n})^{1}$H
scattering is, however, complicated by nuclear physics effects, such
as final-state interactions (FSI), meson exchange currents (MEC),
isobar configurations (IC), and the structure of the deuteron.  The
pioneering study of the sensitivity of the quasielastic
$^{2}$H$(\vec{e},e'\vec{n})^{1}$H reaction to the neutron form
factors, reported by Arenh\"{o}vel \cite{arenhovel87}, revealed that
for perfect quasifree emission of the neutron (i.e., neutron emission
along the three-momentum transfer ${\bf{q}}$),
$P^{(h)}_{t}$ is proportional to $G_{En}$, but is relatively
insensitive to FSI, MEC, IC, and the choice of the $NN$-potential for
the deuteron wavefunction.  A more detailed study of the
$^{2}$H$(\vec{e},e'\vec{n})^{1}$H reaction reported by Arenh\"{o}vel,
Leidemann, and Tomusiak \cite{arenhovel88} found that these results
also apply to $P^{(h)}_{\ell}$.  Similar findings were subsequently
reported by \cite{rekalo89,laget91}.

In Appendix \ref{sec:appendix-a}, we present a detailed discussion of
the formalism for the kinematics and recoil polarization observables
for the quasielastic $^{2}$H$(\vec{e},e'\vec{n})^{1}$H reaction.  In
particular, we provide there a definition for
$\Theta^{\mathrm{c.m.}}_{np}$, the polar angle between the proton
momentum and ${\bf{q}}$ in the recoiling neutron-proton center of mass
frame (hereafter, $n$-$p$ c.m.\ frame), a variable to which we will
refer frequently throughout this paper.  (Perfect quasifree emission
of the neutron is defined by $\Theta^{\mathrm{c.m.}}_{np} =
180^{\circ}$.)  We follow this, in Appendix \ref{sec:appendix-b}, with
a discussion of the sensitivity of the recoil polarization components
to FSI, MEC, IC, and the choice of the $NN$-potential for the deuteron
wavefunction at and away from perfect quasifree emission.

\subsection{Measurements via polarized electron beams and polarized
            targets}
\label{sec:neutron-form-factors-polarizedtarget}

\subsubsection{Elastic $\vec{N}(\vec{e},e'N)$ scattering}
\label{sec:neutron-form-factors-polarizedtarget-elastic}

The cross section in the one-photon exchange approximation for elastic
polarized-electron, polarized-nucleon scattering is well-known to be
of the form \cite{akhiezer58,dombey69,akhiezer74,raskin89}
\begin{equation}
\frac{\mathrm{d}\sigma}{\mathrm{d}\Omega_{e'}} =
\sigma_{0}\left[1 + hA_{eN}(\theta^{*},\phi^{*})\right].
\end{equation}
Here, $\theta^{*}$ and $\phi^{*}$ denote, respectively, the polar and
azimuthal angle between the target nucleon polarization vector and
${\bf{q}}$, and $A_{eN}(\theta^{*},\phi^{*})$ denotes the
polarized-electron, polarized-nucleon beam-target asymmetry, which is
a function of kinematics and the nucleon form factors.  The
sensitivity of $A_{eN}$ to the form factors is enhanced if the target
polarization is oriented in the electron scattering plane either
parallel or perpendicular to ${\bf{q}}$; in the former (latter) case,
the expression for $A_{eN}$ is identical to that for $-P^{(h)}_{\ell}$
($P^{(h)}_{t}$) and will be denoted $A_{\parallel}$ ($A_{\perp}$).
Similar to the recoil polarization technique, measurements of both
$A_{\perp}$ and $A_{\parallel}$ are desirable as the target
polarization (analog to the analyzing power) and beam polarization
cancel in the $A_{\perp}/A_{\parallel}$ ratio, again yielding a
measurement of $G_{E}/G_{M}$ that is relatively free of systematic
uncertainties.

\subsubsection{Quasielastic $^{2}\vec{H}(\vec{e},e'n)^{1}H$ and
               $^{3}\vec{He}(\vec{e},e'n)$ scattering}
\label{sec:neutron-form-factors-polarizedtarget-quasi}

\begin{table*}
\caption{Chronological summary of published data on the neutron form
factors from experiments employing polarization degrees of freedom and
a recent analysis combining data on the deuteron quadrupole form
factor, $G_{Q}$, with data on $t_{20}$ and $T_{20}$.}
\begin{ruledtabular}
\begin{tabular}{llccccc}
Reference& Facility& Published& Type& $Q^{2}$ [(GeV/$c$)$^{2}$]&
  Quantities& Note(s) \\ \hline
Jones-Woodward \textit{et al}.\ \cite{jones-woodward91}& MIT-Bates& 1991&
  $^{3}\vec{\mathrm{He}}(\vec{e},e')$& 0.16&
  $A_{\perp} \rightarrow G_{En}$&
  \footnotemark[1]$^{,}$\footnotemark[2] \\
Thompson \textit{et al}.\ \cite{thompson92}& MIT-Bates& 1992&
  $^{3}\vec{\mathrm{He}}(\vec{e},e')$& 0.2&
  $A_{\perp},~A_{\parallel} \rightarrow G_{En}$&
  \footnotemark[1]$^{,}$\footnotemark[2] \\
Eden \textit{et al}.\ \cite{eden94}& MIT-Bates& 1994&
  $^{2}$H$(\vec{e},e'\vec{n})$& 0.255&
  $P^{(h)}_{t} \rightarrow G_{En}$&
  \footnotemark[3]$^{,}$\footnotemark[4] \\
Gao \textit{et al}.\ \cite{gao94}& MIT-Bates& 1994&
  $^{3}\vec{\mathrm{He}}(\vec{e},e')$& 0.19&
  $A_{\parallel} \rightarrow G_{Mn}$&
  \footnotemark[1]$^{,}$\footnotemark[5] \\
Meyerhoff \textit{et al}.\ \cite{meyerhoff94}& MAMI& 1994&
  $^{3}\vec{\mathrm{He}}(\vec{e},e'n)$& 0.31&
  $A_{\perp},~A_{\parallel} \rightarrow G_{En}$&
  \footnotemark[1]$^{,}$\footnotemark[2] \\
Becker \textit{et al}.\ \cite{becker99}& MAMI& 1999&
  $^{3}\vec{\mathrm{He}}(\vec{e},e'n)$& 0.40&
  $A_{\perp},~A_{\parallel} \rightarrow G_{En}$&
  \footnotemark[2]$^{,}$\footnotemark[6] \\
Ostrick \textit{et al}.\ \cite{ostrick99}, Herberg \textit{et al}.\
  \cite{herberg99}& MAMI& 1999&
  $^{2}$H$(\vec{e},e'\vec{n})$& 0.15, 0.34&
  $P^{(h)}_{t},~P^{(h)}_{\ell} \rightarrow G_{En}$&
  \footnotemark[2]$^{,}$\footnotemark[3] \\
Passchier \textit{et al}.\ \cite{passchier99}& NIKHEF& 1999&
  $^{2}\vec{\mathrm{H}}(\vec{e},e'n)$& 0.21&
  $A_{ed}^{V} \rightarrow G_{En}$&
  \footnotemark[2]$^{,}$\footnotemark[3] \\
Rohe \textit{et al}.\ \cite{rohe99},
  Bermuth \textit{et al}.\ \cite{bermuth03}& MAMI& 1999/2003&
  $^{3}\vec{\mathrm{He}}(\vec{e},e'n)$& 0.67&
  $A_{\perp},~A_{\parallel} \rightarrow G_{En}$&
  \footnotemark[7]$^{,}$\footnotemark[8] \\
Xu \textit{et al}.\ \cite{xu03}& JLab& 2000/2003&
  $^{3}\vec{\mathrm{He}}(\vec{e},e')$& 0.1 -- 0.6&
  $A_{\parallel} \rightarrow G_{Mn}$&
  \footnotemark[1]$^{,}$\footnotemark[9] \\
Schiavilla and Sick \cite{schiavilla01}& ---& 2001&
  analysis& 0.00 -- 1.65&
  $G_{Q} \rightarrow G_{En}$&
  \footnotemark[10] \\
Zhu \textit{et al}.\ \cite{zhu01}& JLab& 2001&
  $^{2}\vec{\mathrm{H}}(\vec{e},e'n)$& 0.495&
  $A_{ed}^{V} \rightarrow G_{En}$&
  \footnotemark[2]$^{,}$\footnotemark[3] \\
Madey \textit{et al}.\ \cite{madey03}, this paper& JLab& 2003&
  $^{2}$H$(\vec{e},e'\vec{n})$& 0.45, 1.13, 1.45&
  $P^{(h)}_{t},~P^{(h)}_{\ell} \rightarrow G_{En}$&
  \footnotemark[3]$^{,}$\footnotemark[11] \\
Warren \textit{et al}.\ \cite{warren04}& JLab& 2004&
  $^{2}\vec{\mathrm{H}}(\vec{e},e'n)$& 0.5, 1.0&
  $A_{ed}^{V} \rightarrow G_{En}$&
  \footnotemark[3]$^{,}$\footnotemark[7] \\
Glazier \textit{et al}.\ \cite{glazier05}& MAMI& 2005&
  $^{2}$H$(\vec{e},e'\vec{n})$& 0.30, 0.59, 0.79&
  $P^{(h)}_{t},~P^{(h)}_{\ell} \rightarrow G_{En}$&
  \footnotemark[3]$^{,}$\footnotemark[12] \\
\end{tabular}
\end{ruledtabular}
\begin{minipage}{\textwidth}
\footnotetext[ 1]{Uncorrected for nuclear physics effects (i.e., for
                  FSI, MEC, or IC).}
\footnotetext[ 2]{Used the dipole parametrization for $G_{Mn}$.}
\footnotetext[ 3]{Applied corrections for FSI, MEC, and IC by averaging
                  calculations of Arenh\"{o}vel \textit{et al}.\
                  \cite{arenhovel87,arenhovel88,tomusiak88,arenhovel95,
                  leidemann91} over the acceptance.}
\footnotetext[ 4]{Used the value for $G_{Mn}$ at
                  $Q^{2} = 0.255$ (GeV/$c$)$^{2}$ as measured by
                  Markowitz \textit{et al}.\ \cite{markowitz93}.}
\footnotetext[ 5]{Used the Galster parametrization \cite{galster71}
                  for $G_{En}$.}
\footnotetext[ 6]{Corrections for FSI and MEC calculated by
                  Golak \textit{et al}.\ \cite{golak01}.}
\footnotetext[ 7]{Used values for $G_{Mn}$ taken from the
                  parametrization of Kubon \textit{et al}.\ \cite{kubon02}.}
\footnotetext[ 8]{Estimated corrections for FSI by scaling calculations of
                  Golak \textit{et al}.\ \cite{golak02} at
                  $Q^{2} = 0.37$ (GeV/$c$)$^{2}$ to
                  $Q^{2} = 0.67$ (GeV/$c$)$^{2}$.}
\footnotetext[ 9]{Used values for $G_{En}$ taken from the
                  parametrization of H\"{o}hler \textit{et al}.\
                  \cite{hohler76}.}
\footnotetext[10]{Theoretical analysis of data on the deuteron quadrupole
                  form factor, $G_{Q}$, tensor moment, $t_{20}$, and tensor
                  analyzing power, $T_{20}$.}
\footnotetext[11]{Used values for $G_{Mn}$ taken from the
                  parametrization of Kelly \cite{kelly02}.}
\footnotetext[12]{Used values for $G_{Mn}$ taken from the
                  parametrization of Friedrich and Walcher \cite{friedrich03}.}
\end{minipage}
\label{tab:table-summary-neutron-formfactors}
\end{table*}

The above formalism is directly applicable to a measurement of
$G_{Ep}/G_{Mp}$ via the elastic $^{1}\vec{\mathrm{H}}(\vec{e},e'p)$
reaction, but an extraction of $G_{En}/G_{Mn}$ from either the
quasielastic $^{2}\vec{\mathrm{H}}(\vec{e},e'n)^{1}$H reaction or the
quasielastic $^{3}\vec{\mathrm{He}}(\vec{e},e'n)$ reaction is again
complicated by nuclear physics effects.  For the case of the
$^{2}\vec{\mathrm{H}}(\vec{e},e'n)^{1}$H reaction, Cheung and Woloshyn
\cite{cheung83} were the first to show that the polarized-electron,
vector-polarized-deuterium beam-target asymmetry, $A_{ed}^{V}$, is
sensitive to $G_{En}$.  More complete calculations of $A_{ed}^{V}$
that accounted for nuclear physics effects were later reported by
Tomusiak and Arenh\"{o}vel \cite{tomusiak88} and others
\cite{arenhovel88,arenhovel95,leidemann91,laget91}.  These
calculations demonstrated that for quasifree neutron kinematics,
$A_{ed}^{V}$ is strongly sensitive to $G_{En}$, but is relatively
insensitive to FSI, MEC, IC, and the choice of the $NN$-potential for
the deuteron wavefunction.

For the case of the $^{3}\vec{\mathrm{He}}(\vec{e},e'n)$ reaction,
Blankleider and Woloshyn \cite{blankleider84} were the first to study
the sensitivity of the inclusive $^{3}\vec{\mathrm{He}}(\vec{e},e')$
asymmetry to $G_{En}$ and $G_{Mn}$.  More detailed studies of the
inclusive asymmetry carried out by others
\cite{ciofidegliatti92,schulze93} suggested that a clean extraction of
$G_{En}$ and $G_{Mn}$ from the inclusive asymmetry would be extremely
difficult due to proton contamination of the inclusive asymmetry.
Such difficulties are, however, mitigated in a
$^{3}\vec{\mathrm{He}}(\vec{e},e'n)$ coincidence experiment; as
further motivation, Laget \cite{laget91} demonstrated that the
exclusive $^{3}\vec{\mathrm{He}}(\vec{e},e'n)$ asymmetry is relatively
insensitive to the effects of FSI and MEC for $Q^{2} \agt 0.3$
(GeV/$c$)$^{2}$.

\subsection{Analysis of the deuteron quadrupole form factor}
\label{sec:neutron-form-factors-GQ}

The unpolarized elastic electron-deuteron cross section is generally
expressed in terms of the elastic structure functions, $A(Q^{2})$ and
$B(Q^{2})$.  These are, in turn, functions of the deuteron's charge,
$G_{Q}$, quadrupole, $G_{Q}$, and magnetic, $G_{M}$, form factors.
$G_{C}$ and $G_{Q}$ are of particular interest for an extraction of
$G_{En}$ as they are both proportional to $(G_{Ep} + G_{En})$.

An unambiguous extraction of $G_{C}$, $G_{Q}$, and $G_{M}$ from a
Rosenbluth separation of $A(Q^{2})$ and $B(Q^{2})$ requires some third
observable.  The tensor moments, $t_{2j}$ ($j=0,1,2$), extracted from
recoil polarization measurements in elastic unpolarized-electron,
unpolarized-deuteron scattering, and the tensor analyzing powers,
$T_{2j}$ ($j=0,1,2$), as measured in elastic unpolarized-electron,
polarized-deuteron scattering, are of particular interest as they are
functions of $G_{C}$, $G_{Q}$, and $G_{M}$ \cite{arnold81,raskin89}.
Indeed, after $G_{C}$, $G_{Q}$, and $G_{M}$ have been separated from
$A(Q^{2})$, $B(Q^{2})$, and the polarization-dependent observables, a
value for $G_{En}$ can be extracted from either $G_{C}$ or $G_{Q}$;
however, as was shown by Schiavilla and Sick \cite{schiavilla01}, an
extraction of $G_{En}$ from data on $G_{Q}$ is particularly
advantageous as the contributions of theoretical uncertainties
associated with short-range two-body exchange operators to $G_{Q}$ are
small.

\subsection{Summary of results}
\label{sec:neutron-form-factors-results}

\begin{figure}
\includegraphics[scale=0.56]{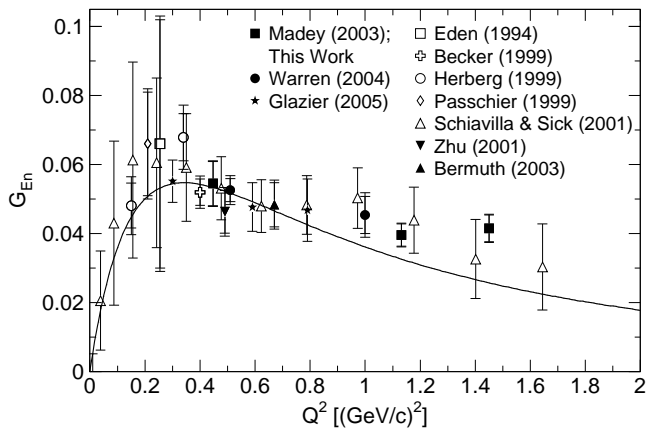}
\caption{Current status of results for $G_{En}$
(\cite{eden94,becker99,herberg99,passchier99,bermuth03,
schiavilla01,zhu01,madey03,warren04,glazier05} and this work).  The
Galster parametrization \cite{galster71} is shown as the solid curve.
See Table \ref{tab:table-summary-neutron-formfactors} for the reaction
types for the individual data points.}
\label{fig:current-world-data-gen}
\end{figure}

In Table \ref{tab:table-summary-neutron-formfactors}, we have compiled
a complete chronological summary of all published data on the neutron
form factors from experiments employing polarization degrees of
freedom and a recent analysis combining data on the deuteron
quadrupole form factor with the polarization-dependent observables
$t_{20}$ and $T_{20}$.  The current status of these results for
$G_{En}$ is shown in Fig.  \ref{fig:current-world-data-gen}.  We have
omitted the results of Jones-Woodward \textit{et al}.\
\cite{jones-woodward91}, Thompson \textit{et al}.\ \cite{thompson92},
and Meyerhoff \textit{et al}.\ \cite{meyerhoff94} from this plot as
these results were not corrected for nuclear physics effects.  It
should be noted that the results of Herberg \textit{et al}.\
\cite{herberg99} and Bermuth \textit{et al}.\ \cite{bermuth03}
supersede those of Ostrick \textit{et al}.\ \cite{ostrick99} and Rohe
\textit{et al}.\ \cite{rohe99}, respectively, as the former set
reported the final results (corrected for nuclear physics effects) for
their respective experiments.

The $Q^{2}$ range of $G_{En}$ is much more limited than those of the
other three nucleon electromagnetic form factors, with only two
results, those of Madey \textit{et al}.\ \cite{madey03} and the
analysis results of Schiavilla and Sick \cite{schiavilla01}, extending
into the $Q^{2} > 1$ (GeV/$c$)$^{2}$ region.  The agreement between
these modern data and the Galster parametrization \cite{galster71}
with its original fitted parameters can be judged only as fortuitous.

\section{Experiment}
\label{sec:experiment}

\subsection{Overview of experiment}
\label{sec:experiment-overview}

Our experiment \cite{madey99}, E93-038, was conducted in Hall C of the
Jefferson Laboratory (JLab) during a run period lasting from September
2000 to April 2001.  Longitudinally polarized electrons extracted from
the JLab electron accelerator \cite{leemann01} scattered from a liquid
deuterium target mounted on the Hall C beamline.  The scattered
electrons were detected and momentum analyzed by the Hall C High
Momentum Spectrometer (HMS) in coincidence with the recoil neutrons.
A stand-alone neutron polarimeter (NPOL) \cite{niculescu98}, designed
and installed in Hall C specifically for this experiment, was used to
measure the up-down scattering asymmetry arising from the projection
of the recoil neutrons' polarization on an axis perpendicular to their
momentum and parallel to the floor of Hall C.  A vertical dipole field
located ahead of NPOL was used to precess the recoil neutrons'
polarization vectors through some chosen spin precession angle in
order to measure this up-down scattering asymmetry from different
projections of the recoil polarization vector on the polarimeter's
sensitive axis.  This vertical dipole field also served as a sweeping
field for the background flux of recoil protons from the deuteron
target.

Data were taken at four central $Q^{2}$ values of 0.447, 1.136, 1.169,
and 1.474 (GeV/$c$)$^{2}$ with associated electron beam energies of
0.884, 2.326, 2.415, and 3.395 GeV, respectively.  The nominal
(central) values of the quasielastic electron and neutron kinematics
and the neutron spin precession angles, $\chi$, for each of these
central $Q^{2}$ points are summarized in Table
\ref{tab:central-kinematics}.  We note that the data acquired at the
separate central $Q^{2}$ values of 1.136 and 1.169 (GeV/$c$)$^{2}$
were combined in our final analysis.  Beam polarizations of 70--80\%
at currents of 20--70 $\mu$A were typical throughout the duration of
the experiment.  The central axis of the neutron polarimeter was fixed
at a scattering angle of $46.0^{\circ}$ relative to the incident
electron beamline for the duration of the experiment.  The scattering
asymmetries measured in our polarimeter were on the order of a few
percent.

\begin{table}
\caption{Nominal (central) values of the quasielastic electron and
neutron kinematics and neutron spin precession angles for each $Q^{2}$
setting in the experiment.  The data from the central $Q^{2}$ values
of 1.136 and 1.169 (GeV/$c$)$^{2}$ were combined in our final
analysis.}
\begin{ruledtabular}
\begin{tabular}{cccccl}
$Q^{2}$& $E_{e}$& $E_{e'}$& & $T_{n}$& Precession \\
$[$(GeV/$c$)$^{2}]$& [GeV]& [GeV]& $\theta_{e'}$& [MeV]&
  Angles $\chi$ \\ \hline
0.447& 0.884& 0.643& 52.65$^{\circ}$& 239& $\pm40^{\circ}$ \\
1.136& 2.326& 1.718& 30.93$^{\circ}$& 606& $0^{\circ}, \pm90^{\circ}$ \\
1.169& 2.415& 1.789& 30.15$^{\circ}$& 624& $\pm40^{\circ}$ \\
1.474& 3.395& 2.606& 23.55$^{\circ}$& 786&
  $0^{\circ}, \pm40^{\circ}, \pm90^{\circ}$ \\
\end{tabular}
\end{ruledtabular}
\label{tab:central-kinematics}
\end{table}

\subsection{Polarized electron source}
\label{sec:experiment-polarized-source}

Polarized electrons were produced at the accelerator source via
optical illumination of a strained GaAs photocathode (GaAs on GaAsP
\cite{poelker01a}) with circularly polarized laser light from a
high-power ($\sim 500$ mW) Ti-sapphire laser
\cite{poelker01a,poelker01b}; the linearly polarized light from the
laser was circularly polarized with a Pockels cell.  The helicity of
the circularly polarized light emerging from the Pockels cell was
flipped at a frequency of 30 Hz (by switching the polarity of the high
voltage applied to the Pockels cell) according to a pseudorandom
scheme in which the helicity of one 33.3 ms window was randomly
chosen, and the helicity of the following 33.3 ms window required to
be that of the opposite helicity (i.e., a sequence of such ``helicity
pairs'' could have been $+-$, $-+$, $-+$, $+-$, etc.).  A $\lambda/2$
plate was intermittently placed in the optics path upstream of the
Pockels cell.  This $\lambda/2$ plate reversed the helicity of the
electron beam that would otherwise have been induced by the Pockels
cell, thereby providing the means for important systematic checks of
any possible helicity-correlated differences.

\subsection{Hall C beamline}
\label{sec:hall-c-beamline}

Beam of the desired energy was extracted from the accelerator and then
transported along the Hall C arc (series of steering/bending magnets)
and beamline.  A number of superharps \cite{yan95} were used to
monitor the beam profile, and four beam position monitors (cavities
with four antennas oriented at angles of $\pm45^{\circ}$ relative to
the horizontal and vertical directions) provided absolute
determinations of the beam position.  The beam current was monitored
with two monitors (cylindrical wave guides with wire loop antennas
coupling to resonant modes of the beam cavity, yielding signals
proportional to the current).

\subsection{Beam polarization measurements}
\label{sec:beam-polarization-moller}

The beam polarization was measured periodically with a M{\o}ller
polarimeter \cite{hauger01} located along the Hall C beamline
approximately 30 m upstream of the cryotarget.  We measured the beam
polarization approximately every one to two days during stable
accelerator operations.  Measurements were also typically conducted
following the insertion or removal of the $\lambda/2$ plate at the
polarized source or other major accelerator changes.  A statistical
precision of $< 1$\% was typically achieved after $\sim$15--20 minutes
of data taking.  Details of the results of our beam polarization
measurements will be discussed later, where it will be seen that the
details of the analysis are relatively insensitive to the exact values
of the beam polarization.  Instead, the beam polarization information
was primarily used to assess systematic uncertainties associated with
temporal fluctuations in the polarization.

It should be noted that although our production scattering asymmetry
data were taken with beam currents as high as 70 $\mu$A, the M{\o}ller
polarimeter was only designed for currents up to $\sim 8$ $\mu$A (due
to the heating and subsequent depolarization of the iron target foil);
therefore, it was necessary to assume that our beam polarization
measurements conducted at currents of 1--2 $\mu$A were valid for the
higher beam currents of our production running.  The validity of this
assumption has been verified for operations in Hall A at JLab where
the results of beam polarization measurements conducted at low
currents (M{\o}ller polarimeter) and high currents (Compton
polarimeter) were found to agree \cite{alcorn04}.

\subsection{Scattering chamber and cryotargets}
\label{sec:scattering-chamber-cryotargets}

The scattering chamber consisted of a vertically-standing cylindrical
aluminum chamber vacuum coupled to the incoming beamline.  Two exit
windows (made of beryllium) faced the HMS and NPOL, while an exit port
faced the downstream beamline leading to the beam dump.  During our
experiment, the scattering chamber housed only one target ladder
divided into a cryogenic target section and a solid target section.
The cryogenic target section consisted of three cryogenic target
``loops''.  Each of these loops consisted of 4-cm and 15-cm long
aluminum target ``cans'', heat exchangers (heat loads from the
electron beam were typically several hundred Watts), high- and
low-power heaters (used to maintain the cryotargets at their specified
temperatures and to correct for fluctuations in the beam current), and
various sensors.  Liquid deuterium and liquid hydrogen, maintained at
(nominal) operating temperatures of 22 K and 19 K, respectively,
circulated through two of these loops; the third loop was filled with
gaseous helium.  Solid (carbon) targets and 4-cm and 15-cm long
``dummy targets'', composed of two aluminum foils spaced 4 cm and 15
cm apart, were mounted on the solid target section of the target
ladder.  As discussed in more detail later, data were taken with the
dummy targets in order to assess the level of contamination due to
scattering from the target cell windows.  The thicknesses of the
liquid deuterium and liquid hydrogen target cell windows were on the
order of 4--6 mils, while those of the dummy targets were much thicker
and on the order of 36--37 mils.

To mitigate the effects of local boiling, the beam was rastered over a
$2 \times 2$ mm$^{2}$ spot on the cryotargets using a fast raster
system \cite{yan02} located $\sim 21$ m upstream of the cryotargets.
Target conditions (e.g., temperatures, heater power levels, etc.)
were monitored continuously throughout the duration of the experiment
using the standard Hall C cryotarget control system.

\subsection{High Momentum Spectrometer}
\label{sec:HMS}

Scattered electrons were detected in the HMS, a three-quadrupole,
single-dipole (QQQD) spectrometer (all magnets are superconducting)
with a solid angle acceptance of 6 msr (defined by an
octogonally-shaped flared collimator), a maximum central momentum of
7.5 GeV/$c$, a $\pm 18$\% momentum acceptance, and a $\sim 27$ m
flight path from the target to the detector package.

\subsubsection{Magnets}
\label{sec:HMS-magnets}

The three quadrupole magnets and the dipole magnet are mounted on a
common carriage that rotates on a rail system about the target.  The
quadrupoles are 1.50~T maximum 20-ton (first, Q$_1$) and 1.56~T
maximum 30-ton (second, Q$_2$, and third, Q$_3$) superconducting coils
with magnetic lengths of 1.89~m and 2.10~m, respectively.  Q$_1$ and
Q$_3$ are used for focusing in the dispersive direction, while Q$_2$
provides transverse focusing.  The dipole is a 1.66~T maximum 470-ton
superconducting magnet with a magnetic length of 5.26~m, a bend angle
of $25^{\circ}$, and a bend radius of 12.06~m.

The magnets were operated in their standard point-to-point tune in
both the dispersive and non-dispersive directions.  For our central
$Q^{2}$ points of 0.447, 1.136, 1.169, and 1.474 (GeV/$c$)$^{2}$, the
nominal field strengths of Q$_1$ were 0.11, 0.31, 0.32, and 0.46~T;
those of Q$_2$ were 0.13, 0.37, 0.38, and 0.55~T; those of Q$_3$ were
0.06, 0.17, 0.18, and 0.26~T; and, finally, those of the dipole were
0.18, 0.47, 0.49, and 0.71~T.

\subsubsection{Detector package}
\label{sec:HMS-detector-package}

\begin{figure}
\includegraphics[angle=270,scale=0.36]{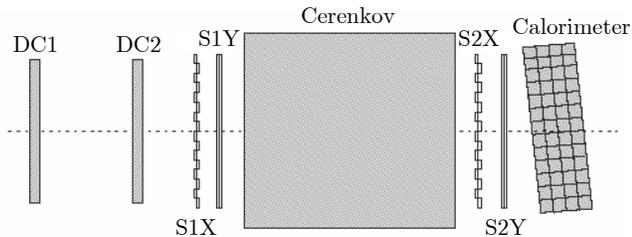}
\caption{Schematic diagram of the ordering of the HMS detector package
elements.  Shown are the two drift chambers (DC1 and DC2), the two
$x$-$y$ hodoscopes (S1X/S1Y and S2X/S2Y), the gas \v{C}erenkov
counter, and the lead-glass calorimeter.}
\label{fig:hms-detector-package}
\end{figure}

The detector package is enclosed within a concrete shielding hut and
includes two drift chambers, two sets of hodoscopes, a gas
\v{C}erenkov counter, and a lead-glass calorimeter.  A schematic
diagram depicting the ordering of the detector package elements is
shown in Fig.\ \ref{fig:hms-detector-package}.

\paragraph{Drift chambers}
The two multiwire drift chambers \cite{baker95}, used for tracking,
each consist of six wire planes: (1) the $X$ and $X'$ planes, which
provide position information on the $x$-coordinate (dispersive
direction); (2) the $Y$ and $Y'$ planes, which provide position
information on the $y$-coordinate (non-dispersive direction); and (3)
the $U$ and $V$ planes, which are inclined at $\pm15^{\circ}$ angles
relative to the orientation of the $X$ and $X'$ planes.  As seen by
incoming particles, the ordering of these planes is $XYUVY'X'$.  The
active area of each plane is 113 ($x$) $\times$ 52 ($y$) cm$^{2}$ with
an alternating sequence of anode wires (25 $\mu$m gold-plated
tungsten) and cathode wires (150 $\mu$m gold-plated copper-beryllium)
spaced $\sim 1$ cm apart.  The individual wire planes are separated by
1.8 cm, and the two drift chambers are separated by 81.2 cm.  The
chambers were filled with equal mixtures (by weight) of argon and
ethane and maintained at a pressure slightly above atmospheric
pressure.  The signals from the anodes were read out in groups of 16
by multi-hit time-to-digital convertors (TDCs).  The fast branch of
the signals from the hodoscope TDCs (to be described shortly) defined
the TDC start for the electron arm trigger, while the delayed signals
from the drift chamber TDCs formed the TDC stop.

\paragraph{Hodoscopes}
The $x$- ($y$-) planes of the two hodoscopes, denoted S1X/S2X
(S1Y/S2Y), consist of 16 (10) 75.5-cm (120.5-cm) long Bicron BC404
plastic scintillator bars with a thickness of 1.0 cm and width of 8.0
cm.  UVT lucite light guides and Philips XP2282B photomultiplier tubes
(PMTs) are coupled to both ends of each scintillator bar.  The S1X/S1Y
and S2X/S2Y planes are separated by $\sim 2.2$ m.  The fast branch of
the PMT signals was routed to leading-edge discriminators.  The
discriminated signals were then split, with one set of outputs
directed to logic delay modules, TDCs, and scalers, and the other set
directed to a logic module.  The overall logic signaling a hit in any
one of the hodoscope planes required a signal above threshold in at
least one of the 16 (10) PMTs mounted on the $x>0$ ($y>0$) side of the
bars and at least one of the 16 (10) PMTs mounted on the opposite
$x<0$ ($y<0$) side.  The slow branch of the PMT signals was directed
to analog-to-digital convertors (ADCs).

\paragraph{\v{C}erenkov detector}
The \v{C}erenkov detector is a cylindrical tank (165-cm length and
150-cm inner diameter) filled with Perfluorobutane (C$_{4}$F$_{10}$,
index of refraction $n = 1.00143$ at STP).  The pressure and
temperature in the tank were monitored on an (approximately) daily
basis and were observed to be highly stable.  Pressures were typically
$\sim$0.401--0.415 atm (indices of refraction $\sim$1.00057--1.00059),
translating into energy thresholds of $\sim 21$ MeV ($\sim 5.6$ GeV)
for pions (electrons).  The tank is viewed by two mirrors, located at
the rear of the tank, which focus the resulting \v{C}erenkov light
into two Burle 8854 PMTs.  The signals from these PMTs were directed
to ADCs.  During this experiment, information from the \v{C}erenkov
detector was used only for electron-hadron discrimination and not for
HMS trigger logic purposes.

\paragraph{Lead-glass calorimeter}
The calorimeter consists of 52 TF1 lead-glass blocks stacked into four
vertical layers of 13 blocks each.  Each block has dimensions of $70
\times 10 \times 10$ cm$^{3}$, corresponding to $\sim$16 radiation
lengths for the total four-layer-thickness of 40 cm.  As is indicated
in Fig.\ \ref{fig:hms-detector-package}, the four layers of the
calorimeter are tilted at an angle of $5^{\circ}$ relative to the
central axis of the detector package to eliminate losses in the gaps
between the individual blocks.  Philips XP3462B PMTs are coupled to
one end of each block, and the signals from these PMTs were routed to
ADCs.  Again, information from the lead-glass calorimeter was not used
for HMS trigger logic purposes during this experiment.

\section{NEUTRON POLARIMETER}
\label{sec:neutron-polarimeter}

\subsection{Overview}
\label{sec:neutron-polarimeter-overview}

\begin{figure}
\includegraphics[scale=0.85]{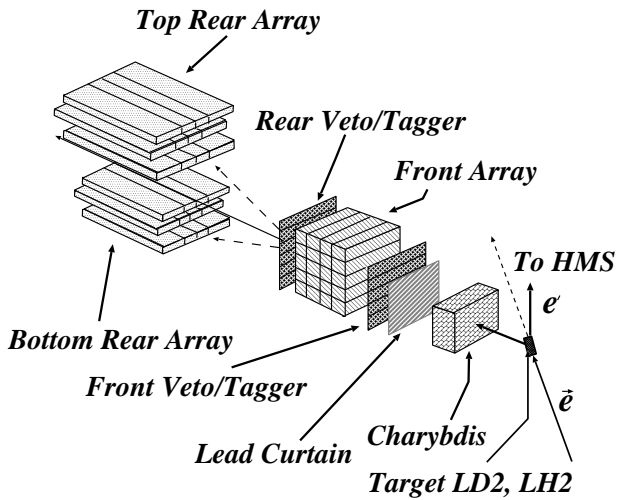}
\caption{Isometric view of the NPOL flight path showing the Charybdis
dipole magnet, the lead curtain, the front veto/tagger array, the
front array, the rear veto/tagger array, and the top and bottom rear
arrays.}
\label{fig:npol-isometric-view}
\end{figure}

A schematic diagram of the experimental arrangement with an isometric
view of the neutron polarimeter is shown in Fig.\
\ref{fig:npol-isometric-view}.  The first element in the NPOL flight
path was a dipole magnet (Charybdis) with a vertically oriented field
that was used to precess the neutrons' spins through an angle $\chi$
in a horizontal plane.  As a by-product, protons and other charged
particles were swept from the acceptance during asymmetry measurements
conducted with the field energized.  The next item in the flight path
was a 10.16-cm thick lead curtain, located directly in front of a
steel collimator (not shown in this figure).  The lead curtain served
to attenuate the flux of electromagnetic radiation and to degrade in
energy the flux of charged particles incident on the polarimeter's
detectors.

The polarimeter consisted of 70 plastic scintillation detectors
enclosed within a steel and concrete shielding hut.  The front array
of the polarimeter functioned as the polarization analyzer (via
spin-dependent scattering from unpolarized protons in hydrogen and
carbon nuclei), while the top and bottom rear arrays, shielded by the
collimator from a direct line-of-sight to the target, were configured
for sensitivity to an up-down scattering asymmetry proportional to the
projection of the recoil polarization on a horizontally-oriented
``sideways'' axis (see next subsection).  Double layers of thin-width
``veto/tagger'' detectors located directly ahead of and behind the
front array tagged incoming and scattered charged particles.  The
flight path from the center of the target to the center of the front
array was 7.0 m, and the distance from the center of the front array
to the center of the rear array (along the polarimeter's central axis)
was $\sim 2.5$ m.

\subsection{Polarimetry}
\label{sec:neutron-polarimeter-polarimetry}

\subsubsection{Coordinate systems}
\label{sec:neutron-polarimeter-polarimetry-coordinates}

Here we establish some necessary notation for a number of different
coordinate systems to which we will refer throughout the remainder of
this paper.

First, calculations of recoil polarization for the quasielastic
$^{2}$H$(\vec{e},e'\vec{n})^{1}$H reaction are usually referred to a
$(\hat{t},\hat{n},\hat{\ell})$ \textit{reaction basis}, defined on an
\textit{event-by-event basis} in the $n$-$p$ c.m.\ frame according to
\begin{equation}
\hat{\ell} \parallel {\bf{p}}^{\mathrm{c.m.}}_{n},~~~~~
\hat{n} \parallel {\bf{q}}^{\mathrm{c.m.}} \times
  {\bf{p}}^{\mathrm{c.m.}}_{n},~~~~~
\hat{t} = \hat{n} \times \hat{\ell},
\label{eq:reaction-basis}
\end{equation}
where ${\bf{p}}^{\mathrm{c.m.}}_{n}$ and ${\bf{q}}^{\mathrm{c.m.}}$
denote, respectively, the incident neutron's momentum and the momentum
transfer in the $n$-$p$ c.m.\ frame.  The reaction basis can best be
visualized by referring to the schematic diagram of the kinematics in
the $n$-$p$ c.m.\ frame shown in Fig.\
\ref{fig:deuteron-electrodisintegration-schematic} of Appendix
\ref{sec:appendix-a}.

Second, we define a \textit{polarimeter basis},
$(\hat{x}_{\mathrm{NPOL}},\hat{y}_{\mathrm{NPOL}},\hat{z}_{\mathrm{NPOL}})$,
\textit{fixed for all events}, defined in the laboratory frame
according to
\begin{subequations}
\label{eq:polarimeter-basis-definition}
\begin{eqnarray}
\hat{z}_{\mathrm{NPOL}} &\parallel& \mathrm{NPOL~central~axis}, \\
\hat{y}_{\mathrm{NPOL}} &\perp& \mathrm{Hall~C~floor}, \\
\hat{x}_{\mathrm{NPOL}} &=& \hat{y}_{\mathrm{NPOL}} \times
  \hat{z}_{\mathrm{NPOL}},
\end{eqnarray}
\end{subequations}
with the center of the target defined to be the origin of this
coordinate system.

Third, the symmetric geometric configuration of the polarimeter's
top/bottom rear arrays suggests the introduction of a
\textit{polarimeter momentum basis}, $(\hat{S},\hat{N},\hat{L})$,
which we again define on an \textit{event-by-event basis} in the
laboratory frame according to
\begin{equation}
\hat{L} \parallel \hat{p}_{n},~~~~~
\hat{S} \parallel \hat{y}_{\mathrm{NPOL}} \times \hat{p}_{n},~~~~~
\hat{N} = \hat{L} \times \hat{S},
\label{eq:polarimeter-momentum-basis-definition}
\end{equation}
where $\hat{p}_{n}$ denotes a unit vector along the incident neutron's
momentum in the laboratory frame.  We will henceforth refer to the
$\hat{S}$ and $\hat{L}$ axes as the polarimeter's ``sideways'' and
``longitudinal'' axes of sensitivity, respectively.  We express the
recoil polarization in terms of the polarimeter momentum basis as
${\bf{P}} = P_{S}\hat{S} + P_{N}\hat{N} + P_{L}\hat{L}$.

\begin{figure}
\includegraphics[scale=0.48,angle=270,clip=]{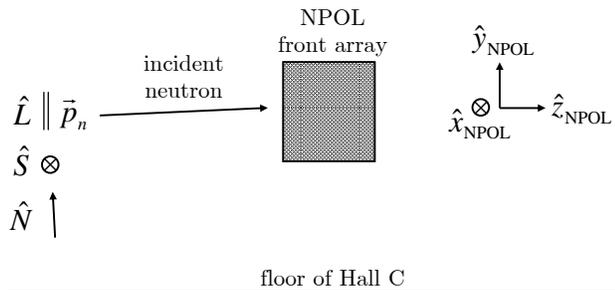}
\caption{Schematic diagram of the
$(\hat{x}_{\mathrm{NPOL}},\hat{y}_{\mathrm{NPOL}},\hat{z}_{\mathrm{NPOL}})$
polarimeter basis (fixed for all events) and the
$(\hat{S},\hat{N},\hat{L})$ polarimeter momentum basis (defined on an
event-by-event basis).  Note that as ${\bf{p}}_{n}$ is not, in general,
restricted to the $\hat{y}_{\mathrm{NPOL}}$-$\hat{z}_{\mathrm{NPOL}}$
plane, $\hat{S}$ is not, in general, parallel to
$\hat{x}_{\mathrm{NPOL}}$.}
\label{fig:coordinate-systems}
\end{figure}

A schematic diagram showing the orientation of the polarimeter basis
and polarimeter momentum basis coordinate systems is shown in Fig.\
\ref{fig:coordinate-systems}.

\subsubsection{Detected scattering asymmetry}
\label{sec:neutron-polarimeter-polarimetry-asymmetry}

We define NPOL polar and azimuthal scattering angles, denoted
$\theta_{\mathrm{scat}}$ and $\phi_{\mathrm{scat}}$, according to
\begin{subequations}
\label{eq:NPOL-scattering-angles}
\begin{eqnarray}
\sin\theta_{\mathrm{scat}} &=& |\hat{p}_{n} \times \hat{p}'_{n}|, \\
\cos\phi_{\mathrm{scat}} &=& \hat{S} \cdot \hat{u},
\end{eqnarray}
\end{subequations}
where $\hat{p}'_{n}$ is a unit vector along the scattered neutron's
three-momentum, and the unit vector $\hat{u}$ is defined according to
$\hat{u} = (\hat{p}_{n} \times \hat{p}'_{n})/ |\hat{p}_{n} \times
\hat{p}'_{n}|$.

The cross section for elastic polarized-nucleon, unpolarized-nucleon
scattering, denoted
$\sigma(\theta_{\mathrm{scat}},\phi_{\mathrm{scat}})$ for short, is of
the form \cite{wolfenstein52}
\begin{eqnarray}
\sigma(\theta_{\mathrm{scat}},\phi_{\mathrm{scat}}) &=&
\sigma_{0}(\theta_{\mathrm{scat}})\left[1 +
A_{y}(\theta_{\mathrm{scat}}){\bf{P}}\cdot\hat{u}\right] \nonumber \\
&\approx& \sigma_{0}(\theta_{\mathrm{scat}})\left[1 +
A_{y}(\theta_{\mathrm{scat}})P_{S}\cos\phi_{\mathrm{scat}}\right], \nonumber \\
\end{eqnarray}
where $\sigma_{0}(\theta_{\mathrm{scat}})$ and
$A_{y}(\theta_{\mathrm{scat}})$ denote the unpolarized cross section
and the analyzing power, respectively.  The above approximation is
valid in the limit that $P_{N}$ is small.  It is then clear that the
asymmetry, $\xi(\theta_{\mathrm{scat}},\phi_{\mathrm{scat}})$, between
scattering ``up'' ($\hat{S}\cdot\hat{u}<0 \Rightarrow
\cos\phi_{\mathrm{scat}}<0$) and scattering ``down''
($\hat{S}\cdot\hat{u}>0 \Rightarrow \cos\phi_{\mathrm{scat}}>0$) into
infinitesimal solid angles
$(\theta_{\mathrm{scat}},\phi_{\mathrm{scat}})$ and
$(\theta_{\mathrm{scat}},\phi_{\mathrm{scat}}+\pi)$, respectively, for
a particular value of $P_{S}$ is
\begin{eqnarray}
\xi(\theta_{\mathrm{scat}},\phi_{\mathrm{scat}}) &=&
\frac{\sigma(\theta_{\mathrm{scat}},\phi_{\mathrm{scat}}) - 
\sigma(\theta_{\mathrm{scat}},\phi_{\mathrm{scat}}+\pi)}
{\sigma(\theta_{\mathrm{scat}},\phi_{\mathrm{scat}}) +
\sigma(\theta_{\mathrm{scat}},\phi_{\mathrm{scat}}+\pi)} \nonumber \\
&=& A_{y}(\theta_{\mathrm{scat}})P_{S}\cos\phi_{\mathrm{scat}}.
\end{eqnarray}

A single value of $P_{S}$ is not, of course, presented to the
polarimeter.  Also, the top and and bottom rear arrays have a finite
geometry; therefore, if the polarimeter is geometrically symmetric in
$\phi_{\mathrm{scat}}$ (i.e., geometrically symmetric top and bottom
rear arrays), the detected scattering asymmetry (i.e., averaged over
kinematics and the top/bottom finite geometry), $\langle \xi \rangle$,
is
\begin{equation}
\langle \xi \rangle = \langle P_{S} \rangle A_{y}^{\mathrm{eff}},
\end{equation}
where $\langle P_{S} \rangle$ and $A_{y}^{\mathrm{eff}}$ denote,
respectively, the acceptance-averaged value of the sideways component
of the polarization and the polarimeter's effective analyzing power
averaged over its geometric acceptance (i.e., over
$\cos\phi_{\mathrm{scat}}$).  Henceforth, when we refer to the
analyzing power $A_{y}$, it should be understood that we are referring
to $A_{y}^{\mathrm{eff}}$.

\subsection{Charybdis dipole magnet and spin precession}
\label{sec:neutron-polarimeter-charybdis}

The Charybdis magnet was a water-cooled, 38-ton, 1.5-m tall, 2.3-m
wide, and 1.7-m long iron dipole magnet installed in Hall C
specifically for this experiment.  The magnet was configured such that
the gap between the pole pieces was 8.25 inches, and the geometric
center of the magnet was located a distance of 2.107 m from the center
of the target.  The two poles were wired in parallel and powered with
a 160 V-1000 A power supply.  Two-inch thick iron field clamps with
apertures machined to match the 8.25-inch pole gap were placed at the
entrance and exit apertures resulting in an effective magnetic length
of $\sim 1.7$ m.

Calculations of the Charybdis field profile were performed with the
\texttt{TOSCA} program \cite{tosca}
for various currents, and values for the field
integral, $\int|{\bf{B}}|~\mathrm{d}\ell$, along the central axis were
derived from these calculations.  The currents were tuned for the
various spin precession angles, $\chi$, according to the relation
\begin{equation}
\chi = \frac{\mu_{N}g}{\beta_{n}}\int|{\bf{B}}|~\mathrm{d}\ell,
\end{equation}
where $\mu_{N}$ is the nuclear magneton, $g/2 = -1.913$ for the
neutron, and $\beta_{n}$ denotes the neutron's velocity.  The field
integrals for the precession angles at each of our $Q^{2}$ points are
tabulated in Table \ref{tab:field-integrals}.

\begin{table}
\caption{Summary of the nominal values of the field integrals
(along the central axis) for the spin precession angles at each
$Q^{2}$ setting.  $\beta_{n}$ denotes the neutron velocity for the
nominal (central) kinematics.}
\begin{ruledtabular}
\begin{tabular}{cccc}
Central $Q^{2}$& & Precession& $\int|{\bf{B}}|~\mathrm{d}\ell$ \\
$[$(GeV/$c$)$^{2}]$& $\beta_{n}$& Angle $\chi$& [T-m] \\ \hline
0.447& 0.604& $\pm40^{\circ}$& 0.6884 \\
1.136& 0.794& $\pm90^{\circ}$& 2.0394 \\
1.169& 0.799& $\pm40^{\circ}$& 0.9123 \\
1.474& 0.839& $\pm40^{\circ}$& 0.9576 \\
1.474& 0.839& $\pm90^{\circ}$& 2.1547 \\
\end{tabular}
\end{ruledtabular}
\label{tab:field-integrals}
\end{table}

The field along the central axis was mapped \cite{charybdis} at the
conclusion of the experiment.  We found that the values for the field
integrals derived from our mapping results and the \texttt{TOSCA}
calculations agreed to better than 0.76\% for $\chi = \pm40^{\circ}$
precession at $Q^{2} = 0.447$ (GeV/$c$)$^{2}$, 0.21\% for $\chi =
+40^{\circ}$ precession at $Q^{2} = 1.169$ (GeV/$c$)$^{2}$, and 0.35\%
for $\chi = +40^{\circ}$ precession at $Q^{2} = 1.474$
(GeV/$c$)$^{2}$.  Small differences in the measured field integrals
for the two magnet polarities (corresponding to a $\pm0.3^{\circ}$
spread) were observed for $\chi = \pm40^{\circ}$ precession at $Q^{2}
= 0.447$ (GeV/$c$)$^{2}$.  Although we did not conduct field
measurements for both polarities at the other $Q^{2}$ points, it is
reasonable to assume that the magnet behaved similarly for other
current settings.

\begin{figure*}
\includegraphics[scale=0.70,angle=270,clip=]{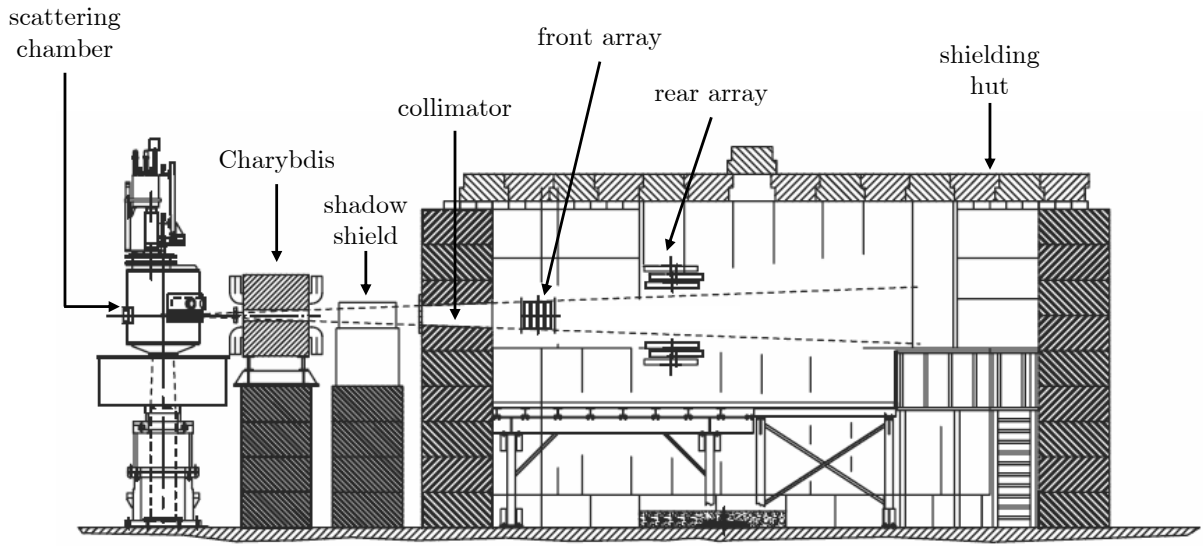}
\caption{Schematic diagram (side view) of the NPOL shielding hut.  The
physical acceptance of the polarimeter, as defined by the collimator,
is indicated by the dashed lines originating in the target.  The rear
array detectors were shielded from a direct line-of-sight to the
target.  The shadow shield, when inserted, was used to assess the room
background rates.}
\label{fig:collimator}
\end{figure*}

\subsection{Neutron polarimeter physical acceptance}
\label{sec:neutron-polarimeter-shielding}


The physical acceptance of the polarimeter was defined by a steel
collimator with entrance and exit apertures located 483.92 cm and
616.00 cm, respectively, from the center of the target.  The
collimator was tapered, with the entrance (exit) port spanning a width
of 72.6 cm (92.4 cm) and a height of 37.3 cm (47.5 cm).  The 10.16-cm
thick lead curtain was located immediately upstream of the
collimator's entrance port.  

A schematic diagram of the polarimeter's shielding hut showing the
shielding of the rear array detectors by the collimator from a direct
line-of-sight to the target appears in Fig.\ \ref{fig:collimator}.

\begin{figure*}
\includegraphics[scale=0.70,angle=270]{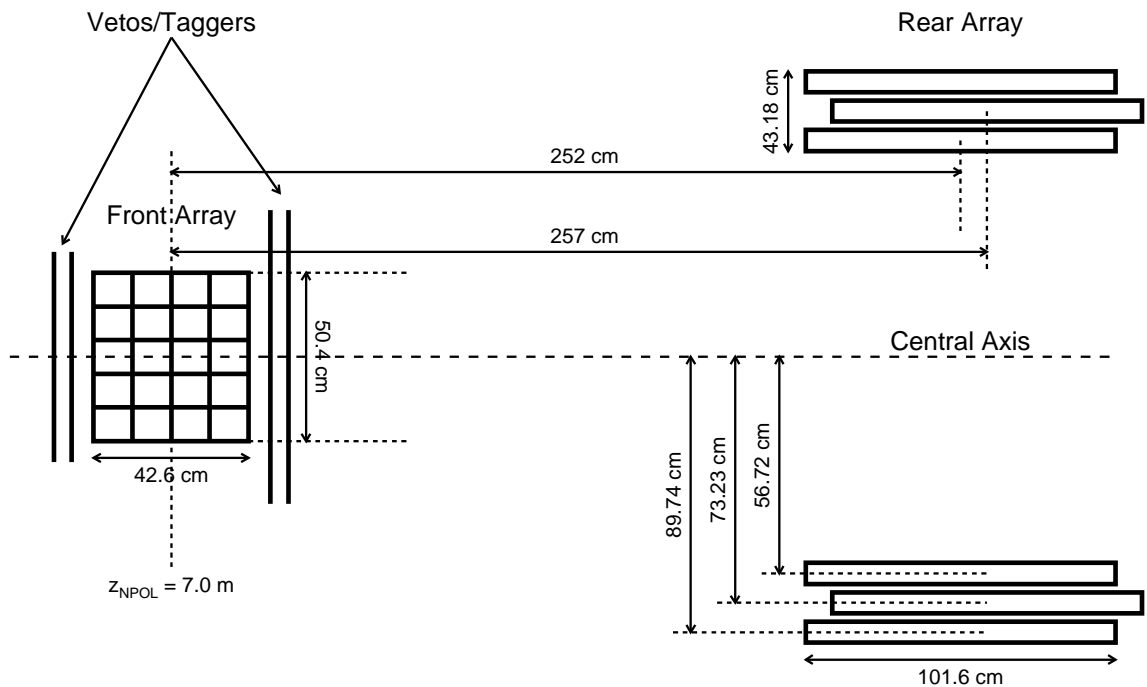}
\caption{Schematic diagram (side view) of the NPOL detector
configuration showing the top and bottom rear subarrays for
measurement of an up-down scattering asymmetry.}
\label{fig:npol-detector-configuration}
\end{figure*}

\subsection{Neutron polarimeter detectors}
\label{sec:neutron-polarimeter-detectors}

The polarimeter consisted of a total of 70 mean-timed BICRON-400
plastic scintillation detectors subdivided into a front veto/tagger
array, a front array, a rear veto/tagger array, and symmetric top and
bottom rear arrays.  The front wall of the polarimeter's shielding hut
was composed of 132.08-cm thick steel blocks; the only opening in this
wall was the lead-shielded collimator.  A schematic diagram of the
polarimeter's detector configuration is shown in Fig.\
\ref{fig:npol-detector-configuration}.

\subsubsection{Front veto/tagger array}
\label{sec:neutron-polarimeter-detectors-front-veto}

The function of the first series of detectors in the neutron flight
path, the front veto/tagger array, was to identify charged particles
incident on the polarimeter.  This veto array consisted of two
vertically-stacked layers of five $160.0 \times 11.0 \times 0.635$
cm$^{3}$ scintillators stacked with their long (160.0 cm) axes
oriented horizontally and perpendicular to the central flight path and
the thin (0.635 cm) dimension oriented along the flight path.  The
vertical spacing between the detectors in each layer was $\sim 1$~mm;
therefore, to eliminate charged particle leakage, the two layers were
offset from each other in the vertical direction by $\sim 1$ cm.  Each
scintillator bar was coupled to two Philips XP2262 2-inch PMTs via
plexiglass light guides.

\subsubsection{Front array}
\label{sec:neutron-polarimeter-detectors-front-array}

The front array was segmented into 20 $100 \times 10 \times 10$
cm$^{3}$ scintillators; segmentation of the front array permitted us
to run with luminosities as high as $3 \times 10^{38}$ cm$^{-2}$
s$^{-1}$ (70 $\mu$A current on a 15-cm liquid deuterium target).  The
long (100 cm) axes of these detectors were oriented horizontally and
perpendicular to the central flight path and were stacked vertically
into four layers of five detectors.  The long ends of each
scintillator were coupled via plexiglass light guides to 2-inch
Hamamatsu R1828-01 PMTs powered by bases designed specifically for
this experiment for purposes of high gain and highly linear output
under conditions of high rate \cite{madey83}.

\subsubsection{Rear veto/tagger array}
\label{sec:neutron-polarimeter-detectors-rear-veto}

Similar to the front veto/tagger array, the purpose of the rear
veto/tagger array was to identify charged particles (e.g., recoil
protons from $np$ interactions in the front array) exiting the front
array.  The detectors in this array were identical to those in the
front veto/tagger array and were vertically stacked in a similar
fashion into two layers of eight detectors each.  [We note that only
one layer of eight detectors existed for the early part of the
experiment during our $Q^{2} = 1.136$ (GeV/$c$)$^{2}$ run).]  As in
the front veto/tagger array, each scintillator was coupled to two
2-inch Philips XP2262 PMTs.

\begin{figure*}
\includegraphics[angle=270,scale=0.65,clip=]{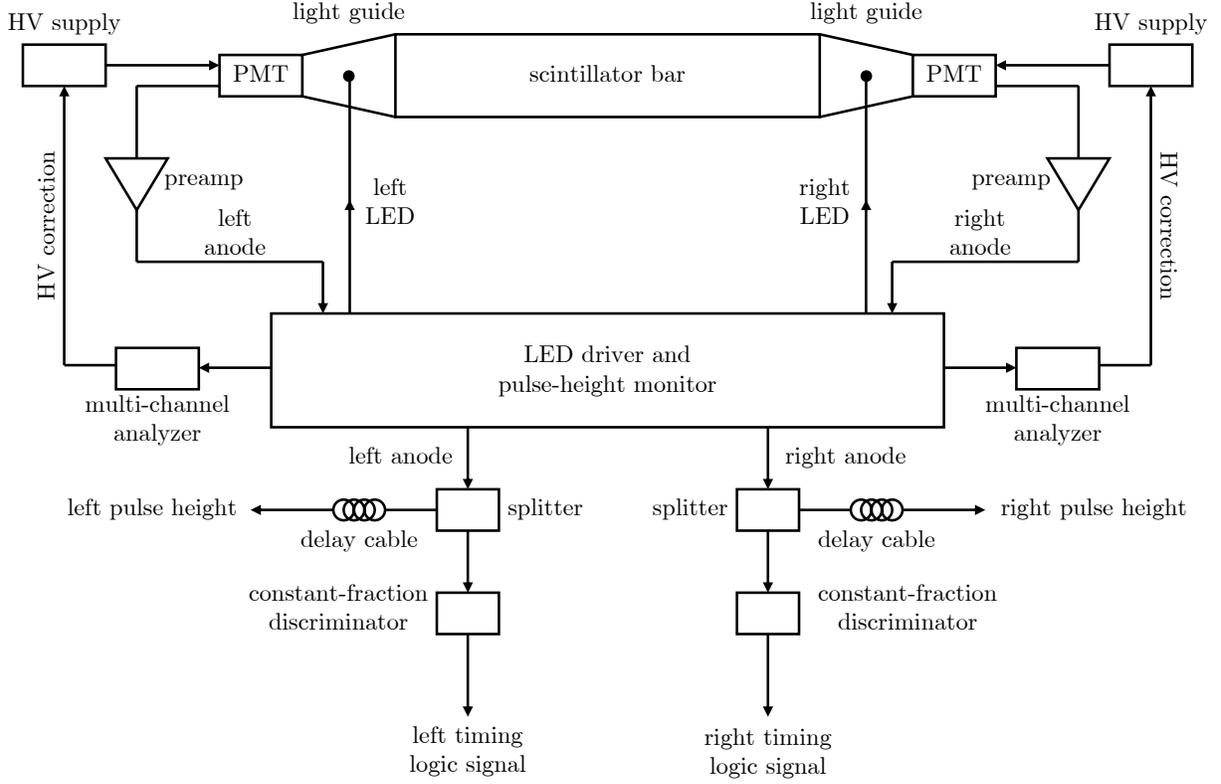}
\caption{Schematic diagram of the configuration of the electronics
in the shielding hut for the front and rear array detectors.  Note
that the anode signals from the rear array detectors were not
preamplified.}
\label{fig:detector-electronics}
\end{figure*}

\subsubsection{Rear array}
\label{sec:neutron-polarimeter-detectors-rear-array}

The top and bottom rear arrays each consisted of twelve detectors
stacked into three layers of four detectors each.  Each layer
contained two ``10-inch'' $25.4 \times 10.16 \times 101.6$ cm$^{3}$
detectors sandwiched in between two larger ``20-inch'' $50.8 \times
10.16 \times 101.6$ cm$^{3}$ detectors.  These detectors were oriented
with their long (101.6 cm) axes parallel to the central flight path
and their 50.8 cm or 25.4 cm dimensions oriented horizontally.  The
centers of the inner, middle, and outer layers were located a vertical
distance of 56.72 cm, 73.23 cm, and 89.74 cm, respectively, above or
below the central axis of the polarimeter and a horizontal distance of
2.52~m, 2.57~m, and 2.52~m, respectively, from the front array
geometric center (see Fig.\ \ref{fig:npol-detector-configuration}).
The long ends of each scintillator were coupled via plexiglass light
guides to 5-inch Hamamatsu R1250 PMTs powered by the same bases built
for the front array.

The vertical positions of the top and bottom arrays relative to the
polarimeter's central axis were optimized for front-to-rear scattering
angles near the peak of the analyzing power for $np$ scattering ($\sim
15^{\circ}$--20$^{\circ}$ for our range of neutron energies).  This
configuration with scattering angles in the vicinity of $\sim
15^{\circ}$--20$^{\circ}$ also guaranteed, for our kinematics, that
only one of the nucleons (for elastic $np$ interactions in the front
array and assuming straight-line trajectories for the recoil proton
through the front array) scattered into either the top or bottom
array.  We also note that the horizontal position of the middle
detector plane was staggered relative to those of the inner and outer
layers so that the majority of the front-to-rear tracks passed through
at least two of the three horizontal planes, reducing the dependence
of the rear array detection efficiency on the scattering angle.

\subsection{Electronics, event logic, and data acquisition}
\label{sec:neutron-polarimeter-electronics}

\subsubsection{Electronics}
\label{sec:neutron-polarimeter-electronics-electronics}

The signals from the 140 NPOL PMTs were processed with electronics
sited in two locations: (1) one set, located inside the shielding hut
was used to form the timing logic signal for each PMT (past experience
with neutron time-of-flight and polarimetry experiments
\cite{eden94nim} revealed that locating the discriminators as close to
the PMTs as practical yielded the best timing resolution); and (2)
another set, located in the counting house, was used to define the
logic for the various event types.

A schematic diagram of the configuration of the electronics in the
shielding hut for each scintillator bar in the front and rear arrays
is shown in Fig.\ \ref{fig:detector-electronics}.  High voltage was
applied to each PMT remotely by an EPICS-controlled 64-channel
high-voltage CAEN mainframe crate located in the counting house.
Modest levels of high voltage were applied to the PMTs for the front
array detectors, as deterioration in the performance of these PMTs was
of concern because of the high count rates in these scintillators;
however, no deterioriation in their performance was observed during
the experiment (instead, gains were stable to within $\sim 10$\%).  To
compensate for the resulting lower levels of gain obtained directly
from these PMTs, the anode signals were preamplified by fast
preamplifiers with a gain of eight, custom-designed and assembled for
this experiment.  The anode signals from the PMTs in the rear array
and the front and rear veto/tagger arrays were not preamplified.

The anode signals from the front and rear arrays were then directed to
an LED driver and pulse height monitor.  When desired, this device was
used to assess the response of each PMT to a flashing blue LED mounted
on its light guide.  The centroid channels of the LED spectra were
monitored periodically, and any necessary changes to the high voltage
levels were performed remotely.  The gains of the front and rear
veto/tagger array PMTs were not monitored with this system.

The anode signals from all four detector arrays were then split.  The
signals in the fast branch (for the event trigger and timing
measurements) were directed to either constant-fraction discriminators
(front and rear arrays) or leading-edge discriminators (front and rear
veto/tagger arrays) located inside the shielding hut and then sent to
the electronics in the counting house.  We did not employ
constant-fraction discrimination for the veto/tagger array detectors
for the following reasons: (1) the dynamic range of energy deposition
in these detectors was small for those events of interest, so the
time-walk was tolerable; and (2) the timing measurements from these
detectors were not used for energy determinations, so resolutions of a
few ns were sufficient for charged particle tagging.  Those signals
diverted to the slow branch were routed through delays located inside
the shielding hut and then sent to the counting house.

Upon arrival in the counting house, both the analog and timing signals
were directed through filters/transformers designed to eliminate
low-frequency noise.  The analog signals were then sent directly to
ADCs, while the timing signals were first sent to discriminators and
then routed to two branches of a timing circuit.  In one branch, the
output from these discriminators were directed through level
translators, delays, discriminators, and then further split and
directed to TDCs and scalers.  In the other branch of this timing
circuit (used to form the event triggers), the timing signals from the
PMTs on all of the detectors, except those in the rear veto/tagger
array, were first sent to logic modules which were used to generate
logic signals for coincidences between the timing signals for the two
PMTs on each detector.  Logical \texttt{OR}s were generated for each
of the twenty front array detector two-PMT coincidences.  These
signals were then sent to a fan-in with one set of outputs directed to
scalers and the other through a discriminator; the output from this
discriminator was then directed to the trigger circuit.  The logical
\texttt{OR}s for the rear array detectors and the front
veto/tagger-array detectors were routed through a fan-in and then
directed to the trigger circuit.  The timing signals from the rear
veto/tagger-array detectors were not used for trigger purposes.

\subsubsection{Event logic and triggers}
\label{sec:neutron-polarimeter-electronics-logic}

All event trigger logic was performed by two LeCroy 8LM 2365 Octal
Logic Matrix modules.  Pretrigger logic signals from the HMS
(coincident hits in at least three of the four hodoscope planes), the
NPOL front array, the NPOL rear array, and the NPOL front veto/tagger
array were routed to the 8LM modules.  In addition to these logic
signals, triggers from the polarized electron source were also input
to these modules.  As previously discussed, the helicity of the
electron beam was flipped pseudorandomly at 30 Hz.  Electronics at the
polarized source generated a logic signal for readout of
helicity-gated scalers for each 33.3 ms helicity window.  Further,
these modules also generated a helicity-transition logic signal which
was used to veto otherwise valid data triggers that occured during
transitions at the polarized source from one helicity state to
another.  The duration of this helicity-transition logic pulse was
$\sim 600$ $\mu$s, resulting in an effective data-taking helicity
window of $\sim 32.7$~ms.

An electronic module known as the Trigger Supervisor (TS) functioned
as the interface between the 8LM logic modules and the data
acquisition system (DAQ).  The TS generated a logic signal indicating
the status of the DAQ (e.g., busy or not busy) that was input to the
logic modules.  The logic modules then determined whether the logic
for any of the eight possible physics triggers (e.g., electron
singles, electron/front array coincidences, electron/front array/rear
array coincidences, etc.) was satisfied.  If the logic for any
particular trigger was satisfied, the TS generated an accept signal
leading to generation of the appropriate ADC gate and TDC common
signals.  The ADCs, TDCs, and scalers were then read out with
real-time \texttt{UNIX}-based processors.

The event triggers of interest were three-fold coincidences between
hits in the electron arm, the front array, and the rear array.  These
events constituted $\sim 80$--85\% of the event triggers, as the
higher rate events, such as electron singles or two-fold coincidences
between the electron arm and the front array, were prescaled.


\subsubsection{Data acquisition}
\label{sec:neutron-polarimeter-electronics-daq}

The DAQ was controlled by the CEBAF Online Data Acquisition System
(CODA) \cite{jlabcoda}.  CODA includes an event-builder subsystem
programmed to assemble the individual ADC channel, TDC channel, and
scaler read-out data fragments into an event.  The data for the events
were then written to disk in CODA format by another subsystem.

Typical data acquisition rates were one million events in $\sim 1.0$
($\sim 0.5$) hours with the Charybdis dipole field energized
(de-energized).

\section{Data analysis}
\label{sec:data-analysis}

\subsection{Electron reconstruction and tracking}
\label{sec:data-analysis-electron}

\subsubsection{Overview of analysis code}
\label{sec:data-analysis-electron-code}

The raw ADC, TDC, and scaler data written to disk and encoded by the
DAQ in CODA format were decoded with a modified version of the
standard Hall C \texttt{ENGINE} analysis code (see, e.g.,
\cite{arrington98} for a discussion of the standard version) employed
for the analysis of nearly all experiments conducted in Hall C.
Modifications to the standard version were necessary to accommodate
the raw data stream from the 70 NPOL detectors; hereafter, whenever we
refer to the \texttt{ENGINE} analysis code, it should be assumed that
we are referring to our modified version of this code.

For each event, the scattered electron's track through the HMS was
reconstructed, and various kinematic quantities (e.g., momentum,
energy, focal plane distributions, etc.) were computed.
\texttt{ENGINE} was not configured to reconstruct the track of the
nucleon through the polarimeter; instead, the NPOL detector data were
simply written to new data files for later processing by other
analysis tools.

\subsubsection{Extraction of electron information}
\label{sec:data-analysis-electron-electroninfo}

\paragraph{Tracking}
The overall strategy of the tracking algorithm \cite{arrington98} was
to use the hit information from the drift chambers and reference start
times provided by TDC information from the scintillators in the
hodoscope planes to reconstruct the trajectory of the particle through
the drift chambers.  TDC information from those scintillators in the
hodoscope planes recording hits was used to establish reference start
times.  This information, coupled with TDC information from the drift
chambers, was then used to determine the location of the hit in the
drift chamber planes.  ``Left-right ambiguities" in the drift chambers
(i.e., whether a particle passed to the left or right of any given
wire) were resolved by fitting a (straight-line) track to each
left-right hit combination in the six planes of each drift chamber.
The full track through both drift chambers with the overall smallest
track reconstruction $\chi^{2}$ was defined to be the final
reconstructed track through the drift chamber planes.

\begin{figure*}
\includegraphics[angle=270,scale=0.70]{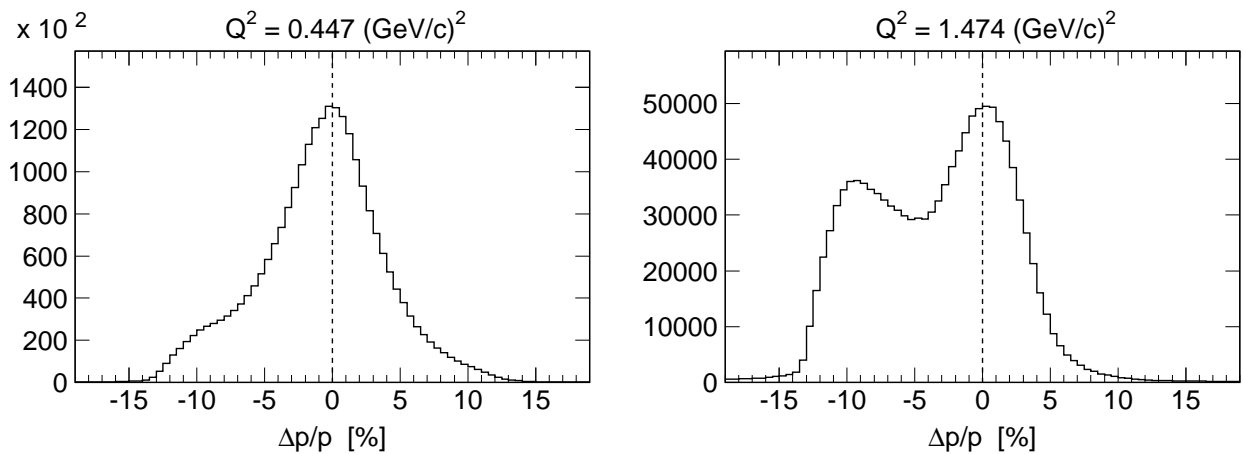}
\caption{Distributions of $\Delta p/p$ for the full HMS acceptance at
 $Q^{2} = 0.447$ and 1.474 (GeV/$c$)$^{2}$.}
\label{fig:hms-raw-delta}
\end{figure*}

\paragraph{Transport}
\texttt{ENGINE} then attempted to relate the positions and angles at
the focal plane (determined from the track through the drift chambers)
to target quantities.  In standard coordinate notation for transport
through a spectrometer, $\hat{z}_{\mathrm{fp}}$ is taken to point
along the central ray of the spectrometer, $\hat{x}_{\mathrm{fp}}$ in
the dispersive direction (by convention, taken to point ``downwards"),
and $\hat{y}_{\mathrm{fp}} = \hat{z}_{\mathrm{fp}} \times
\hat{x}_{\mathrm{fp}}$.  It should be noted that HMS focal plane
variables are traditionally referred to the detector focal plane,
defined to be perpendicular to the central ray (i.e., parallel to the
drift chamber planes) with the origin of the
$x_{\mathrm{fp}}$-$y_{\mathrm{fp}}$ plane defined to be that point in
space where the central ray of the spectrometer intersects the true
(magnetic) focal plane.  In addition to the dispersive and
non-dispersive variables, two other standard transport variables,
$x'_{\mathrm{fp}}$ and $y'_{\mathrm{fp}}$, are defined to be the
slopes of the rays at the focal plane, $x'_{\mathrm{fp}} \equiv
\mathrm{d}x_{\mathrm{fp}}/\mathrm{d}z$ and $y'_{\mathrm{fp}} \equiv
\mathrm{d}y_{\mathrm{fp}}/\mathrm{d}z$, respectively.  The focal plane
variables $x_{\mathrm{fp}}$, $y_{\mathrm{fp}}$, $x'_{\mathrm{fp}}$,
and $y'_{\mathrm{fp}}$ were converted to target quantities
$x'_{\mathrm{tar}} \equiv \mathrm{d}x_{\mathrm{tar}}/\mathrm{d}z$,
$y_{\mathrm{tar}}$, $y'_{\mathrm{tar}} \equiv
\mathrm{d}y_{\mathrm{tar}}/\mathrm{d}z$, and $\delta \equiv
(|{\bf{p}}_{e'}| -
|\overline{{\bf{p}}_{e'}}|)/|\overline{{\bf{p}}_{e'}}|$, where
$|\overline{{\bf{p}}_{e'}}|$ denotes the central momentum setting, via
computation of transport matrix elements derived from optics studies.
For this choice of target coordinates, $x_{\mathrm{tar}}$ was not
reconstructed but was, instead, defined to be $x_{\mathrm{tar}} = 0$
for all events.

\subsubsection{Sample electron reconstruction results}
\label{sec:data-analysis-electron-results}

\begin{figure}
\includegraphics[angle=270,scale=0.70]{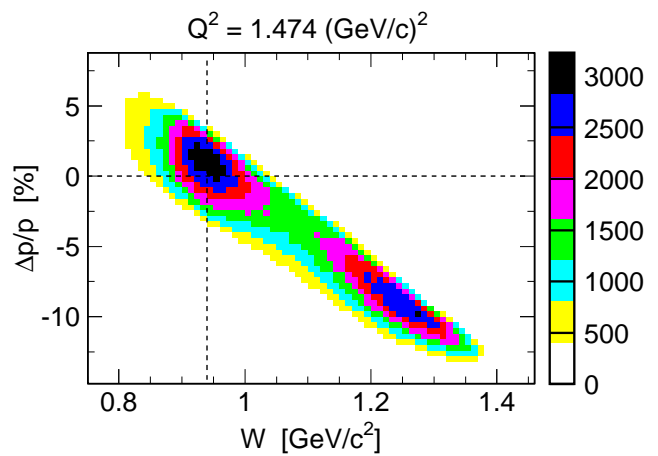}
\caption{(Color online) Correlation plot of $\Delta p/p$ versus $W$
for the full HMS acceptance at $Q^{2} = 1.474$ (GeV/$c$)$^{2}$.}
\label{fig:hms-raw-delta-vs-w}
\end{figure}

Sample histograms of the reconstructed $\delta$-distribution,
hereafter referred to as the ``$\Delta p / p$-distribution", at our
lowest and highest $Q^{2}$ points are shown in Fig.\
\ref{fig:hms-raw-delta}.  The quasielastic peak is clearly visible in
both spectra, but a large accompanying background of inelastic events
associated with pion-production in the target is present in the $Q^{2}
= 1.474$ (GeV/$c$)$^{2}$ spectrum.  Inelastic peaks were also clearly
visible in the $Q^{2} = 1.136$ and 1.169 (GeV/$c$)$^{2}$ spectra but
are not shown here.  A sample two-dimensional histogram of $\Delta p /
p$ plotted versus the invariant mass, $W$, calculated from the
electron kinematics according to
\begin{equation}
W = \sqrt{(\omega + m_{N})^{2} - |{\bf{q}}|^{2}},
\end{equation}
where $m_{N}$ is the nucleon mass, is shown in Fig.\
\ref{fig:hms-raw-delta-vs-w} for our $Q^{2} = 1.474$ (GeV/$c$)$^{2}$
point.  The $\Delta(1232)$ resonance is prominent in this
distribution.

\begin{figure}
\includegraphics[angle=270,scale=0.70]{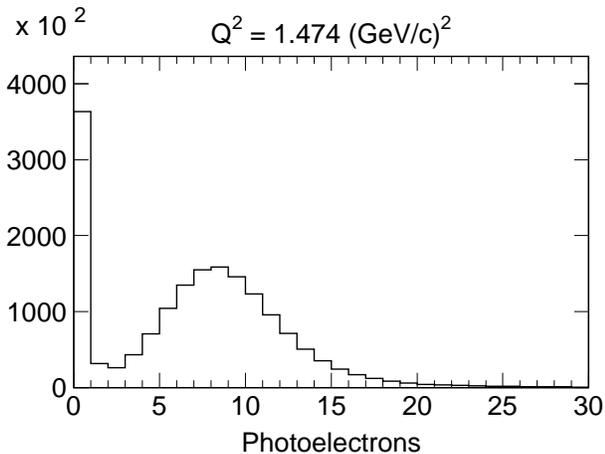}
\caption{\v{C}erenkov photoelectron spectrum for the full HMS
acceptance at $Q^{2} = 1.474$ (GeV/$c$)$^{2}$.  A prominent hadron
peak appears at zero photoelectrons.}
\label{fig:hms-raw-cerenkov}
\end{figure}

Hadrons in the HMS were identified via examination of the \v{C}erenkov
photoelectron spectrum.  As expected, a hadron peak was not visible in
the $Q^{2} = 0.447$ (GeV/$c$)$^{2}$ spectrum; however, prominent
hadron peaks (at zero photoelectrons) were observed at the three
higher $Q^{2}$ settings.  An example of such a photoelectron spectrum
from our $Q^{2} = 1.474$ (GeV/$c$)$^{2}$ data is shown in Fig.\
\ref{fig:hms-raw-cerenkov}.  Cuts on the number of photoelectrons,
coupled with cuts on the energy deposition in the calorimeter, were
sufficient for electron-hadron discrimination.

\subsection{Neutron polarimeter energy calibration}
\label{sec:data-analysis-energycalibration}

The (charge-integrating) ADCs for the front and rear array detector
PMTs were calibrated with the Compton spectra from a $^{228}$Th source
(2.61 MeV $\gamma$-rays); the front and rear veto/tagger array
detectors were not calibrated as ADC information was not used for
charged particle tagging.  These calibrations were parametrized in
terms of an equivalent electron energy (denoted ``eVee''), where the
relation between the light output of recoil protons and
Compton-scattered electrons in organic scintillator was found by Madey
\textit{et al}.\ \cite{madey78} to be well described by the
parametrization
\begin{equation}
T_{e} = a_{1}\left[1 - \exp(-a_{2}\left(T_{p}\right)^{a_{3}}\right] +
a_{4}T_{p}.
\end{equation}
Here, $T_{p}$ denotes the energy deposition of a recoil proton,
$T_{e}$ denotes the energy deposition of an electron that yields the
equivalent light output, and the $a_{i}$ are empirically determined
parameters.

Unfortunately, the range of electron energies (2.38 MeV Compton edge)
was not sufficient, as typical energy depositions for the recoil
protons were estimated to be $\agt$ several MeVee \cite{plaster03};
further, the hardware thresholds for the front (rear) array detectors
were set at 4 (10) MeVee.  To remedy these shortcomings, a
custom-designed linear amplifier with a gain of ten was placed in the
timing circuit during calibration runs.  The resulting ADC spectra
were fitted to the sum of the Klein-Nishina distribution (smeared by a
Gaussian resolution function) and an exponential background tail.
Pulse-height calibrations were performed at three different times
during the experiment (roughly at the start, middle, and conclusion);
minor differences ($\sim 10$\%) in the extracted calibration
parameters were observed but were deemed to be relatively unimportant
as the selection of quasielastic $^{2}$H$(e,e'n)^{1}$H events did not
rely heavily on pulse height information.

\subsection{Neutron polarimeter timing calibration}
\label{sec:data-analysis-timingcalibration}

To optimize track reconstruction and background rejection in the
neutron polarimeter, the relative timing relationships between the
NPOL detectors and the HMS were carefully calibrated with a series of
algorithms designed to: (1) generate position calibrations for each
detector; (2) generate relative timing calibrations for each detector
in the front array and discern the relationship between the mean time
for each front array detector and the trigger mean time; (3) calibrate
the timing between the HMS and the front array (yielding a coincidence
time-of-flight); (4) generate relative timing calibrations for each
detector in the rear array and calibrate the time-of-flight between
the front array and the rear array; and (5) generate position and
timing calibrations for the front and rear veto/tagger detectors.

\subsubsection{Front and rear array position calibrations}
\label{sec:data-analysis-timingcalibration-position}

The position calibration algorithm for the front and rear array
detectors employed data acquired with the Charybdis magnet
de-energized, such that charged particles illuminated the front array
almost uniformly.  The relationship between the hit position and the
difference (in channels) between the TDCs from the PMTs mounted on the
two ends of each scintillator was parametrized in a linear form with
an unknown slope and offset.  Histograms of these TDC channel
differences were accumulated for each detector and then
boxcar-smoothed.  The algorithm identified the channel of maximum
content and then scanned away in both directions until channels with
10\% of the maximum content were identified.  Slope and offset
parameters were then chosen such that these 10\%-content channels were
aligned with the physical edges of each detector; the resulting
calibrated position spectra displayed sharp edges near the physical
detector edges.

\subsubsection{Front-array timing and trigger calibrations}
\label{sec:data-analysis-timingcalibration-fronttiming}

The first goal of the front-array timing calibration was to align the
mean times of all the detectors in the front array using events with a
single hit in the front array.  Data acquired with the Charybdis
magnet energized (for suppression of background processes) were
employed for this step of the timing calibration, and events with $>0$
($>1$) hits in the front veto/tagger array (front array) were
discarded.  An offset was chosen for each detector such that the mean
value of its mean-time spectrum was aligned on zero.

The second goal of the front array timing calibration was to construct
a variable that could be used to identify which hit generated the
trigger (for events with multiple front array hits), as the trigger
circuit did not identify the triggering hit.  Proper identification of
the triggering hit via examination of the correlation between the TDC
channels for the two PMTs on each detector and the position dependence
of the mean times yielded self-timing spectra with FWHM of $\sim 0.4$
ns.

\subsubsection{Coincidence time-of-flight calibrations}
\label{sec:data-analysis-timingcalibration-ctof}

To maximize our signal-to-noise ratio, we constructed a coincidence
time-of-flight variable that accounted for the quasielastic
$^{2}$H$(e,e'n)^{1}$H kinematics, pathlength variations through the
HMS and NPOL, and variations in the delay between an interaction in a
detector and the arrival of its timing signal at the TDC.  For this
step of the calibration, a minimal set of cuts were applied to the
data for purposes of (loose) quasielastic event selection (e.g., cuts
on the calorimeter energy deposition, $\Delta p/p$, etc.).  Again,
front array single-hit events (with no hits in the front veto/tagger
array) acquired with the Charybdis magnet energized were used for this
step of the calibration.

The algorithm first predicted the neutron time-of-flight from the
target to the front array using only position information (i.e., the
reconstructed vertex information for the primary scattering event in
the target cell and the position of the front array hit) and electron
kinematics.  For a three-body final state (i.e., no pion production),
four-momentum conservation demands
\begin{subequations}
\label{eq:quasielastic-momentum-conservation}
\begin{eqnarray}
m_{d} + \omega &=& \sqrt{|{\bf{p}}_{n}|^{2} + m_{n}^{2}} +
  \sqrt{|{\bf{p}}_{p}|^{2} + m_{p}^{2}}, \\
{\bf{q}} &=& {\bf{p}}_{n} + {\bf{p}}_{p}.
\end{eqnarray}
\end{subequations}
From this, it follows that a value for $|{\bf{p}}_{n}|$ (and, then,
the predicted neutron time-of-flight) can be derived from the solution
to the quadratic equation $A|{\bf{p}}_{n}|^{2} + B|{\bf{p}}_{n}| + C =
0$, where
\begin{subequations}
\label{eq:quasielastic-quadratic}
\begin{eqnarray}
A &=& (m_{d} + \omega)^{2} - ({\bf{q}} \cdot \hat{p}_{n})^{2}, \\
B &=& -2({\bf{q}} \cdot \hat{p}_{n})D, \\
C &=& m_{n}^{2}(m_{d} + \omega)^{2} - D^{2}, \\
2D &=& m_{d}^{2} + m_{n}^{2} - m_{p}^{2} - Q^{2} +
  2m_{d}\omega.
\end{eqnarray} 
\end{subequations}

A value for the actual measured time-of-flight was then extracted from
information in the signal output of a TDC started by a signal
generated by the NPOL trigger and stopped by the HMS trigger, a
correction for pathlength variations and delays between interactions
and signals in the HMS computed by \texttt{ENGINE}, and the mean time
of the front array detector recording the hit.  This measured
time-of-flight was then compared with the predicted time-of-flight,
and the resulting difference, the coincidence time-of-flight
(hereafter, referred to as cTOF), was computed for each event.  The
resulting cTOF spectra were fairly narrow with FWHM of $\sim 1.25$ ns
and signal-to-noise ratios of $\sim 6$:1--10:1.  Sample cTOF spectra
are shown later in this paper.

\subsubsection{Rear-array timing calibrations}
\label{sec:data-analysis-timingcalibration-reartiming}

The algorithm for the rear-array timing calibration selected
single-hit events (with no hits in both the front and rear veto/tagger
arrays) acquired with the Charybdis magnet energized and then filtered
these hits according to a set of cuts designed to select quasielastic
events.  In addition, a $|\mathrm{cTOF}| \leq 2$ ns cut was enforced.

In the first step, the algorithm aligned the mean time spectra of the
rear array detectors relative to each other.  As for the front array,
histograms of mean times were accumulated for each detector.  The
channel of maximum content was identified, and an offset parameter for
each detector was then chosen such that the peak channel was aligned
on zero.

In the second step, the algorithm performed an absolute timing
calibration of the rear array detectors relative to the front array
detectors via a front-to-rear velocity calibration.  The scattering
angle for the front-to-rear track was computed using the incident
neutron's three-momentum and the position information for the hits in
the front and rear array.  The algorithm then predicted the
front-to-rear velocity for elastic $np$ scattering in the front array
via computation of the scattered neutron's kinetic energy, $T_{np}$,
where
\begin{equation}
T_{np} = \frac{2T_{n}\cos^{2}\theta_{\mathrm{scat}}}
{(\gamma_{n}+1) - (\gamma_{n}-1)\cos^{2}\theta_{\mathrm{scat}}}.
\label{eq:Tnp}
\end{equation}
Here, $T_{n}$ denotes the incident neutron's kinetic energy,
$\theta_{\mathrm{scat}}$ denotes the neutron scattering angle in the
polarimeter, $\gamma_{n}$ is the usual Lorentz factor for the incident
neutron, and the proton and neutron masses are assumed to be equal.
Relative time-of-flight (hereafter, referred to as rTOF) histograms,
defined to be the difference between the predicted and measured values
of the front-to-rear time-of-flight, were accumulated, and offsets
were then chosen for each detector such that the peak channel was
aligned on zero.  Again, sample rTOF spectra are shown later in this
paper.

\subsubsection{Front and rear veto/tagger-array calibrations}
\label{sec:data-analysis-timingcalibration-vetos}

The position and timing calibration of the front and rear
veto/tagger-array detectors consisted of three steps.  Data for
charged particle tracks acquired with the Charybdis magnet
de-energized were employed for this calibration; hits were required in
each layer of the front veto/tagger array, the front array, and the
rear veto/tagger array.

First, as leading-edge discrimination was employed for these
detectors, the algorithm began by computing corrections for walk.  The
relationship between the observed TDC and ADC channels,
$\texttt{TDC}_{\mathrm{obs}}$ and $\texttt{ADC}_{\mathrm{obs}}$, was
parametrized as $\texttt{TDC}_{\mathrm{obs}} = \texttt{TDC} +
\gamma\log(\texttt{ADC}_{\mathrm{obs}}/\texttt{ADC}_{\mathrm{peak}})$,
where $\texttt{TDC}$ denotes the TDC channel in the absence of walk
effects, $\gamma$ is an empirical parameter, and
$\texttt{ADC}_{\mathrm{peak}}$ denotes the peak ADC channel.  A value
for $\gamma$ was then computed via the method of least squares.

Second, the veto/tagger array detectors were position calibrated using
a different algorithm than that employed for the position calibration
of the front and rear array detectors due to the facts that the
collimator partly obscured the edges of the front veto/tagger array
detectors and that the outer rear veto/tagger array detectors did not
receive adequate illumination from front-to-rear charged tracks.  (The
front and rear veto/tagger arrays were designed to provide more than
adequate coverage of target-to-front and front-to-rear charged
tracks.)  As such, position calibration parameters for these detectors
were deduced via a comparison of the recorded hit position with the
nearest hit position in the front array, and offset parameters were
determined via a $\chi^{2}$ minimization of the difference between the
predicted and recorded hit positions.  To improve the statistics for
the outer rear array veto/tagger detectors, the algorithm searched for
$(n,p)$ charge-exchange events in the front array.  Tracks from these
events were used to predict hit locations in the rear veto/tagger
array detectors, and position calibration parameters were then deduced
from another $\chi^{2}$ minimization of the difference between the
predicted and recorded hit positions.  The resulting calibrated
position spectra were well aligned about the physical center of each
detector with somewhat more rounded spectra than observed in the front
and rear array spectra due to the use of leading-edge discrimination.

Last, the mean times were aligned relative to each other via the same
procedure employed for the mean-time calibration of all the other
detectors.

\subsection{Nucleon reconstruction and tracking}
\label{sec:data-analysis-neutron}

\begin{figure*}[t]
\includegraphics[angle=270,scale=0.65]{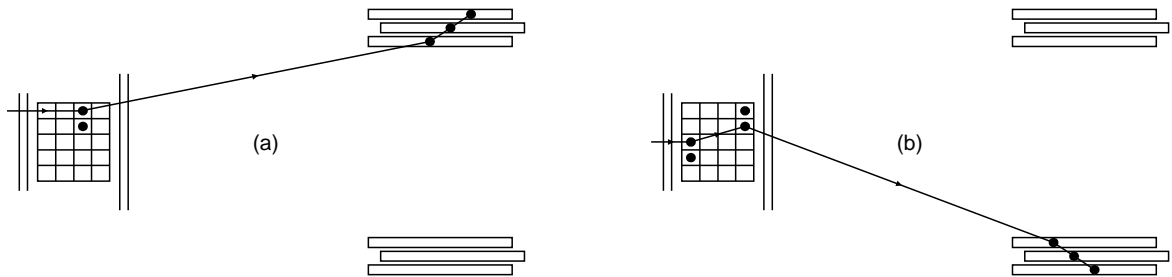}
\caption{Examples of reconstructed tracks for: (a) an event with a
single cluster in the front array, no missing layers, and multiple
hits in the top rear array; and (b) an event with two clusters in the
front array (separated by two missing layers) and multiple hits in the
bottom rear array.}
\label{fig:track-reconstruction}
\end{figure*}

\subsubsection{Overview of analysis code}
\label{sec:data-analysis-neutron-codeoverview}

The algorithm we developed for reconstruction and tracking in the
neutron polarimeter began by translating the raw NPOL detector data
decoded by \texttt{ENGINE} into hit positions and times.  The code
then attempted to determine which hit in the front array generated the
trigger.  All hits were then filtered according to a number of
different selection criteria, with the surviving hits grouped into
recognizable patterns.  The code then attempted to determine the
primary hits in the front and rear arrays and the charges of the
incident particle and the particle detected in the rear array.
Finally, kinematic quantities and time-of-flight variables were then
computed for those events satisfying all tracking criteria.

\subsubsection{Trigger selection and hit filtering}
\label{sec:data-analysis-neutron-triggerfiltering}

The algorithm assigned the location of the triggering front array hit
to the detector with the smallest absolute self timing value.  All
hits were then filtered according to a number of selection criteria
designed to discard hits with unphysical reconstructed detector
positions or mean times falling outside of specified windows.  These
mean time windows were chosen sufficiently wide for purposes of
quasielastic event selection, elastic/quasielastic scattering in the
front array, and charged particle tagging in the veto/tagger arrays.
In particular, the mean-time windows for both the front and rear
veto/tagger arrays safely bracketed the entire peak regions with the
borders extending into the regions of flat background.

\subsubsection{Pattern grouping and track reconstruction}
\label{sec:data-analysis-neutron-patternreconstruction}

\paragraph{Incomplete and simple events}
The algorithm began by identifying incomplete and simple events.
First, events with either no surviving hits in the front and/or rear
array or events with hits in both the top and bottom rear array were
discarded.  Second, simple events with exactly one hit in the front
array, one hit in the rear array, and no hits in both the front and
rear veto/tagger arrays were identified.  For these events, the
incident particle and the particle detected in the rear array were,
obviously, designated neutral particles, and reconstruction of the
track was deemed complete.

\paragraph{Multiple hit events}
The majority of the events were more complicated than these simple
events because of propagation of the recoil protons through adjacent
scintillator bars or multiple scattering of the neutron.  For these
more complicated events, the code began by identifying which layer in
the front array (i.e., first, second, third, or fourth) was hit first;
henceforth, we will refer to the hit(s) in this layer as the ``first
cluster".  If the first cluster contained more than one hit, the
(vertically) highest and lowest hits were identified; such hit
patterns were assumed to be the result of an $np$ or $pp$ interaction
in one detector followed by the penetration of the recoil proton into
a vertically adjacent detector.  Accordingly, if the hits occurred in
non-contiguous detectors within the same vertical layer (i.e.,
existence of a vertical ``gap"), the event was discarded.

The code then searched for evidence of one or more ``missing layers"
in the front array (e.g., an event with hits in the first layer and
the fourth layer); a missing layer was taken to be evidence for
multiple scattering of the incident neutron.  If such a ``second
cluster" of hits was not found, the location of the front array
scattering vertex was assigned to the highest (lowest) hit in the
first cluster if the top (bottom) rear array recorded one or more
hits.  If, instead, a second cluster of hits was found, the code
determined whether the second cluster contained a gap; again, events
with gaps in the second cluster were discarded.  The algorithm then
attempted to discern whether the second cluster was located above or
below the first cluster; if the second cluster was above (below) the
first cluster, the location of the first cluster scattering vertex was
assigned to the highest (lowest) hit in the first cluster.  Then, if
the top (bottom) rear array was hit, the location of the second
cluster scattering vertex was assigned to the highest (lowest) hit in
the second cluster.  Finally, if more than one hit was recorded in
either the top or bottom rear array, the rear array scattering vertex
was assigned to that hit closest in distance to the final front array
scattering vertex.

Illustrative examples of two possible types of reconstructed tracks
are shown in Fig.\ \ref{fig:track-reconstruction}.  We note here, and
discuss later in Section \ref{sec:data-analysis-datasets-cuts}, that
events with a ``second cluster'' were reconstructed but were not used
in our extraction of scattering asymmetries.

\subsubsection{Charge identification}
\label{sec:data-analysis-neutron-chargeidentification}

After the track through the front and rear arrays was reconstructed,
the code then checked for hits in the veto/tagger arrays.  The charge
of the incident particle was determined via the following algorithm.
(1) If there were no hits in any of the front veto/tagger detectors,
the particle was designated a neutral particle.  (2) If there were
hits in the front veto/tagger detectors, the radial distance between
the location of the veto/tagger hit and the location of the first
scattering vertex was computed according to $d =
\sqrt{(x_{\mathrm{vt}}-x_{\mathrm{fr}})^{2} +
(y_{\mathrm{vt}}-y_{\mathrm{fr}})^{2}}$, where the coordinates refer
to the polarimeter basis, defined in
Eq.~(\ref{eq:polarimeter-basis-definition}).  If at least one hit in
each veto/tagger layer satisfied $d \leq 30$ cm, the incident particle
was designated a charged particle.  If no hits in either veto/tagger
layer satisfied $d \leq 30$ cm, the incident particle was designated a
neutral particle.  Finally, if a hit in one of the front/veto tagger
layers satisfied this distance requirement but no hits in the other
layer satisfied this condition, the charge of the incident particle
was declared to be ambiguous.

The algorithm for the determination of the charge of the particle
detected in the rear array was essentially identical to that described
above.  The only difference was that the code predicted where the
hits in the rear veto/tagger arrays should have occurred assuming
a straight-line trajectory from the final front array scattering
vertex to the rear array scattering vertex.  The computed value of the
radial distance between the location of the actual hit and the predicted
hit was then used, in an identical manner, for rear array
neutral/charged tagging.

The choice of the 30-cm radial track-distance threshold was based on
an examination of track-distance spectra for the front and rear
veto/tagger arrays.  The spectra for the front veto/tagger array were
found to be relatively narrow with an abrupt change in slope around 30
cm, believed to be related to these scintillators' position
resolution.  The spectra for the rear veto/tagger array did not
contain such a feature as the recoil protons arising from interactions
in the front array were widely distributed in angle; nevertheless, the
same 30-cm condition was employed as the position resolutions for
these detectors were similar to those in the front veto/tagger array.

\begin{figure*}[t]
\includegraphics[angle=270,scale=0.70]{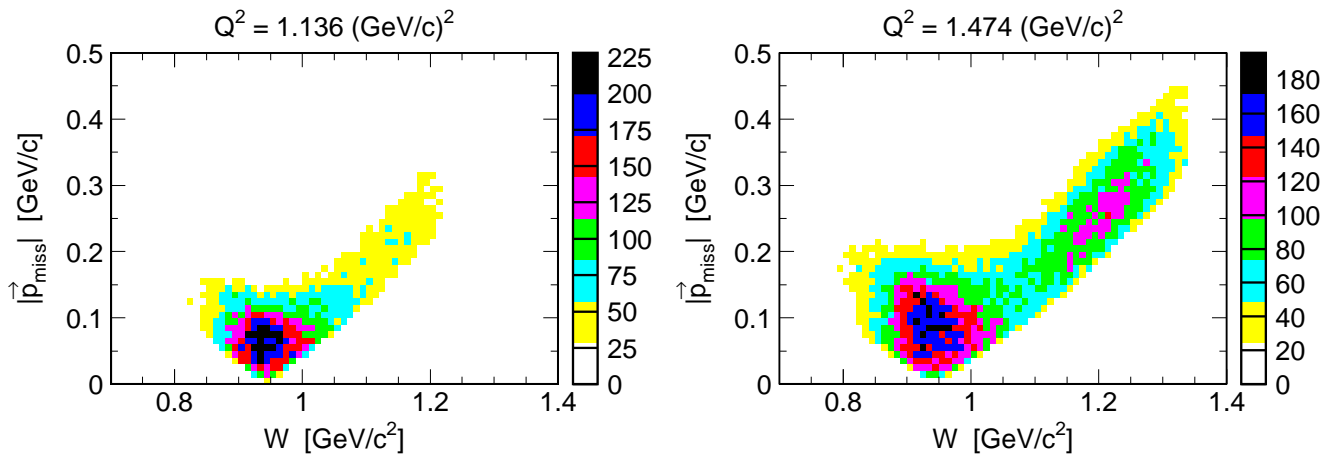}
\caption{(Color online) Correlation plot of
$|{\bf{p}}_{\mathrm{miss}}|$ versus $W$ for the full NPOL acceptance
at $Q^{2} = 1.136$ and 1.474 (GeV/$c$)$^{2}$.}
\label{fig:npol-raw-pmiss-vs-w}
\end{figure*}

\subsubsection{Kinematic distributions and time-of-flight variables}
\label{sec:data-analysis-neutron-kinematictof}

Following reconstruction of the track through the polarimeter,
kinematic and time-of-flight quantities were computed for fully
reconstructed events.  First, the incident particle's momentum was
computed using only position information for the reconstructed target
vertex, position information for the first scattering vertex in the
front array, and the four-momentum transfer $(\omega,{\bf{q}})$, via
solution of the quadratic equation for $|{\bf{p}}_{n}|$ given
previously in Eq.~(\ref{eq:quasielastic-quadratic}).  The momentum was
then used to predict the target-to-front array time-of-flight; the
difference between the predicted and measured time-of-flight was then
stored as the cTOF variable.  Laboratory frame polar and azimuthal
neutron scattering angles with respect to ${\bf{q}}$, $\theta_{nq}$
and $\phi_{nq}$, were computed from information on ${\bf{q}}$ and
${\bf{p}}_{n}$.  Second, front-to-rear polar and azimuthal scattering
angles, $\theta_{\mathrm{scat}}$ and $\phi_{\mathrm{scat}}$, were
computed using information on ${\bf{p}}_{n}$ and the scattering
vertices in the front and rear arrays.  This information was used to
compute a value for $T_{np}$, Eq.\ (\ref{eq:Tnp}), which was then used
to predict the front-to-rear time-of-flight; the difference between
the predicted and measured time-of-flight was then stored as the rTOF
variable.  Finally, the missing momentum, ${\bf{p}}_{\mathrm{miss}}$,
missing energy, $E_{\mathrm{miss}}$, and missing mass,
$m_{\mathrm{miss}}$, were computed according to
\begin{subequations}
\begin{eqnarray}
{\bf{p}}_{\mathrm{miss}} &=& {\bf{q}} - {\bf{p}}_{n}, \\
E_{\mathrm{miss}} &=& (m_{d}+\omega) - (T_{n}+m_{n}), \\
m_{\mathrm{miss}} &=& \sqrt{E_{\mathrm{miss}}^{2} -
  |{\bf{p}}_{\mathrm{miss}}|^{2}}.
\end{eqnarray}
\end{subequations}

\subsubsection{Sample nucleon reconstruction results}
\label{sec:data-analysis-neutron-results}

To illustrate the full range of the polarimeter's acceptance, sample
two-dimensional histograms of $|{\bf{p}}_{\mathrm{miss}}|$ plotted
versus the invariant mass $W$ at our $Q^{2} = 1.136$ and 1.474
(GeV/$c$)$^{2}$ points are shown in Fig.\
\ref{fig:npol-raw-pmiss-vs-w}.  A minimal set of cuts designed to
eliminate scattering from the target cell walls, hadrons in the HMS,
and protons incident on NPOL were applied to these spectra.  Our
acceptance was sensitive to missing momenta ranging up to $\sim 450$
MeV/$c$ at our highest $Q^{2}$ point.  As can clearly be seen in these
correlation plots, quasielastic events were associated with missing
momenta in the range $\alt 150$ MeV/$c$.  Larger values of
$|{\bf{p}}_{\mathrm{miss}}|$ are, of course, seen to correspond to
inelastic events, with the $\Delta(1232)$ resonance prominent at large
missing momenta in the $Q^{2} = 1.474$ (GeV/$c$)$^{2}$ spectrum.  The
correlation plot for $Q^{2} = 1.169$ (GeV/$c$)$^{2}$ was essentially
identical to that at $Q^{2} = 1.136$ (GeV/$c$)$^{2}$, while the $Q^{2}
= 0.447$ (GeV/$c$)$^{2}$ distribution was restricted to considerably
smaller ranges of $|{\bf{p}}_{\mathrm{miss}}|$ ($\alt 100$ MeV/$c$).

\subsection{Data selection criteria, data sets, and cuts}
\label{sec:data-analysis-datasets}

\subsubsection{Data selection criteria and data sets}
\label{sec:data-analysis-datasets-criteria}

Only those data runs satisfying the following criteria were employed
for the final production data analysis: (1) no problems with the HMS
equipment (e.g., magnet trips, detector failures, etc.); (2) no
problems with delivery of the electron beam (e.g., unstable beam
parameters); (3) no problems with the DAQ; (4) no problems with the
cryogenic target (e.g., large temperature fluctuations, monitoring
system failures, etc.); and (5) no problems with the Charybdis magnet
or the NPOL detectors (e.g., fluctuations in the magnet current,
detector high-voltage trips, etc.).  We note that additional problems
may have resulted in the designation of a run as unsuitable for the
production analysis.

\begin{table}
\caption{Quantity of data (accumulated charge) employed for the final
production analysis.  A total of 194 Coulombs of charge was delivered
to the experiment for production running with the deuterium target.}
\begin{ruledtabular}
\begin{tabular}{ccd}
Central $Q^{2}$& Precession& \multicolumn{1}{c}{Charge} \\
$[$(GeV/$c$)$^{2}]$& Angle $\chi$& \multicolumn{1}{c}{$[$Coulombs$]$} \\ \hline
0.447& $-40^{\circ}$& 25.122 \\
0.447& $+40^{\circ}$& 14.569 \\ \hline
1.136& $0^{\circ}$&   27.587 \\
1.136& $-90^{\circ}$&  4.701 \\
1.136& $+90^{\circ}$&  4.158 \\ \hline
1.169& $-40^{\circ}$&  7.006 \\
1.169& $+40^{\circ}$&  6.321 \\ \hline
1.474& $0^{\circ}$&   26.239 \\
1.474& $-90^{\circ}$&  4.097 \\
1.474& $+90^{\circ}$&  4.098 \\ \hline
1.474& $-40^{\circ}$& 20.803 \\
1.474& $+40^{\circ}$& 16.762 \\ \hline
Total& &        161.463 \\
\end{tabular}
\end{ruledtabular}
\label{tab:production-data-summary}
\end{table}

The quantity of data satisfying the above selection criteria is
summarized in Table \ref{tab:production-data-summary}.  There, we list
the accumulated charge for each of the individual $Q^{2}$ points and
neutron spin precession angles.

\subsubsection{Cuts for extraction of time-of-flight spectra}
\label{sec:data-analysis-datasets-cuts}

A summary of the final set of cuts applied to the production data sets
for extraction of the cTOF and rTOF time-of-flight spectra is as
follows.

\paragraph{Target variables}
Scattering from the target cell windows was suppresed via the
requirement that the reconstructed target vertex lie within $\pm 7$ cm
of the center of the target (for the 15-cm target) along the incident
beamline.  Further, events with unreasonable reconstructed values for
$x'_{\mathrm{tar}}$ and $y'_{\mathrm{tar}}$ were discarded.

\paragraph{HMS variables}
The reconstructed electron track was required to fall within the the
collimator acceptance, and events with unreasonably large track
reconstruction $\chi^{2}$ values were discarded.  Hadrons in the HMS
were suppressed via cuts on the number of \v{C}erenkov photoelectrons
and the energy deposition in the calorimeter.  Events away from the
quasielastic peak were suppressed via a tight $\Delta p/p \in
[-3\%,+5\%]$ cut.

\paragraph{NPOL variables}
Software thresholds of 8 (20) MeVee designed to suppress low-energy
backgrounds were applied to the front (rear) array pulse height
distributions.  Also, to suppress lower-energy neutrons originating
from charge-exchange Pb$(p,n)$ reactions in the lead curtain
(discussed in more detail later), the mean times for front array hits
were required to lie within a $[-5,5]$ ns window, due to the expected
degradation in the energy of the incident protons prior to the
charge-exchange reaction.  Events with more than one scattering vertex
in the front array (i.e., existence of a second cluster) were
discarded to eliminate the effects of depolarization following the
first interaction in the front array.

The front-to-rear polarimeter scattering angle,
$\theta_{\mathrm{scat}}$, was required to satisfy
$\theta_{\mathrm{scat}} \in [5^{\circ},35^{\circ}]$ at $Q^{2} = 0.447$
(GeV/$c$)$^{2}$ and $\in [5^{\circ},30^{\circ}]$ for the other $Q^{2}$
points.  The lower cut of $5^{\circ}$ eliminated unreasonably small
scattering angles, while the upper cut of $30^{\circ}$ or $35^{\circ}$
was used to suppress zero (or negative) values of the analyzing power
at larger scattering angles (as predicted by \texttt{SAID}
\cite{arndt03}).

\paragraph{$^{2}H(e,e'n)^{1}H$ reaction variables}
Pion-production events were suppressed via tight cuts on the missing
momentum and invariant mass of $|{\bf{p}}_{\mathrm{miss}}| \leq 100$
MeV/$c$ and $W \leq 1.04$ GeV/$c^{2}$.

\subsection{Extraction of time-of-flight spectra and scattering asymmetries}
\label{sec:data-analysis-TOFasym}

\subsubsection{Polarimeter event types}
\label{sec:data-analysis-TOFasym-codeoverview}

An analysis code developed to extract the physical scattering
asymmetries subjected each event to the cuts discussed previously.  In
addition, each event was also subjected to a more stringent test for
the determination of the incident particle's charge.  As we used
single-hit TDCs, an early accidental hit in a front veto/tagger
detector falling outside the mean-time window for the front
veto/tagger array would have prevented that TDC from recording any
later (on-time) hits, leading to the incorrect tagging of a charged
particle as a neutral particle.

Histograms of cTOF were accumulated for two types of front array
scattering events, $(n,n)$ and $(n,p)$ events, corresponding (for a
neutral particle incident on the polarimeter) to the detection of a
neutral and charged particle, respectively, in the rear array.  We
identified $(n,n)$ events with the scattering of the neutron from the
front array to the rear array, while we identified $(n,p)$ events with
forward scattering of the recoil proton with sufficient energy for
penetration of the front array.  It should be noted that for the
incident neutron kinetic energies of interest, the analyzing power for
elastic $np$ scattering becomes negative for neutron scattering angles
greater than $\sim 40^{\circ}$; therefore, the signs of the detected
asymmetries for $(n,n)$ and $(n,p)$ events were the same.
Events with charges deemed ambiguous in either the front or rear array
were rejected.

Histograms of rTOF summed over all front-to-rear tracks were
accumulated for those events falling within a prescribed cTOF window.
To compensate for variations in the flight path between the front
array and the rear array, the rTOF values were normalized to a nominal
250-cm flight path.  The accumulated rTOF spectra were decomposed into
the following event types: (1) ``RU events" (positive beam helicity
and scattering from the front array to the top rear array); (2) ``LU
events" (negative beam helicity, top rear array); (3) ``RD events"
(positive beam helicity, bottom rear array); and (4) ``LD events"
(negative beam helicity, bottom rear array).  The scattering
asymmetries were then extracted from the yields in these four spectra.

\subsubsection{HMS-NPOL coincidence event types}
\label{sec:data-analysis-TOFasym-eventtypes}

\begin{figure}
\includegraphics[scale=0.45]{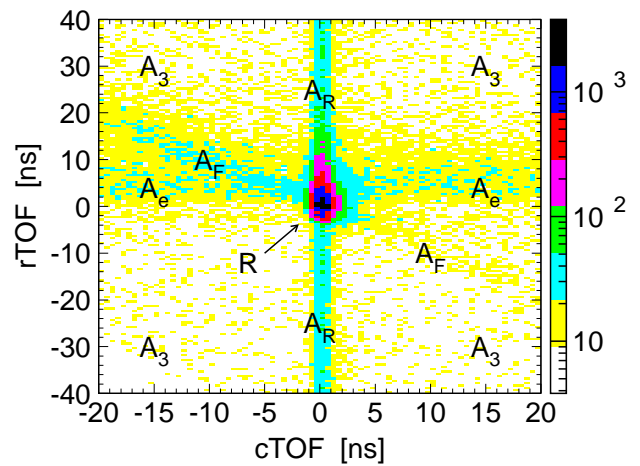}
\caption{(Color online) Correlation between cTOF and rTOF
at $Q^{2} = 1.474$ (GeV/$c$)$^{2}$ with the various event types
(see text) identified.}
\label{fig:ctof-rtof-correlation-1}
\end{figure}

A two-dimensional histogram of the correlation between cTOF and rTOF
summed over $(n,n)$ and $(n,p)$ events at $Q^{2} = 1.474$
(GeV/$c$)$^{2}$ is shown in Fig.\ \ref{fig:ctof-rtof-correlation-1}.

\begin{figure*}
\includegraphics[angle=270,scale=0.70]{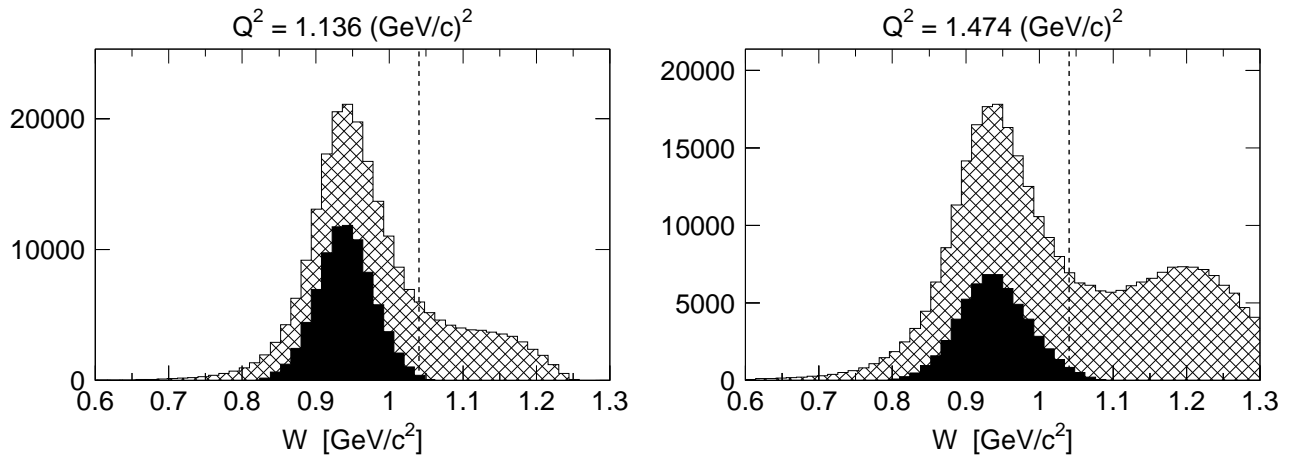}
\caption{Distributions of the invariant mass $W$ before
(cross-hatched) and after (solid) all cuts except for those on $\Delta
p/p$, $|{\bf{p}}_{\mathrm{miss}}|$, and cTOF at $Q^{2} = 1.136$ and
1.474 (GeV/$c$)$^{2}$.  The vertical dashed lines denote the final $W
< 1.04$ GeV/$c^{2}$ cut.}
\label{fig:final-cuts-w}
\end{figure*}

Five different event types can readily be identified in this
correlation plot.  (1) Real three-fold HMS/front-array/rear-array
coincidence events are denoted ``$R$" and form the peak centered at
$\mathrm{cTOF} = \mathrm{rTOF} = 0$ ns.  (2) Three-fold accidental
coincidences, denoted ``$A_{3}$", require a random electron in the
HMS, a random neutral particle in the front array, and a random
particle in the rear array, and are distributed uniformly over the
entire plot area.  (3) Real two-fold front-array/rear-array
coincidences with an accidental electron are denoted ``$A_{e}$" and
are associated with the ``horizontal band" defined by $\mathrm{rTOF} =
0$ ns.  (4) Real two-fold electron/front-array coincidences with an
accidental rear array particle are denoted ``$A_{R}$" and are
identified with the ``vertical band" defined by $\mathrm{cTOF} = 0$
ns.  (5) Real two-fold electron/rear-array coincidences with an
accidental front-array particle are denoted ``$A_{F}$".  These events
are located along a diagonal band defined (approximately) by
$\mathrm{cTOF} = -\mathrm{rTOF}$.  Such events are attributed to the
corruption of an otherwise $R$-type event by an accidental front array
hit occuring some time $\Delta t_{A}$ before or after the true
interaction.  The values of cTOF and rTOF extracted from the data will
then be $\mathrm{cTOF} = \mathrm{cTOF}_{\mathrm{uncorr}} - \Delta
t_{A}$ and $\mathrm{rTOF} = \mathrm{rTOF}_{\mathrm{uncorr}} + \Delta
t_{A}$, where the ``uncorr" subscript denotes the (true) uncorrupted
values.  For uncorrupted values centered on zero, it then follows that
$\mathrm{cTOF} = -\mathrm{rTOF}$, in accordance with the observed
result.

\subsubsection{Quasielastic event selection}
\label{sec:data-analysis-TOFasym-quasielastic}

\begin{figure*}
\includegraphics[angle=270,scale=0.70]{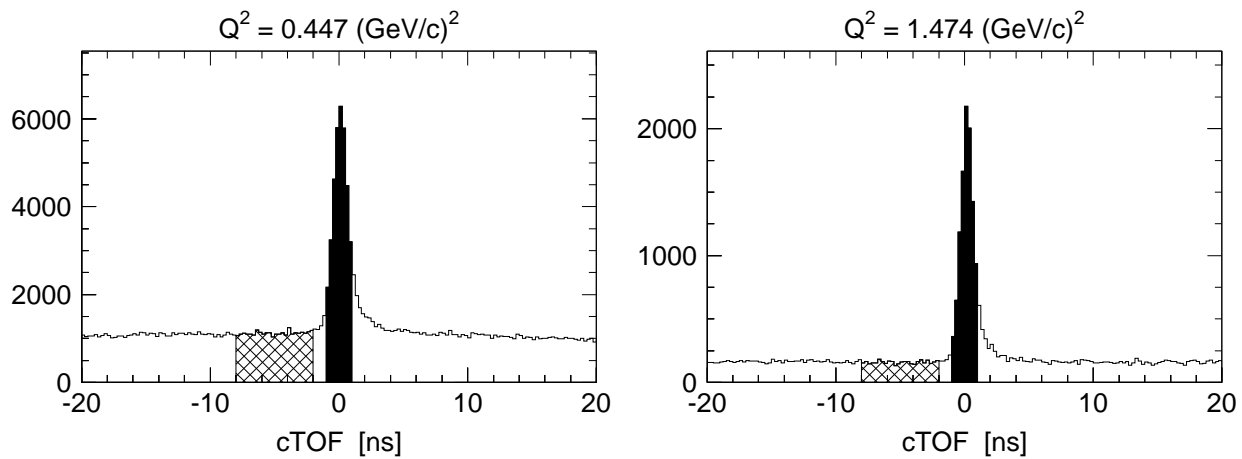}
\caption{Distributions of cTOF after application of the final set of
cuts at $Q^{2} = 0.447$ and 1.474 (GeV/$c$)$^{2}$.  The dark shaded
regions indicate the selected peak window, while the cross-hatched
regions indicate the sampled background region.}
\label{fig:ctof-final-cuts}
\end{figure*}

Real $R$-type coincidence events were selected via tight
$\mathrm{cTOF} \in [-1,1]$ ns and $\mathrm{rTOF} \in [-1,8]$ ns cuts.
As evidence our cuts selected quasielastic $^{2}$H$(e,e'n)^{1}$H
events, comparisons of invariant mass spectra, $W$, obtained before
and after cuts on $\Delta p/p$, $|{\bf{p}}_{\mathrm{miss}}|$, cTOF,
and rTOF are shown in Fig.\ \ref{fig:final-cuts-w} for our $Q^{2} =
1.136$ and 1.474 (GeV/$c$)$^{2}$ points.  After all cuts (except for
the additional cut on $W < 1.04$ GeV/$c^{2}$ itself), these
distributions converged to fairly narrow peaks centered on the neutron
mass.

\subsubsection{Extraction of asymmetries from time-of-flight spectra}
\label{sec:data-analysis-TOFasym-asymmetryextraction}

\begin{figure*}
\includegraphics[angle=270,scale=0.70]{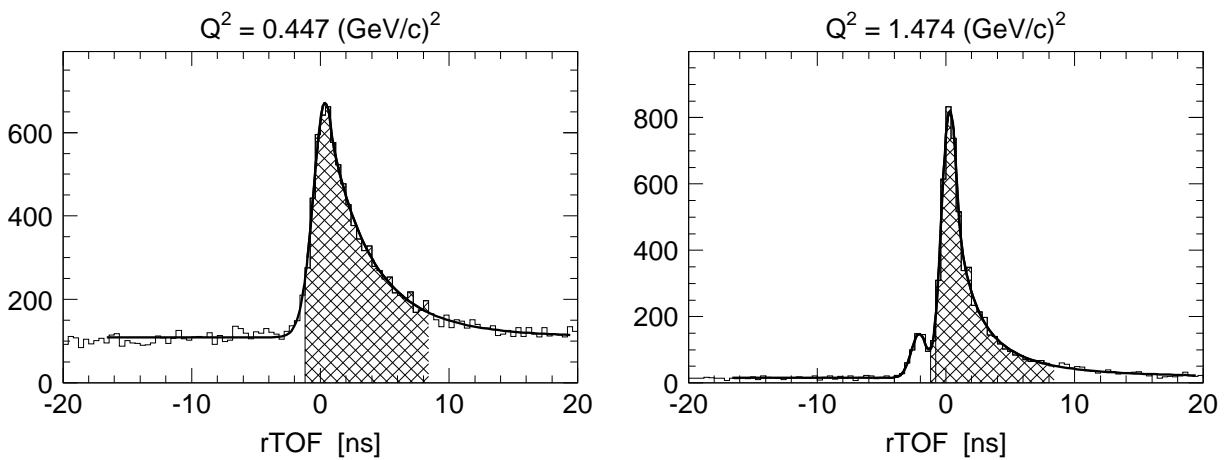}
\caption{Distributions of rTOF for cTOF peak events at $Q^{2} = 0.447$
and 1.474 (GeV/$c$)$^{2}$.  The cross-hatched regions indicate the
accepted window.  The solid curves are the results of our fits to
these spectra.}
\label{fig:rtof-final-cuts}
\end{figure*}

One-dimensional projections of cTOF are shown in Fig.\
\ref{fig:ctof-final-cuts} for our lowest and highest $Q^{2}$ points.
Histograms of rTOF were accumulated for those events falling within
the $[-1,1]$ ns peak cTOF window.  In addition, histograms of rTOF
were accumulated also for a sampled background region of $[-8,-2]$ ns
in the cTOF spectrum.  The signal-to-noise ratios were independent of
the state of the Charybdis magnet at each of our $Q^{2}$ points.

Sample rTOF spectra summed over all RU, LU, RD, and LD events for cTOF
peak events at our lowest and highest $Q^{2}$ points are shown in
Fig.\ \ref{fig:rtof-final-cuts}.  The asymmetric tails on the slow
sides are due to scattering from protons bound in carbon nuclei and
other nuclear reactions, and the small satellite peak observed in the
$Q^{2} = 1.474$ (GeV/$c$)$^{2}$ spectrum on the fast side at $\sim
-2.5$ ns is attributed to quasifree $\pi^{0}$-production in the
scintillators, followed by decay and detection of a photon in the rear
array.  Indeed, front-to-rear velocity spectra for these events are
centered on $c$.  This $\pi^{0}$-production peak was observed in the
$Q^{2} = 1.136$, 1.169, and 1.474 (GeV/$c$)$^{2}$ rTOF spectra but was
absent in the $Q^{2} = 0.447$ (GeV/$c$)$^{2}$ spectrum, as the
energies of those neutrons were below threshold.

The yields for those events falling within the $[-1,8]$~ns rTOF window
were obtained via peak fitting, with contributions from the
$\pi^{0}$-production peak and the flat background excluded.  These
yields were then further corrected for the contents of the rTOF
spectra accumulated for the sampled cTOF background region.  The
desired quantities, the physical scattering asymmetries, $\xi$, were
extracted from the final background-subtracted yields in the four
decomposed rTOF spectra via the cross-ratio technique \cite{ohlsen73}.
In obvious notation, the cross ratio, $r$, is defined to be the ratio
of two geometric means,
\begin{equation}
r = \sqrt{\frac{N_{RU}N_{LD}}{N_{RD}N_{LU}}},
\label{eq:cross-ratio-definition}
\end{equation}
and is related to the asymmetry $\xi$ via
\begin{equation}
\xi = \frac{r-1}{r+1} = \frac{\sqrt{N_{RU}N_{LD}} - \sqrt{N_{RD}N_{LU}}}
{\sqrt{N_{RU}N_{LD}} + \sqrt{N_{RD}N_{LU}}}.
\label{eq:xi-cross-ratio}
\end{equation}
The merit of the cross-ratio technique is that $\xi$ is insensitive to
\cite{ohlsen73}: (1) the number of particles incident on the
polarimeter (i.e., target luminosities) for the two beam helicity
states; and (2) the relative efficiencies and acceptances of the
polarimeter's top and bottom rear arrays.

\subsection{Asymmetry results}
\label{sec:data-analysis-asymmetryresults}

\subsubsection{Electron beam polarization normalization}
\label{sec:data-analysis-asymmetryresults-beampolarization}

Unlike recoil polarization measurements in which both polarization
components, $P^{(h)}_{t}$ and $P^{(h)}_{\ell}$, can be extracted
simultaneously from the data (e.g., recoil polarization experiments
with focal-plane polarimeters), our polarimeter was sensitive to only
one of these components (or a combination thereof).  As such, it was
necessary to normalize our run-by-run scattering asymmetries to some
common value of the beam polarization.

Our normalization procedure was as follows.  As the beam polarization
was measured only periodically with the M{\o}ller polarimeter, we
defined the beam polarization for a run to be the result of the most
recent prior M{\o}ller measurement (if the accelerator parameters were
unchanged in the interim).  All of our run-by-run scattering
asymmetries and their statistical errors were then normalized to a
common value of 80\%.

\begin{figure}
\includegraphics[scale=0.55]{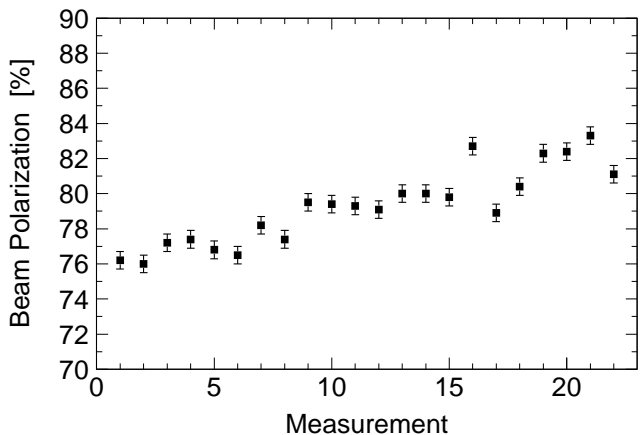}
\caption{Results of 23 successive M{\o}ller beam polarization
measurements conducted during the $Q^{2} = 1.474$ (GeV/$c$)$^{2}$
$\chi = \pm40^{\circ}$ running period spanning the days of February
20, 2001 through March 5, 2001.  The errors shown are statistical.}
\label{fig:moller-q147pm}
\end{figure}

We found that the beam polarization was fairly stable, with small (few
percent) fluctuations observed in successive measurements during
periods of continuous beam delivery to our experiment.  To illustrate,
the results of 23 successive M{\o}ller measurements conducted during
our $Q^{2} = 1.474$ (GeV/$c$)$^{2}$ $\chi = \pm40^{\circ}$ running
period spanning the days of February 20, 2001 through March 5, 2001
are shown in Fig.\ \ref{fig:moller-q147pm}.

\subsubsection{Corrections for charge-exchange in the lead curtain}
\label{sec:data-analysis-asymmetryresults-leadcurtain}

Contamination from the two-step $^{2}$H$(\vec{e},e'\vec{p})$ +
Pb$(\vec{p},\vec{n})$ charge-exchange reaction in the lead curtain
could either dilute the ``real'' $^{2}$H$(\vec{e},e'\vec{n})$
asymmetry or contribute to a false asymmetry if the flux of
charge-exchange neutrons was unpolarized or polarized, respectively.
A significant advantage of our neutron flight path setup in which the
lead curtain was located downstream of the Charybdis dipole field was
that the majority of the quasielastic protons were swept from the
front face of the lead curtain.

Accounting for such nuclear reactions, the measured asymmetry,
$\xi_{M}$, can be parametrized as
\begin{equation}
\xi_{M} = f_{R}\xi_{R} + f_{B}\xi_{B},
\label{eq:xiM}
\end{equation}
where $f_{B}$ denotes the contamination level from the two-step
charge-exchange process, $\xi_{B}$ denotes the asymmetry for
charge-exchange neutrons, $f_{R} = 1 - f_{B}$ denotes the fraction of
$^{2}$H$(\vec{e},e'\vec{n})$ neutrons, and $\xi_{R}$ denotes the
asymmetry for the $^{2}$H$(\vec{e},e'\vec{n})$ reaction.  The
asymmetry for the background process can further be written as
\begin{equation}
\xi_{B} = \left(P^{p}_{S}\cos\chi_{p} + P^{p}_{L}\sin\chi_{p}\right)
D_{SS}^{\mathrm{Pb}}A_{y},
\label{eq:xiB}
\end{equation}
where $P^{p}_{S}$ and $P^{p}_{L}$ denote, respectively, the
projections of the $^{2}$H$(\vec{e},e'\vec{p})$ recoil proton's
polarization on the polarimeter momentum basis $\hat{S}$- and
$\hat{L}$-axis; $\chi_{p}$ is the proton spin precession angle in the
Charybdis field; and $D_{SS}^{\mathrm{Pb}}$ denotes the polarization
transfer coefficient for the Pb$(\vec{p},\vec{n})$ reaction.  It then
follows that if $f_{B}$, $P^{p}_{S}$, $P^{p}_{L}$, $\chi_{p}$,
$D_{SS}^{\mathrm{Pb}}$, and $A_{y}$ are all known or measured,
$\xi_{R}$ can be determined.

To estimate the contamination levels, $f_{B}$, we took data with a
liquid hydrogen target.  The rates for $(n,n)$ and $(n,p)$ events
extracted from these data were compared with those extracted from our
liquid deuterium data and corrected for differences in the two
targets' densities and atomic numbers.  We found that the
contamination levels were negligible ($\alt 0.3$\%) at all of our
$Q^{2}$ points when the Charybdis field was energized for $\chi = \pm
40^{\circ}$ and $\pm 90^{\circ}$ precession and also when the field
was de-energized at $Q^{2} = 1.136$ (GeV/$c$)$^{2}$ for
$\chi=0^{\circ}$ precession; therefore, we did not apply corrections
to any of these asymmetries.  Non-negligible event rates were observed
when the Charybdis field was de-energized for $\chi = 0^{\circ}$
precession at $Q^{2} = 1.474$ (GeV/$c$)$^{2}$, with observed
contamination levels of $\sim 2.2$\% and $\sim 4.2$\% for $(n,n)$ and
$(n,p)$ events, respectively.  Corrections were applied to these
asymmetries assuming $D_{SS}^{\mathrm{Pb}} = 0$ for our kinematics of
$T_{p} \sim 786$ MeV.  $D_{SS}^{\mathrm{Pb}}$ was measured at $T_{p} =
795$ MeV and found to be consistent with zero ($0.014 \pm 0.013$)
\cite{prout00}.

\subsubsection{Summary of asymmetry results}
\label{sec:data-analysis-asymmetryresults-summary}

\begin{figure}
\includegraphics[angle=270,scale=0.75]{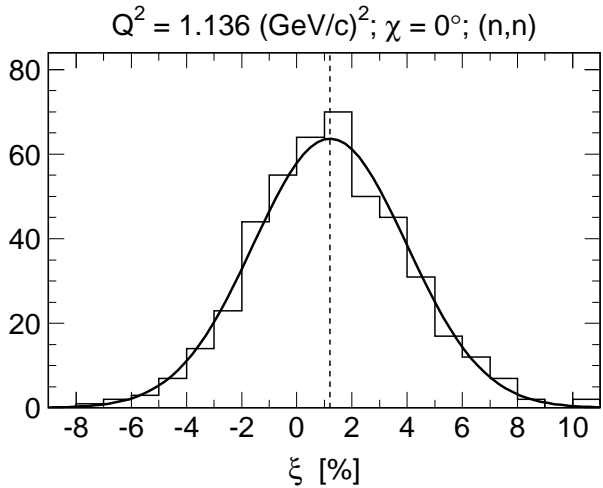}
\caption{Histogram of the $Q^{2} = 1.136$ (GeV/$c$)$^{2}$ $\chi =
0^{\circ}$ $(n,n)$ asymmetries.  The solid curve is a Gaussian fit,
and the vertical dashed line denotes the mean value for the asymmetry
given in Table \ref{tab:final-asymmetry-data}.}
\label{fig:q113z-asymmetry-gaussian}
\end{figure}

Our final asymmetry data for $(n,n)$ and $(n,p)$ events at each of our
$Q^{2}$ points and precession angles are tabulated in Table
\ref{tab:final-asymmetry-data}.  To illustrate the quality of our
asymmetry data, a histogram of the $(n,n)$ asymmetries for the $Q^{2}
= 1.136$ (GeV/$c$)$^{2}$ $\chi = 0^{\circ}$ data set is shown in Fig.\
\ref{fig:q113z-asymmetry-gaussian}; the distribution is of an
appropriate Gaussian shape.

\begin{table}
\caption{Final $(n,n)$ and $(n,p)$ asymmetry data normalized to a beam
polarization of 80\%.  The $Q^{2} = 1.474$ (GeV/$c$)$^{2}$ $\chi =
0^{\circ}$ asymmetries were corrected for contamination from
charge-exchange in the lead curtain.}
\begin{ruledtabular}
\begin{tabular}{ccrr}
Central $Q^{2}$& Precession&
  \multicolumn{1}{c}{$(n,n)$}& \multicolumn{1}{c}{$(n,p)$} \\
$[$(GeV/$c)^{2}]$& Angle $\chi$&
  \multicolumn{1}{c}{$\xi$ [\%]}&
  \multicolumn{1}{c}{$\xi$ [\%]} \\ \hline
0.447& $-40^{\circ}$&  $-4.51 \pm 0.22$&  $-2.97 \pm 0.19$  \\
0.447& $+40^{\circ}$&  $ 6.38 \pm 0.28$&  $ 4.98 \pm 0.29$  \\ \hline
1.136& $  0^{\circ}$&  $ 1.20 \pm 0.13$&  $ 0.57 \pm 0.10$  \\
1.136& $-90^{\circ}$&  $-5.71 \pm 0.32$&  $-3.11 \pm 0.25$  \\
1.136& $+90^{\circ}$&  $ 5.67 \pm 0.35$&  $ 3.18 \pm 0.25$  \\
1.169& $-40^{\circ}$&  $-2.92 \pm 0.29$&  $-1.42 \pm 0.22$  \\
1.169& $+40^{\circ}$&  $ 4.75 \pm 0.31$&  $ 2.76 \pm 0.25$  \\ \hline
1.474& $  0^{\circ}$&  $ 1.29 \pm 0.19$&  $ 0.64 \pm 0.17$  \\
1.474& $-40^{\circ}$&  $-2.26 \pm 0.20$&  $-0.88 \pm 0.18$  \\
1.474& $+40^{\circ}$&  $ 4.03 \pm 0.24$&  $ 2.11 \pm 0.21$  \\
1.474& $-90^{\circ}$&  $-4.64 \pm 0.47$&  $-2.92 \pm 0.50$  \\
1.474& $+90^{\circ}$&  $ 5.07 \pm 0.49$&  $ 2.14 \pm 0.43$  \\
\end{tabular}
\end{ruledtabular}
\label{tab:final-asymmetry-data}
\end{table}

\begin{figure}
\includegraphics[angle=270,scale=0.72]{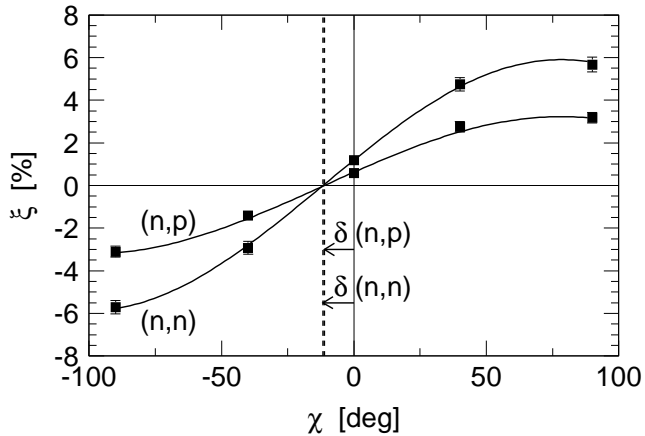}
\caption{Sinusoidal fits of the $Q^{2}=1.136/1.169$ (GeV/$c$)$^{2}$
$(n,n)$ and $(n,p)$ asymmetries as a function of the precession
angle.}
\label{fig:q113-uncorrected-phase-shift}
\end{figure}

\subsection{Extraction of uncorrected values for $\bm{G_{En}/G_{Mn}}$}
\label{sec:data-analysis-uncorrectedGEnGMn}

\begin{table*}
\caption{Values of $\delta = \tan^{-1}(P^{(h)}_{t}/P^{(h)}_{\ell})$
and the uncorrected results for $G_{En}/G_{Mn}$ at each of the $Q^{2}$
points.}
\begin{ruledtabular}
\begin{tabular}{cccccc}
Central $Q^{2}$& \multicolumn{2}{c}{$\delta$ [deg]}&
  \multicolumn{2}{c}{$G_{En}/G_{Mn}$}& $G_{En}/G_{Mn}$ \\
$[$(GeV/$c$)$^{2})]$& $(n,n)$& $(n,p)$& $(n,n)$& $(n,p)$&
  Combined\footnotemark[1] \\ \hline
0.447& $8.2 \pm 1.5$& $12.0 \pm 1.9$&
  $-0.0580 \pm 0.0106$& $-0.0854 \pm 0.0138$& $-0.0681 \pm 0.0084$ \\
1.136/1.169\footnotemark[2]& $11.7 \pm 1.2$& $11.2 \pm 1.7$&
  $-0.124 \pm 0.013$& $-0.118 \pm 0.019$& $-0.122 \pm 0.011$ \\
1.474&                       $14.0 \pm 1.6$& $16.9 \pm 2.9$&
  $-0.166 \pm 0.020$& $-0.203 \pm 0.037$& $-0.174 \pm 0.017$ \\
\end{tabular}
\end{ruledtabular}
\begin{minipage}{\textwidth}
\footnotetext[1]{Weighted average of $G_{En}/G_{Mn}$ from $(n,n)$ and
                 $(n,p)$ events.}
\footnotetext[2]{Result obtained via averaging of the nominal (central)
                 electron kinematics for the two $Q^{2}$ points.}
\end{minipage}
\label{tab:delta-uncorrected-GEnGMn}
\end{table*}

We extracted values for $G_{En}/G_{Mn}$ from our asymmetry data
assuming elastic scattering from a free neutron and infinitesimal
pointlike HMS and NPOL acceptances and neglecting nuclear physics
corrections for FSI, MEC, and IC.  To do so, we fitted the asymmetries
as a function of the precession angle to the functional form
$\xi(\chi) \propto \sin(\chi + \delta)$, where the phase-shift
parameter $\delta = \tan^{-1}\big(P^{(h)}_{t}/P^{(h)}_{\ell}\big)$
was defined in terms of form factors and kinematics in
Eq.~(\ref{eq:phase-shift-delta}).  To illustrate the quality of these
fits, our $Q^{2} = 1.136/1.169$ (GeV/$c$)$^{2}$ $(n,n)$ and $(n,p)$
asymmetry data are plotted as a function of the precession angle in
Fig.\ \ref{fig:q113-uncorrected-phase-shift}.  These data are fitted
well by sinusoids with excellent agreement seen between the
independent fits to the $(n,n)$ and $(n,p)$ asymmetry data.  We could
not fit the $Q^{2} = 0.447$ (GeV/$c$)$^{2}$ asymmetries to a sinusoid
as asymmetry data were taken only at two precession angles.

The values for $G_{En}/G_{Mn}$ we derived from our values for $\delta$
using the nominal (central) values for the kinematics listed in Table
\ref{tab:central-kinematics} are summarized in Table
\ref{tab:delta-uncorrected-GEnGMn}.

\subsection{Simulation programs}
\label{sec:data-analysis-simulations}

We developed two independent simulation programs, \texttt{GENGEN} and
the \texttt{Acceptance} program, to extract acceptance-averaged and
nuclear physics corrected values for $G_{En}/G_{Mn}$ from our measured
experimental asymmetries.  The \texttt{GENGEN} simulation program, a
pure Monte Carlo simulation program, included realistic models for the
primary $^{2}$H$(\vec{e},e'\vec{n})^{1}$H reaction in the target, the
HMS acceptance, neutron spin precession in the Charybdis dipole field,
spin-dependent neutron scattering in the lead curtain, elastic and
quasielastic $np$ scattering in the front and rear arrays of NPOL,
tracking of the incident neutron and recoil proton from the front
array to the rear array, and the detector response of the polarimeter
to $np$ interactions in the front and rear array.  The
\texttt{Acceptance} program was not a Monte Carlo simulation program,
but was, instead, designed to extract the corrections for the finite
experimental acceptance and nuclear physics effects directly from our
experimental data.

\subsubsection{\texttt{GENGEN} simulation program}
\label{sec:data-analysis-simulations-gengen-program}

\paragraph{Event sampling technique}
A uniform sampling scheme was employed in which events were generated
uniformly over the available kinematic phase space, with an event
weight computed according to a model cross section.  The vertex
position for the primary $^{2}$H$(\vec{e},e'\vec{n})^{1}$H interaction
in the extended target was sampled uniformly within the raster
pattern, and the scattered electron's kinematics were sampled
uniformly over specified ranges.  The physical acceptance of the HMS
was enforced via inclusion of an HMS transport model taken from the
\texttt{SIMC} simulation code \cite{simc}.  In-plane and out-of-plane
scattering angles for the recoil neutron were sampled uniformly over
specified ranges, permitting computation of the magnitude of the
neutron's momentum according to Eq.\
(\ref{eq:quasielastic-quadratic}).  Complete specification of the
electron and neutron kinematics permitted computation of those
variables of particular interest for the quasielastic
$^{2}$H$(e,e'n)^{1}$H reaction, such as $\Theta^{\mathrm{c.m.}}_{np}$,
${\bf{p}}_{\mathrm{miss}}$, etc.

\paragraph{Cross section and recoil polarization}
We employed the Arenh\"{o}vel formalism
\cite{arenhovel87,arenhovel88,arenhovel95} for computation of the
$^{2}$H$(\vec{e},e'\vec{n})^{1}$H differential cross section and
recoil polarization.  These calculations modeled the deuteron as a
non-relativistic $n$-$p$ system and employed the Bonn $R$-space $NN$
potential \cite{machleidt87} for the deuteron wavefunction and the
inclusion of FSI; further, leading-order relativistic contributions
(RC) to the wavefunctions and one-body current were added via
inclusion of the most important kinematic part of the wavefunction
boost.  In the current operator, explicit MEC contributions beyond the
Siegert operators (essentially from $\pi$- and $\rho$-exchange) and IC
were included.  The treatment of IC permitted permitted consideration
of kinematic regions away from the quasielastic ridge and excitations
up to the $\Delta$ region.

Acceptance-averaging of those calculations performed within the Born
approximation (hereafter, termed the ``PWBA model'') permitted
extraction of the corrections for the finite experimental acceptance
(over the pointlike results discussed in Section
\ref{sec:data-analysis-uncorrectedGEnGMn}), while averaging of the
full calculations which included FSI, MEC, IC, and RC (hereafter,
termed the ``FSI+MEC+IC+RC model'') permitted application of
corrections for nuclear physics effects.  In order to implement the
Arenh\"{o}vel formalism within \texttt{GENGEN}, lookup tables for the
structure functions for the $^{2}$H$(\vec{e},e'\vec{n})^{1}$H reaction
were constructed over a sufficiently dense kinematic grid indexed by
$(E_{e'},\theta_{e'}, \Theta^{\mathrm{c.m.}}_{np}$), and tri-cubic
spline interpolation among the grid elements was used to compute the
cross section and recoil polarization for the kinematics of each
simulated event according to the formalism outlined in Appendix
\ref{sec:appendix-a}.

\paragraph{Nucleon form factors}
All of the structure function calculations assumed the dipole
parametrization for $G_{Mn}$, $G_{Ep}$, and $G_{Mp}$.  For the form
factor of interest, $G_{En}$, the structure function calculations were
first performed for various multiplicative factors of the standard
Galster parametrization, $G_{En} = -S\mu_{n}\tau G_{D}/(1 + 5.6\tau)$,
where the scale factor $S \in \{0.50,0.75,1.00,1.25,1.50\}$.  To
investigate the influence of a different $Q^{2}$ dependence for
$G_{En}$, structure function calculations were performed also for
multiplicative factors of a modified Galster parametrization, $G_{En}
= -Sa\mu_{n}\tau G_{D}/(1 + b\tau)$, with $a = 0.894$, $b = 3.55$
(which choice will be explained later), and the same set of $S$
factors.

\paragraph{Charybdis field transport}
The recoil polarization was transported point-by-point through a grid
of the Charybdis field, with the time derivative of the spin vector
computed at each grid point according to standard relativistic
electrodynamics.  The precession angle was computed from information
on the initial and final spatial orientations of the spin vector.

\paragraph{Lead curtain interactions}
Neutron interactions in the lead curtain were simulated with a
spin-dependent multiple scattering algorithm that employed quasifree
scattering from a lead nucleus modeled as a Fermi gas, with the Fermi
momentum for $^{208}$Pb taken to be 265 MeV/$c$ \cite{frullani84}.
The probability for an interaction of the neutron with a lead nucleus
was determined via interpolation (or extrapolation) of existing data
on total $n + \mathrm{Pb}$ cross sections \cite{finlay93}.  A polar
scattering angle was sampled from cumulative probability distributions
for the polar scattering angle as a function of neutron energy, and an
azimuthal scattering angle was chosen via an acceptance-rejection
algorithm for the spatial scattering asymmetry resulting from non-zero
analyzing power.  Pauli blocking was enforced.  For those neutrons
suffering an interaction, the scattered neutron's and recoil nucleon's
polarization components were constructed via computation of the
depolarization and polarization-transfer tensors for $NN$ scattering
using helicity amplitude routines obtained from \texttt{SAID}
\cite{arndt03}.

\begin{figure*}
\includegraphics[scale=0.85]{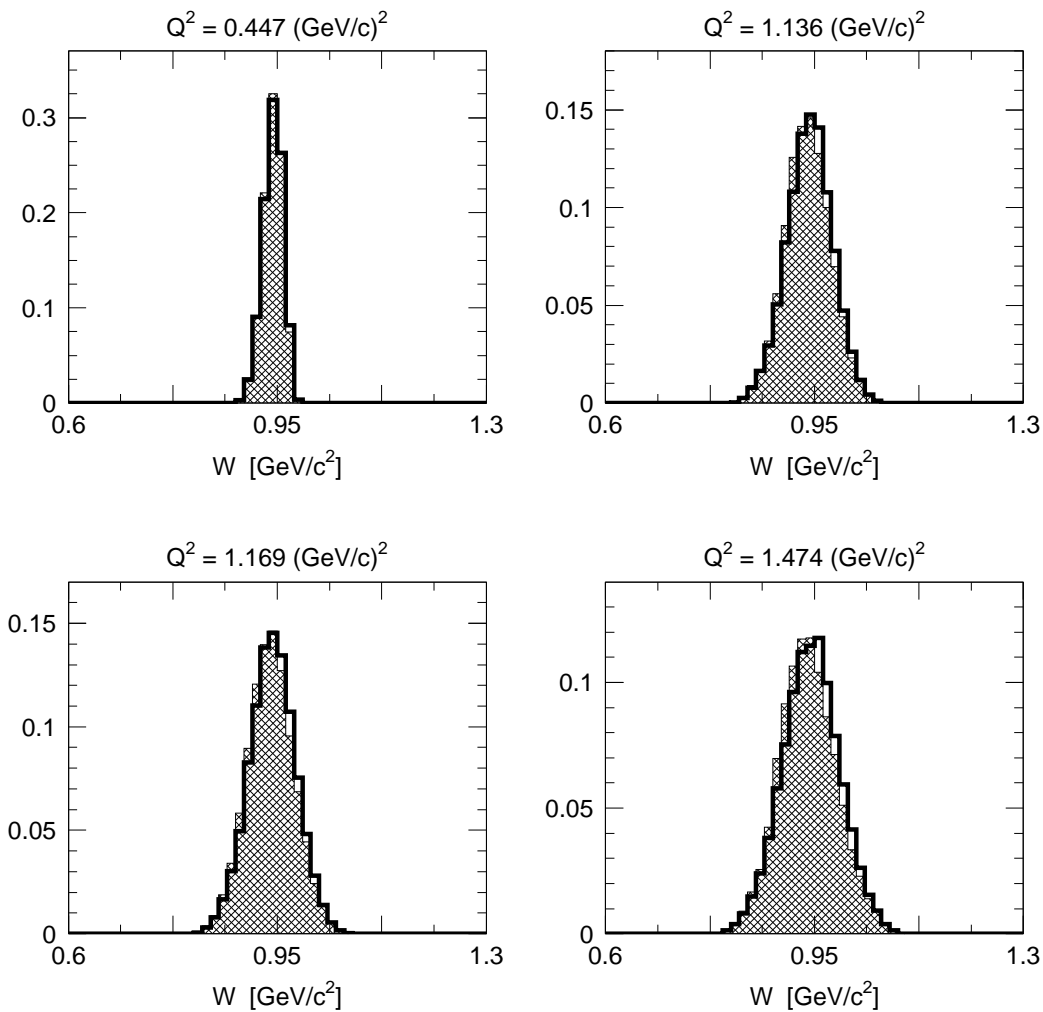}
\caption{Comparison of \texttt{GENGEN} simulated (unfilled histograms
with thick solid line borders) and experimental (cross-hatched filled
histograms) distributions of $W$ for the four central $Q^{2}$ points.
Identical cuts were applied to both the simulated and experimental
data.  The simulated results shown here employed the
FSI+MEC+IC+RC model and the Galster parametrization for $G_{En}$.}
\label{fig:compare-gengen-data-1}
\end{figure*}

\paragraph{Polarimeter interactions}
Finally, following (successful) transport of the neutron through the
steel collimator into the front array, interactions in NPOL were
simulated.  A scattering vertex was chosen randomly assuming a fixed
value for the mean free path of neutrons in the plastic scintillator,
and both the elastic (scattering from free protons) and quasielastic
(scattering from protons bound in carbon nuclei) channels were
simulated.  The scattering angles in the polarimeter were determined
using the same algorithms employed for $NN$ scattering in the lead
curtain.  We employed a rather simple model for the propagation of the
recoil proton, with the energy deposition and range (assuming a
straight-line trajectory) computed according to the Cecil, Anderson,
and Madey \cite{cecil79} range-energy formulas for protons in the
hydrocarbon scintillator.

\begin{figure*}
\includegraphics[scale=0.85]{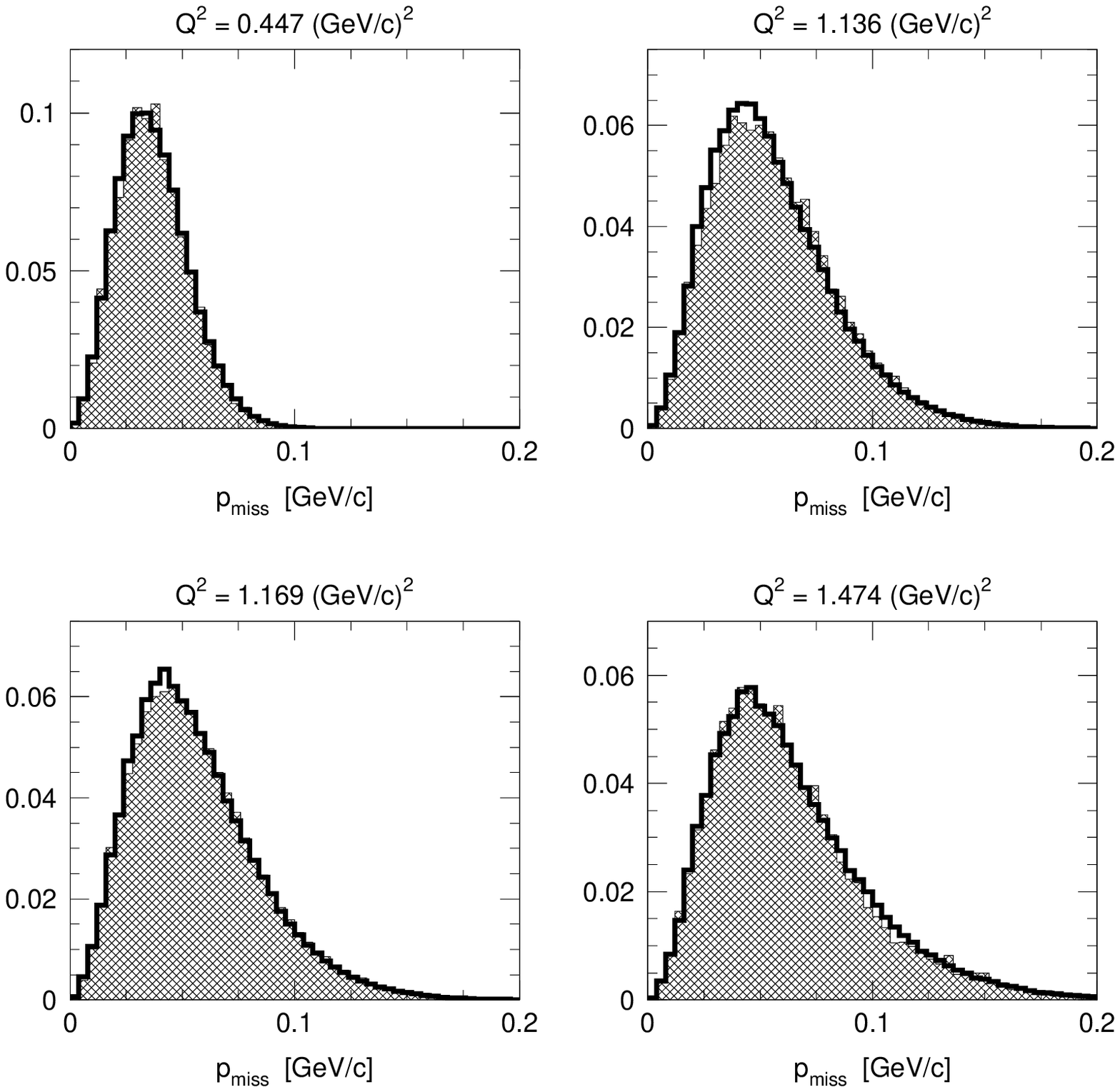}
\caption{Comparison of \texttt{GENGEN} simulated (unfilled histograms
with thick solid line borders) and experimental (cross-hatched filled
histograms) distributions of $|{\bf{p}}_{\mathrm{miss}}|$ for the four
central $Q^{2}$ points.  Identical cuts were applied to both the
simulated and experimental data.  The simulated results shown here
employed the FSI+MEC+IC+RC model and the Galster
parametrization for $G_{En}$.}
\label{fig:compare-gengen-data-2}
\end{figure*}

\subsubsection{\texttt{GENGEN} performance}
\label{sec:data-analysis-simulations-gengen-performance}

A rigorous and reliable extraction of the corrections for the finite
experimental acceptance and nuclear physics effects from simulated
data is feasible if the simulated acceptance reasonably matches the
experimental acceptance; therefore, we now document the performance of
\texttt{GENGEN} by comparing: (1) simulated distributions of important
kinematic quantities with those derived from experimental data; and
(2) the behavior of the acceptance-averaged simulated polarizations
and the experimental asymmetries as a function of the cut on some
kinematic variable (here, taken to be the invariant mass $W$).

\paragraph{Kinematic distributions}
Sample comparisons of experimental and simulated kinematic
distributions of two important kinematic variables, $W$ and
$|{\bf{p}}_{\mathrm{miss}}|$, are shown in Figs.\
\ref{fig:compare-gengen-data-1} and \ref{fig:compare-gengen-data-2}.
Reasonable agreement is seen between the \texttt{GENGEN} distributions
and those extracted from experimental data.  Although not shown here,
reasonable agreement was also obtained between simulated and
experimental distributions of variables related to $np$ scattering in
NPOL (e.g., scattering angles, velocity spectra, etc.).

\begin{figure}
\includegraphics[angle=270,scale=0.70]{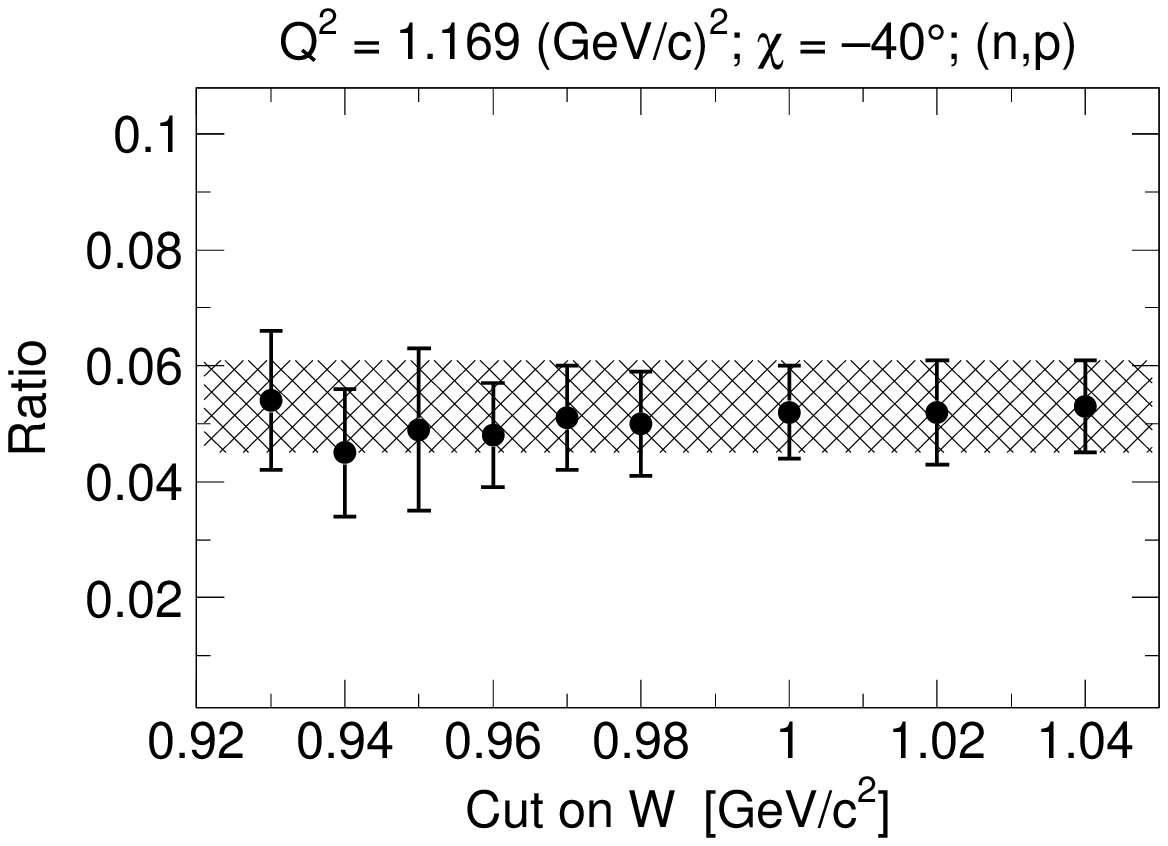}
\caption{Ratio of the asymmetries extracted from the experimental data
to the \texttt{GENGEN} simulated polarizations as a function of the
cut on $W$ for $(n,p)$ events and $\chi = -40^{\circ}$ precession at
our $Q^{2} = 1.169$ (GeV/$c$)$^{2}$ point.  The shaded band indicates
the statistical error on the ratio for the nominal cut on $W$ of $<
1.04$ GeV/$c^{2}$.  The simulated results shown here employed the
FSI+MEC+IC+RC model and the Galster parametrization for
$G_{En}$.}
\label{fig:kin2c-m40-asym-pol-vs-w}
\end{figure}

\paragraph{Experimental asymmetries and simulated polarizations}
A sample comparison of the behavior of the experimental asymmetries
and acceptance-averaged simulated polarizations following transport
through the Charybdis dipole field is shown in Fig.\
\ref{fig:kin2c-m40-asym-pol-vs-w}.  There, we plot the ratio of the
experimental asymmetries to the simulated polarizations as a function
of the upper cut on $W$ for $(n,p)$ events and $\chi = -40^{\circ}$
precession at our $Q^{2} = 1.169$ (GeV/$c$)$^{2}$ point.  Within
statistical errors, the experimental asymmetries and simulated
polarizations are seen to scale similarly with the cut on $W$.
Similar results were observed for our other $Q^{2}$ points and
precession angles.

It should be noted that in this figure the simulated
acceptance-averaged polarizations were computed assuming some certain
parametrization for $G_{En}$ (here, the Galster parametrization);
therefore, the ratios of the asymmetries to the simulated
polarizations shown in this figure are not equivalent to the
polarimeter's analyzing power.

\subsubsection{\texttt{Acceptance} program}
\label{sec:data-analysis-simulations-acceptance}

The \texttt{Acceptance} program was developed as an alternative to the
\texttt{GENGEN} Monte Carlo simulation program.  This program used the
kinematics of the reconstructed quasielastic events from the actual
experimental data to compute, on an event-by-event basis, the recoil
polarization presented to the polarimeter for each event employed in
our final data analysis (i.e., for those events satisfying all final
analysis cuts).  The \texttt{Acceptance} program used the same
$^{2}$H$(\vec{e},e'\vec{n})^{1}$H interpolation and Charybdis spin
transport algorithms developed for \texttt{GENGEN}.  Although the
\texttt{Acceptance} program was, technically, not a true Monte Carlo
simulation, a significant advantage of this method was that it did not
require a model for the experimental acceptance; however, the
disadvantage of this method was that the reconstruction of the
event-by-event kinematics is, of course, subject to measurement
uncertainties, leading to uncertainties in the computation of the
recoil polarization.

\section{Final results for $\bm{G_{En}/G_{Mn}}$ and $\bm{G_{En}}$}
\label{sec:final-results}

\subsection{Distributions of $\bm{\Theta^{\mathrm{c.m.}}_{np}}$}
\label{sec:final-results-thetacmnp}

\begin{figure}
\includegraphics[angle=270,scale=0.70]{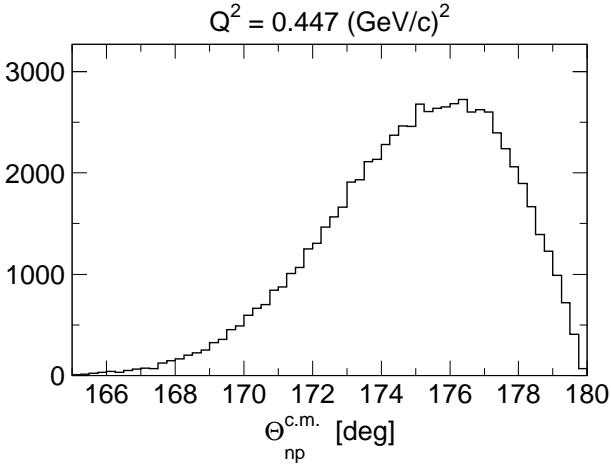}
\caption{Distributions of $\Theta^{\mathrm{c.m.}}_{np}$ after
application of the final set of analysis cuts at $Q^{2} = 0.447$
(GeV/$c$)$^{2}$.}
\label{fig:final-cuts-thetanpcm}
\end{figure}

Distributions of $\Theta^{\mathrm{c.m.}}_{np}$ for those events
surviving all analysis cuts at our lowest $Q^{2}$ point are shown in
Fig.\ \ref{fig:final-cuts-thetanpcm}.  The majority of the accepted
events are seen to fall within $\sim 10$--15$^{\circ}$ of perfect
quasifree emission.  The distributions of
$\Theta^{\mathrm{c.m.}}_{np}$ at our other $Q^{2}$ points are similar,
but are restricted to somewhat smaller ranges, $170^{\circ} <
\Theta^{\mathrm{c.m.}}_{np} < 180^{\circ}$.

Even for perfect quasifree emission, $\Theta^{\mathrm{c.m.}}_{np} =
180^{\circ}$, the PWBA and FSI+MEC+IC+RC calculations of the
$P^{(h)}_{t}/P^{(h)}_{\ell}$ polarization ratio differ by 4.2\% for
the central kinematics of our lowest $Q^{2} = 0.447$ (GeV/$c$)$^{2}$
point and 1.6\% at our highest $Q^{2} = 1.474$ (GeV/$c$)$^{2}$ point.
As the differences between the PWBA and FSI+MEC+IC+RC calculations
increase away from $\Theta^{\mathrm{c.m.}}_{np} = 180^{\circ}$, these
numbers provide essentially lower bounds for the expected magnitude of
corrections for nuclear physics effects.

\subsection{Extraction of acceptance-averaged and nuclear physics corrected
            values for $\bm{G_{En}/G_{Mn}}$}
\label{sec:final-results-extraction}

\subsubsection{Overview of acceptance-averaging analysis procedure}
\label{sec:final-results-extraction-overview}

The recoil polarization component we were interested in was the
projection of the polarization vector on the polarimeter momentum
basis $\hat{S}$-axis following transport through the Charybdis field
and the lead curtain.  We denote this polarization component as
$P'_{S}$, where the prime denotes transport through the dipole field
and lead curtain.  Acceptance-averaged and nuclear physics corrected
values for $G_{En}/G_{Mn}$ were extracted from our experimental
asymmetries and simulations at each $Q^{2}$ point via the following
procedure:

(1) Acceptance-averaged polarizations $\langle P'_{S} \rangle$
computed according to the PWBA and FSI+MEC+IC+RC models were extracted
from simulated data for each precession angle at each $Q^{2}$ point
and for each scale factor $S$ of the Galster parametrization (see
Section \ref{sec:data-analysis-simulations-gengen-program}).

(2) In our ``pairwise analysis method'', for each $S$ factor, we
compared the ratio of the experimental asymmetries to the ratio of the
simulated polarizations for the different precession angle
combinations (i.e., $\chi = 0^{\circ}, \pm90^{\circ}$ and $\chi =
\pm40^{\circ}$) and then computed a $\chi^{2}$ value for each
precession angle combination and each event type [i.e., $(n,n)$ or
$(n,p)$ events] according to
\begin{equation}
\chi^{2} = \frac{\left(\eta_{\mathrm{sim}} - \eta_{\mathrm{exp}}\right)^{2}}
{\left(\Delta\eta_{\mathrm{sim}}\right)^{2} +
\left(\Delta\eta_{\mathrm{exp}}\right)^{2}},
\end{equation}
where $\eta_{\mathrm{sim}} = \langle P'_{S}(0^{\circ})\rangle /
\langle P'_{S}(\pm90^{\circ})\rangle$ for the $\chi = 0^{\circ},
\pm90^{\circ}$ precession angle combination and $\langle
P'_{S}(-40^{\circ}) \rangle / \langle P'_{S}(+40^{\circ}) \rangle$ for
the $\chi = \pm40^{\circ}$ precession angle combination.  The
expressions for $\eta_{\mathrm{exp}}$ are identical, with the
acceptance-averaged polarizations replaced by the experimental
asymmetries.  $\Delta\eta_{\mathrm{sim}}$ and
$\Delta\eta_{\mathrm{exp}}$ denote the statistical errors.  The
resulting $\chi^{2}$ values were fitted as a function of the scale
factor $S$ to a parabolic function, with the optimal value of $S$
defined by the zero of the parabolic fitting function.

(3) In our ``global analysis method'', we compared the experimental
asymmetries with the simulated polarizations via minimization of a
global $\chi^{2}$ value computed according to
\begin{eqnarray}
\chi^{2}\left(A_{y}^{(n,n)},A_{y}^{(n,p)}\right) =
\sum \frac{\left(\xi -
A_{y}^{(n,n),~(n,p)}\langle P'_{S} \rangle\right)^{2}}
{\left(\Delta\xi\right)^{2}+(\Delta\langle P'_{S} \rangle)^{2}}. \nonumber \\
\end{eqnarray}
Here, the sum runs over all 10 asymmetries, $\xi$, and simulated
polarizations, $\langle P'_{S} \rangle$, for each $Q^{2}$ point [i.e.,
5 different precession angles, and $(n,n)$ and $(n,p)$ events], and
$A_{y}^{(n,n)}$ and $A_{y}^{(n,p)}$ denote the polarimeter's analyzing
power for $(n,n)$ and $(n,p)$ events.  $\Delta\xi$ and $\Delta \langle
P'_{S} \rangle$ denote the statistical errors.  In this analysis, the
analyzing powers and scale factor $S$ were treated as free parameters,
with the optimal values extracted from the minimal $\chi^{2}$ value.

We note that the simulation statistical errors were generally an order
of magnitude smaller than the experimental statistical errors.

\subsubsection{Acceptance-averaged values of $Q^{2}$}
\label{sec:final-results-extraction-Q2}

The acceptance-averaged values of $Q^{2}$, denoted $\langle Q^{2}
\rangle$, were determined to be $\langle Q^{2} \rangle = 0.447$,
1.126, 1.158, and 1.450 (GeV/$c$)$^{2}$ for the central $Q^{2} =
0.447$, 1.136, 1.169, and 1.474 (GeV/$c$)$^{2}$ points, respectively.
The distribution of $Q^{2}$ values for the $\langle Q^{2} \rangle =
0.447$ (GeV/$c$)$^{2}$ point was sharply peaked around the central
value of 0.447 (GeV/$c$)$^{2}$, while the distributions of $Q^{2}$
values for the $\langle Q^{2} \rangle = 1.126/1.158$ and 1.450
(GeV/$c$)$^{2}$ points were integrated from $\sim 1.0$ to $\sim 1.3$
(GeV/$c$)$^{2}$ and from $\sim 1.2$ to $\sim 1.7$ (GeV/$c$)$^{2}$,
respectively.

Henceforth, we will use $\langle Q^{2} \rangle = 1.132$
(GeV/$c$)$^{2}$ to denote the sample-size weighted average of the
$\langle Q^{2} \rangle = 1.126$ and 1.158 (GeV/$c$)$^{2}$ data sets.

\subsubsection{Acceptance-averaging analysis iterations}
\label{sec:final-results-extraction-iterations}

We performed two iterations of the above-described analysis procedure
with both the \texttt{Acceptance} and \texttt{GENGEN} simulation
programs.

In the first iteration, the simulations were conducted with the PWBA
and FSI+MEC+IC+RC calculations that assumed different multiplicative
factors of the standard Galster parametrization for the $Q^{2}$
dependence of $G_{En}$.  The optimal values for the scale factors $S$
were then used to compute the optimal values for $G_{En}/G_{Mn}$
according to $G_{En}/G_{Mn} = -S_{\mathrm{optimal}} \times
\langle\tau\rangle / (1 + 5.6\langle\tau\rangle)$, where
$\langle\tau\rangle = \langle Q^{2} \rangle / 4m_{n}^{2}$.  Values for
$G_{En}$ were then extracted from our optimal values for
$G_{En}/G_{Mn}$ using the best-fit values for $G_{Mn}$ taken from the
parametrization of Kelly \cite{kelly02}.  Then we fitted our
first-iteration results for $G_{En}$ together with the then-available
world data on $G_{En}$ (as of early 2003) to the modified Galster
parametrization described previously in Section
\ref{sec:data-analysis-simulations-gengen-program}; the best-fit
parameters we found at that time were $a = 0.894 \pm 0.023$ and $b =
3.55 \pm 0.37$.  This fit included the then-available data on $G_{En}$
extracted from measurements using polarization degrees of freedom
\cite{eden94,becker99,herberg99,passchier99,bermuth03,zhu01} and an
analysis of the deuteron quadrupole form factor \cite{schiavilla01},
and also data on the slope of $G_{En}$ as measured via low-energy
neutron scattering from electrons in heavy atoms \cite{kopecky97}.
Since this analysis, new data on $G_{En}$ have been published
\cite{warren04,glazier05}, and a new modified Galster parametrization
has been published \cite{kelly04}.

In our second analysis iteration, a second set of the PWBA and
FSI+MEC+IC+RC calculations were performed that assumed this modified
Galster parametrization for the $Q^{2}$ dependence of $G_{En}$.  The
\texttt{Acceptance} and \texttt{GENGEN} simulations were both repeated
using these new calculations, and the procedure for the extraction of
the optimal $G_{En}/G_{Mn}$ values was identical to that of the first
iteration.

The differences between the first and second analysis iterations were
negligible.  This result is not surprising, because: (1) both
parametrizations have small second derivatives in the vicinity of our
$Q^{2}$ points; and (2) the acceptance was fairly symmetric about the
acceptance-averaged values of $Q^{2}$.

\subsubsection{Acceptance-averaging analysis results}
\label{sec:final-results-extraction-results}

The pairwise analysis method was employed for the extraction of our
$G_{En}/G_{Mn}$ values at $\langle Q^{2} \rangle = 0.447$
(GeV/$c$)$^{2}$ (only two precession angles), while the global
analysis method was employed for the analysis of our $\langle Q^{2}
\rangle = 1.132$ and 1.450 (GeV/$c$)$^{2}$ data sets.  The final
acceptance-averaged and nuclear physics corrected values for
$G_{En}/G_{Mn}$ we obtained with the \texttt{Acceptance} program and
\texttt{GENGEN} agreed to better than 1\% at $\langle Q^{2} \rangle =
0.447$ and 1.132 (GeV/$c$)$^{2}$ and 2\% at $\langle Q^{2} \rangle =
1.450$ (GeV/$c$)$^{2}$, well within the statistical errors; therefore,
the values for $G_{En}/G_{Mn}$ we report later in Table
\ref{tab:summary-final-results} are the average of the central values
obtained with our two simulation programs.  The analyzing powers we
extracted from our acceptance-averaging analysis procedures are
summarized in Table \ref{tab:analyzing-powers}.

\begin{table}
\caption{Analyzing powers for $(n,n)$ and $(n,p)$ events at each
of our $Q^{2}$ points.  The errors are statistical.}
\begin{ruledtabular}
\begin{tabular}{cccc}
Event& \multicolumn{3}{c}{$\langle Q^{2} \rangle$ [(GeV/$c$)$^{2}$]} \\
Type& 0.447& 1.132& 1.450 \\ \hline
$(n,n)$& $0.141 \pm 0.004$& $0.137 \pm 0.010$& $0.144 \pm 0.013$ \\
$(n,p)$& $0.103 \pm 0.005$& $0.075 \pm 0.007$& $0.071 \pm 0.011$
\end{tabular}
\end{ruledtabular}
\label{tab:analyzing-powers}
\end{table}

\subsection{Systematic uncertainties}
\label{sec:final-results-systematics}

An itemized summary of estimates for the magnitudes of our relative
systematic uncertainties in $G_{En}/G_{Mn}$ appears in Table
\ref{tab:systematic-uncertainties}.  Our final values for the total
relative systematic uncertainties, 2--3\%, are much smaller than our
relative statistical uncertainties.  Brief discussions of each
itemized systematic uncertainty (and others deemed negligibly small)
appear below.

\subsubsection{Beam polarization}
\label{sec:final-results-systematics-beampol}

The beam polarization cancels in the form factor ratio only if it does
not vary during sequential measurements of the scattering asymmetries.
Consequently, fluctuations in the beam polarization measurements
introduce a systematic uncertainty.  We estimated the temporal
uncertainty in the beam polarization via the following procedure.
First, polarization measurements conducted under similar conditions at
the polarized source were grouped into clusters.  Second, the mean
value of the polarization for each cluster was computed and then
recentered about the nominal 80\% polarization.  Next, the statistical
error for the entire data set (i.e., all identified clusters) was
computed, and the overall uncertainty was then increased by the square
root of $\chi^{2}$ (to account for the observed fluctuations).
Finally, our total estimated uncertainty in the polarization was
propagated through the expression for the form factor ratio, Eq.\
(\ref{eq:phase-shift-delta}).

\begin{table}
\caption{Compilation of our estimated relative systematic
uncertainties in $G_{En}/G_{Mn}$ [\%].  The total systematic error that
is quoted for each $Q^{2}$ point and precession angle combination
is the quadrature sum of the itemized systematic uncertainties.}
\begin{ruledtabular}
\begin{tabular}{lccccc}
& \multicolumn{5}{c}{$\langle Q^{2} \rangle$ [(GeV/$c$)$^{2}$]} \\
Source&
  0.447\footnotemark[1]&
  1.132\footnotemark[1]&
  1.132\footnotemark[2]&
  1.450\footnotemark[1]&
  1.450\footnotemark[2] \\ \hline
Beam polarization& 1.6& 0.7& 0.4& 1.2& 0.3 \\
Charge-exchange& $<$0.1& $<$0.1& 0.1& $<$0.1& 0.2 \\
Depolarization& $<$0.1& 0.1& $<$0.1& $<$0.1& 0.6 \\
Positioning/traceback& 0.2& 0.3& 0.3& 0.4& 0.4 \\
Precession angle& 1.1& 0.3& 0.1& 0.5& 0.1 \\
Radiative corrections& 0.7& 0.1& 0.1& 0.1& 0.1 \\
Timing calibration& 2.0& 2.0& 2.0& 2.0& 2.0 \\ \hline
Total& 2.9& 2.2& 2.1& 2.4& 2.2
\end{tabular}
\end{ruledtabular}
\footnotetext[1]{$\chi = \pm40^{\circ}$ precession.}
\footnotetext[2]{$\chi = 0^{\circ}, \pm90^{\circ}$ precession.}
\label{tab:systematic-uncertainties}
\end{table}

\subsubsection{Charge-exchange in the lead curtain}
\label{sec:final-results-systematics-charge-exchange}

Estimates of the contamination levels from the two-step
$^{2}$H$(\vec{e},e'\vec{p})$ + Pb$(\vec{p},\vec{n})$ charge-exchange
reaction were given previously in Section
\ref{sec:data-analysis-asymmetryresults-leadcurtain}.  To estimate the
systematic uncertainty in $G_{En}/G_{Mn}$ due to contamination from
this background process, we computed values for the recoil proton's
polarization using values for $G_{Ep}$ and $G_{Mp}$ taken from the
parametrization of \cite{brash02}.  These polarization components were
then transported through the Charybdis dipole field using estimates
for the proton spin precession angles.  As there are very few data on
the lead polarization transfer coefficient, $D_{SS}^{\mathrm{Pb}}$, we
calculated the correction to the asymmetries (using information on the
analyzing powers extracted from our acceptance-averaging analysis and
the values for, and the uncertainties in, the charge-exchange
contamination levels) for various (reasonable) choices of
$D_{SS}^{\mathrm{Pb}}$.  Spreads in the resulting values of
$G_{En}/G_{Mn}$ were then defined to be the systematic uncertainties.

\subsubsection{Neutron depolarization in the lead curtain}
\label{sec:final-results-systematics-depolarization}

The total $n + \mathrm{Pb}$ cross section is fairly flat at $\sim 3$
barns over the range of neutron kinetic energies in our experiment
(slow rise with energy) \cite{finlay93}.  For our 10.16-cm thick lead
curtain, our \texttt{GENGEN} simulations indicated a 30.8\%, 42.5\%,
43.0\%, and 46.7\% interaction probability for the neutron energies at
our $\langle Q^{2} \rangle = 0.447$, 1.126, 1.158, and 1.450
(GeV/$c$)$^{2}$ points, respectively.  We found that the contamination
levels within our $[-1,1]$ ns cTOF window from neutrons suffering one
or more interactions in the lead curtain were 0.04\%, 3.8\%, 4.2\%,
and 9.3\% at $\langle Q^{2} \rangle = 0.447$, 1.126, 1.158, and 1.450
(GeV/$c$)$^{2}$, respectively.  The fact that our simulations
predicted a much more rapid increase in the contamination levels with
energy as compared to the interaction probabilities is because the
angular distributions for $nn$ and $np$ scattering peak at large
(small) scattering angles for the neutron kinetic energies at our
lowest (highest) $Q^{2}$ point (as computed by \texttt{SAID}
\cite{arndt03}).  Further, our simulations suggested that interactions
in the lead curtain may have been partly responsible for the small
tail observed on the slow side of our experimental cTOF distributions
at our highest $Q^{2}$ point (see Fig.\ \ref{fig:ctof-final-cuts}).

\begin{figure}
\includegraphics[angle=270,scale=0.72]{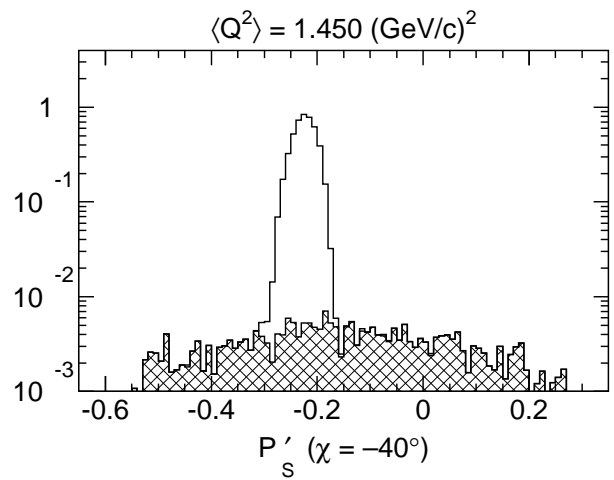}
\caption{Sample \texttt{GENGEN} simulated $\langle P'_{S} \rangle$
spectrum for $\chi = -40^{\circ}$ precession at $\langle Q^{2} \rangle
= 1.450$ (GeV/$c$)$^{2}$.  The unfilled histogram is summed over all
simulated events, while the cross-hatched histogram is summed over
those events suffering one or more interactions in the lead curtain.
The units of the ordinate are arbitrary.}
\label{fig:kin3-pm-lead-polarization}
\end{figure}

The quantity of interest was the spectrum of the polarization
presented to the polarimeter front array for neutrons which did and
did not interact with the lead curtain.  A sample result comparing
polarization spectra for these two types of events for $\chi =
-40^{\circ}$ precession at $\langle Q^{2} \rangle = 1.450$
(GeV/$c$)$^{2}$ is shown in Fig.\ \ref{fig:kin3-pm-lead-polarization}.
Our simulations indicated that the distribution of polarizations for
neutrons suffering an interaction in the lead curtain is a broad
continuum, yielding a depolarization of the neutron flux presented to
the polarimeter.  Similar results were observed at our other $Q^{2}$
points.  We found, though, that the effects of depolarization in the
lead curtain tend to cancel in the polarization ratio, leading to
small systematic uncertainties in the $G_{En}/G_{Mn}$ ratio.  The
magnitudes of the residual non-cancellations were taken to be the
uncertainties listed in Table \ref{tab:systematic-uncertainties}.

\subsubsection{Positioning and traceback}
\label{sec:final-results-systematics-positioning-traceback}

Two contributions to an uncertainty in the electron scattering angle
were considered: positioning (offset in the scattering angle from the
nominal value) and traceback (reconstruction from the focal plane to
the target).  For the purposes of this analysis, we assumed the
uncertainties in the electron scattering angle, $\Delta \theta_{e'}$,
were $\Delta \theta_{e'} = 1.2$ mrad and 1.3 mrad for the positioning
and traceback uncertainties, respectively; these values were derived
from a systematic analysis of kinematic data taken during this
experiment.  The systematic uncertainties in $G_{En}/G_{Mn}$ were
obtained via propagation of these values for $\Delta \theta_{e'}$
through Eq.\ (\ref{eq:phase-shift-delta}) for the form factor ratio.

\subsubsection{Precession angle}
\label{sec:final-results-systematics-precession-angle}

\begin{table*}
\caption{Summary of our final results for $G_{En}/G_{Mn}$ and
$G_{En}$.  The first (second) set of errors is statistical
(systematic).  The results reported here are the weighted average of
$(n,n)$ and $(n,p)$ events in the polarimeter.}
\begin{ruledtabular}
\begin{tabular}{lcccc}
& & \multicolumn{3}{c}{$\langle Q^{2} \rangle$ [(GeV/$c$)$^{2}$]} \\
Analysis& Quantity& 0.447& 1.132& 1.450 \\ \hline
$n(\vec{e},e'\vec{n})$& $G_{En}/G_{Mn}$& $-0.0681 \pm 0.0084 \pm 0.0020$&
  $-0.122 \pm 0.011 \pm 0.003$& $-0.174 \pm 0.017 \pm 0.004$ \\
$^{2}$H$(\vec{e},e'\vec{n})^{1}$H PWBA& $G_{En}/G_{Mn}$&
  $-0.0713 \pm 0.0086 \pm 0.0021$& $-0.126 \pm 0.010 \pm 0.003$&
  $-0.183 \pm 0.018 \pm 0.004$ \\
$^{2}$H$(\vec{e},e'\vec{n})^{1}$H FSI+MEC+IC+RC& $G_{En}/G_{Mn}$&
  $-0.0755 \pm 0.0089 \pm 0.0022$& $-0.131 \pm 0.011 \pm 0.003$&
  $-0.189 \pm 0.018 \pm 0.004$ \\ \hline
Values from \cite{kelly04}& $G_{Mn}/\mu_{n}G_{D}$&
  $1.003 \pm 0.005$& $1.067 \pm 0.012$& $1.064 \pm 0.016$ \\ \hline
$n(\vec{e},e'\vec{n})$& $G_{En}$& $0.0492 \pm 0.0061 \pm 0.0015$&
  $0.0370 \pm 0.0032 \pm 0.0009$& $0.0383 \pm 0.0038 \pm 0.0011$ \\
$^{2}$H$(\vec{e},e'\vec{n})^{1}$H PWBA& $G_{En}$&
  $0.0515 \pm 0.0062 \pm 0.0015$& $0.0381 \pm 0.0032 \pm 0.0009$&
  $0.0403 \pm 0.0039 \pm 0.0011$ \\
$^{2}$H$(\vec{e},e'\vec{n})^{1}$H FSI+MEC+IC+RC& $G_{En}$&
  $0.0545 \pm 0.0064 \pm 0.0016$& $0.0396 \pm 0.0032 \pm 0.0010$&
  $0.0415 \pm 0.0039 \pm 0.0011$
\end{tabular}
\end{ruledtabular}
\label{tab:summary-final-results}
\end{table*}

Uncertainties in the neutron spin precession angle were estimated
\cite{charybdis} via a calculational scheme that employed the
reconstructed kinematics from the experimental data as the source of
the neutron momentum vectors incident on the Charybdis dipole field.
Spin vectors were transported through the field using the same
magnetic spin transport algorithms developed for our two simulation
programs.  This technique provided a measure of the sensitivity of the
precession angle to details of the field map.  The uncertainties in
the mean values of the precession angles derived from these studies
(at the level of $\pm 0.2^{\circ}$) were combined in quadrature with
two other sources of uncertainty.  First, as discussed in Section
\ref{sec:neutron-polarimeter-charybdis}, we observed small differences
between the measured field integrals for opposite magnet polarities
and also between the field integrals derived from our measured maps
and the calculated \texttt{TOSCA} maps.  These uncertainties were
estimated to be on the level of $\pm 0.3^{\circ}$.  Second, as also
discussed in Section \ref{sec:neutron-polarimeter-charybdis}, the
field was mapped only along the central axis; therefore, we assigned
further uncertainties (at the level of $\pm 0.2^{\circ}$) for
incomplete knowledge of the field beyond the central axis.  Our best
estimates of the total uncertainties in the precession angle were then
propagated through the form factor ratio, Eq.\
(\ref{eq:phase-shift-delta}).

\subsubsection{Radiative corrections}
\label{sec:final-results-systematics-radiative}

Radiative corrections were calculated specifically for the kinematics
of this experiment by Afanasev \textit{et al}.\ \cite{afanasev01}.
The primary effect of radiative corrections on the recoil polarization
components $P^{(h)}_{t}$ and $P^{(h)}_{\ell}$ was found to be
depolarization of the electron such that both components of the recoil
polarization should be increased by $\sim 1.9$\%, $\sim 3.7$\%, and
$\sim 4.4$\% at $\langle Q^{2} \rangle = 0.447$, 1.132, and 1.450
(GeV/$c$)$^{2}$, respectively; however, these corrections nearly
cancel in the form factor ratio such that the net effect is small at
$\langle Q^{2} \rangle = 0.447$ (GeV/$c$)$^{2}$ and negligible at the
two higher $Q^{2}$ points.  The residual non-cancellations of the
corrections in the form factor ratio were taken to be the systematic
uncertainties we quote in Table \ref{tab:systematic-uncertainties}.

\subsubsection{Timing calibration of the polarimeter}
\label{sec:final-results-systemtics-timing-calibration}

The timing calibrations we deemed suitable for certain running
conditions (e.g., periods in between changes to the high-voltages for
the PMTs) were obtained using a subset of the data for that particular
running period.  To assess the dependence of our results for the
scattering asymmetries on the choice of the subset of data employed
for the timing calibration, various calibrations were generated from
different subsets of the available data.  Excellent agreement was
always found between the results for the scattering asymmetries
obtained from analyses using these different calibrations; however, we
did find a $\sim 2$\% sensitivity of our results to the choice of the
subset of data employed for the timing calibration.

\subsubsection{Other uncertainties}
\label{sec:final-results-systematics-others}

We deemed two other possible sources of systematic uncertainties to be
negligible.  First, we demonstrated quantitatively that our scattering
asymmetries were insensitive (within statistical errors) to a possible
geometric asymmetry in the polarimeter (i.e., a spin-averaged
``top-bottom'' asymmetry) by varying our software energy thresholds on
the top (bottom) rear array while maintaining a constant threshold on
the bottom (top) rear array.  Second, analysis of our data taken with
the ``dummy targets'' (see Section
\ref{sec:scattering-chamber-cryotargets}) showed that the level of
contamination within our $[-1,1]$ ns cTOF window from scattering in
the target cell windows was negligible ($<0.05$\%).

\subsection{Summary of final $\bm{G_{En}/G_{Mn}}$ and $\bm{G_{En}}$ results}
\label{sec:final-results-summary}

\begin{figure}
\includegraphics[scale=0.55]{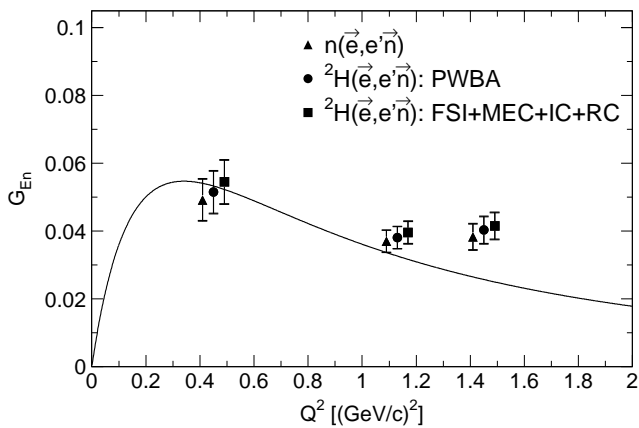}
\caption{Comparison of our results for $G_{En}$ at $\langle Q^{2}
\rangle = 0.447$, 1.132, and 1.450 (GeV/$c$)$^{2}$ extracted from the
various analyses summarized in Table \ref{tab:summary-final-results}.
The data points shown for the three analyses at each $\langle Q^{2}
\rangle$ point have been slightly displaced about the actual $\langle
Q^{2} \rangle$ value for clarity.  The solid curve is the Galster
parametrization \cite{galster71}.}
\label{fig:point-pwba-fsi-gen-results}
\end{figure}

Our final results for $G_{En}/G_{Mn}$ and $G_{En}$ extracted from
three different analyses are tabulated in Table
\ref{tab:summary-final-results} and compared in Fig.\
\ref{fig:point-pwba-fsi-gen-results}.  The three analyses are for: (1)
elastic $n(\vec{e},e'\vec{n})$ scattering and infinitesimal HMS and
NPOL point acceptances; (2) quasielastic
$^{2}$H$(\vec{e},e'\vec{n})^{1}$H scattering and acceptance-averaging
of the PWBA model; and (3) quasielastic
$^{2}$H$(\vec{e},e'\vec{n})^{1}$H scattering and acceptance-averaging
of the FSI+MEC+IC+RC model.

We note that in our first publication \cite{madey03} we used values
for $G_{Mn}$ taken from \cite{kelly02}; here we use slightly different
values for $G_{Mn}$ taken from \cite{kelly04}.  The total systematic
errors we quote for $G_{En}$ are the quadrature sum of the
experimental sytematic errors and the relative uncertainties in
$G_{Mn}$.


\subsection{Comparison of nucleon form factor data with selected
            theoretical model calculations}
\label{sec:final-results-data-theory}

The availability of precise new data on nucleon form factors has
stimulated much more theoretical work in the last few years than we
can review here; our selection of models is not intended to be
complete.  Although the present experiment is limited to $G_{En}$, we
believe that comparison with models must consider all four Sachs form
factors.  In this section, we compare representative models with
selected data.  The data from this experiment are shown as filled
circles in Figs.\ \ref{fig:vmd}--\ref{fig:qm} while data from other
experiments are shown as open circles.  We selected
\cite{simon80,price71} for $G_{Ep}$,
\cite{jones00,gayou02,pospischil01,milbrath99} for $G_{Ep}/G_{Mp}$,
\cite{hohler76,brash02} for $G_{Mp}$,
\cite{passchier99,herberg99,becker99,zhu01,bermuth03,warren04,glazier05}
for $G_{En}$, and \cite{anklin94,anklin98,lung93,kubon02,xu03} for
$G_{Mn}$.

\subsubsection{Models based upon vector meson dominance}
\label{sec:final-results-data-theory-vmd}

\begin{figure*}
\includegraphics[scale=0.71,angle=90]{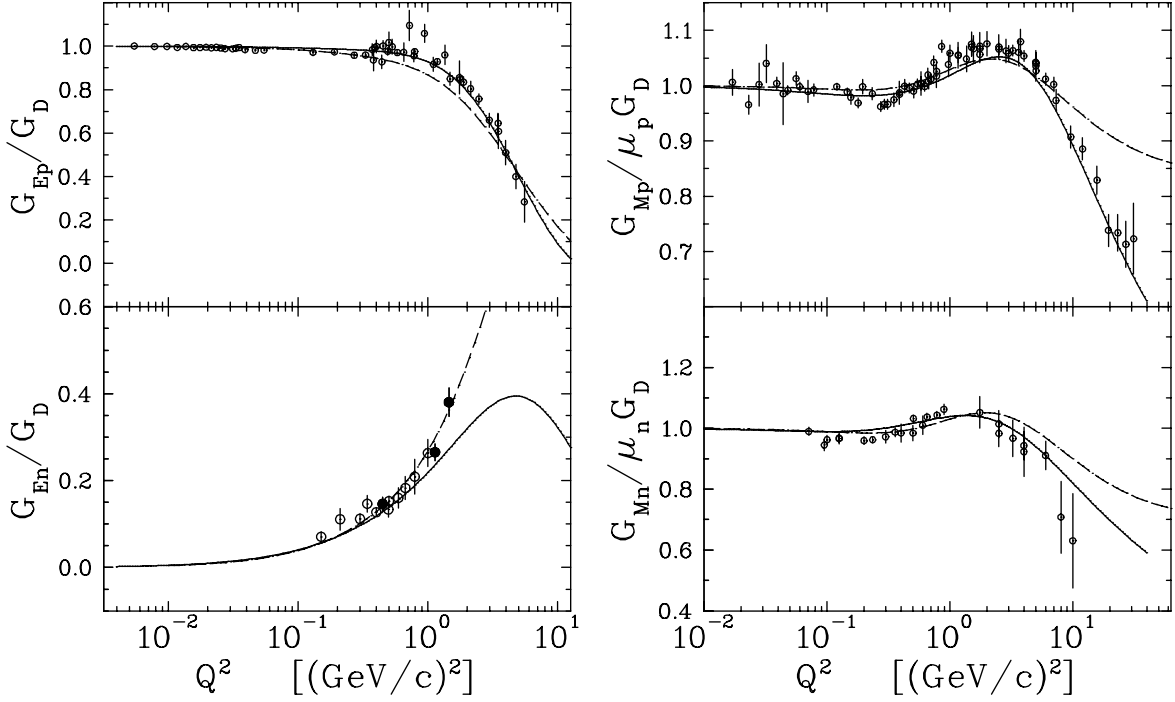}
\caption{Comparison of representative VMD models with nucleon form
factor data ($G_{Ep}$ from \cite{simon80,price71}; $G_{Ep}/G_{Mp}$
from \cite{jones00,gayou02,pospischil01,milbrath99}; $G_{Mp}$ from
\cite{hohler76,brash02}; $G_{En}$ from
\cite{passchier99,herberg99,becker99,zhu01,bermuth03,warren04,glazier05};
$G_{Mn}$ from \cite{anklin94,anklin98,lung93,kubon02,xu03}).  Dashed
curve: Bijker and Iachello \cite{bijker04}.  Solid curve: version
GKex(02S) of Lomon \cite{lomon02}.}
\label{fig:vmd}
\end{figure*}

Models based upon vector meson dominance (VMD) postulate that the
virtual photon couples either directly to an intrinsic nucleon core or
through the mediation of a neutral vector meson, initially limited to
the lowest $\omega$, $\rho$, and $\phi$ mesons.  The earliest versions
assumed that the core is a structureless Dirac particle.  Iachello
\textit{et al}.\ \cite{iachello73} assigned the core a form factor and
employed a model of the $\rho$ width.  Gari and Kr\"umpelmann
\cite{gari85,gari92a} then permitted different Dirac and Pauli
intrinsic form factors and introduced modifications that ensured
consistency with pQCD scaling at large $Q^2$ and logarithmic running
of the strong coupling constant.  Bijker and Iachello \cite{bijker04}
adopted the Gari and Kr\"umpelmann (GK) pQCD prescriptions and refit
their model to modern data, still using a common intrinsic form
factor.  This fit, using a total of 6 free parameters, is compared
with the data in Fig.\ \ref{fig:vmd}.  Finally, Lomon
\cite{lomon01,lomon02} produced a more flexible set of fits using a
model described as ``GK extended''; the GKex(02S) version is also
shown in Fig.\ \ref{fig:vmd}.  The Lomon model uses two intrinsic form
factors, the GK prescription for the pQCD limit, and includes
$\rho^\prime(1450)$ and $\omega^\prime(1419)$ couplings in addition to
the customary $\rho$, $\omega$, and $\phi$ couplings.  The $\rho$
width is included but the $\rho^\prime$ and $\omega^\prime$ structures
are not.  The fit achieved by this extended model, with 13 free
parameters, is clearly superior, especially at large $Q^2$.  The
Bijker and Iachello model describes the qualitative behavior of
$G_{Ep}$, but its transition between $G_{Ep}/G_D \approx 1$ at low
$Q^2$ and the nearly linear decrease for $1 < Q^2 < 6$ (GeV/$c$)$^2$
is too gradual.  Nor does it reproduce the slope in
$G_{Mp}/\mu_{p}G_D$ for $Q^2 > 10$ (GeV/$c$)$^2$.  Both of these
features are fit well by the Lomon model.  Unfortunately, the neutron
data do not discriminate between these models very strongly.  The
Bijker and Iachello model provides a slightly better fit to the
present $G_{En}$ data, but the Lomon fit was performed before these
data became available; it is likely that only a slight parameter
adjustment would be needed to achieve a comparable fit without
sacrificing the fits to other form factors.  It will be interesting to
see whether the rather large values for $G_{En}/G_D$ for $Q^2 > 2$
(GeV/$c$)$^2$ predicted by the Bijker and Iachello model are confirmed
by upcoming experiments \cite{e02013,e04110} that will probe $G_{En}$
to $Q^{2} = 4.3$ (GeV/$c$)$^{2}$.  Note, however, that the Bijker and
Iachello fit is systematically above the $G_{Mn}$ data for the same
kinematics, $Q^2 > 2$ (GeV/$c$)$^2$.

\subsubsection{Models emphasizing the pion cloud}
\label{sec:final-results-data-theory-pion}

\begin{figure*}
\includegraphics[scale=0.71,angle=90]{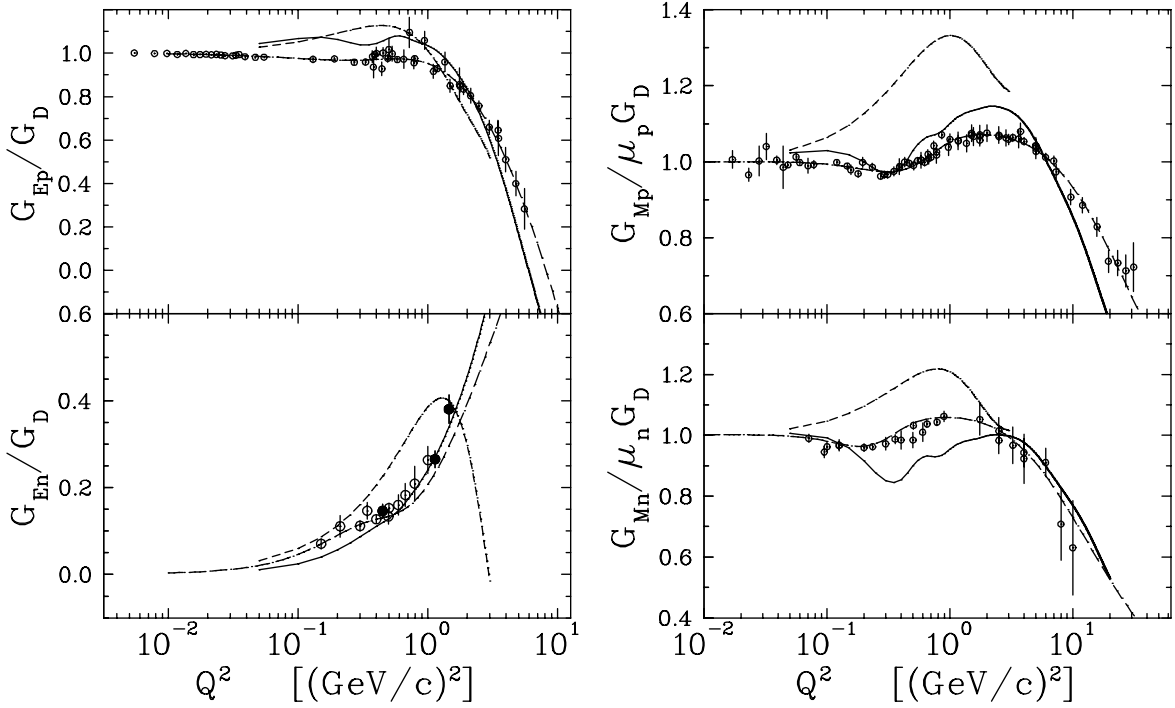}
\caption{Comparison of representative pion cloud models with nucleon
form factor data ($G_{Ep}$ from \cite{simon80,price71};
$G_{Ep}/G_{Mp}$ from \cite{jones00,gayou02,pospischil01,milbrath99};
$G_{Mp}$ from \cite{hohler76,brash02}; $G_{En}$ from
\cite{passchier99,herberg99,becker99,zhu01,bermuth03,warren04,glazier05};
$G_{Mn}$ from \cite{anklin94,anklin98,lung93,kubon02,xu03}).
Short-dashed curve: QMC model \cite{lu98a}.  Solid curve: LFCBM
\cite{miller02}. Long-dashed curve: Friedrich and Walcher
parametrization \cite{friedrich03}.}
\label{fig:pion}
\end{figure*}

The role of the pion in mediation of the long-range nucleon-nucleon
interaction clearly demonstrates its importance in understanding form
factors for low $Q^2$.  Typical pion cloud models describe nucleon
form factors using diagrams in which the virtual photon couples to
either a bare nucleon core or to the nucleon or the pion loop in a
single-pion loop.  Some models also permit excitation of the
intermediate state and include additional contact terms.  A relatively
simple example is the Adelaide version \cite{lu98a} of the cloudy bag
model (CBM) in which the core is based upon the bag model,
intermediate excitation is neglected, and relativistic corrections are
made using a simple ansatz for Lorentz contraction \cite{licht70a}.
Predictions from Lu {\it et al}.\ \cite{lu98a} using a bag radius of
0.8 fm are compared with the data in Fig.\ \ref{fig:pion}.  Although
density-dependent extensions of this model, described as the
quark-meson coupling (QMC) model, have been used to study the
sensitivity of recoil polarization in nucleon electromagnetic knockout
to medium modifications of the nucleon form factors
\cite{lu98b,lu99,dieterich01,strauch03}, its description of free form
factors is rather poor and one must hope that the density dependence
of $G_{E}/G_{M}$ ratios is more accurate.

Alternatively, the light front cloudy bag model (LFCBM) of Miller
\cite{miller02} maintains Poincar\'e invariance by formulating wave
functions using the light-front approach.  This version should then be
applicable to higher $Q^2$.  There are only 4 adjustable parameters
and the results for Set 1 are compared with data in Fig.\
\ref{fig:pion}.  A previous version of this model \cite{frank96}
provided one of the earliest predictions of the sharp slope in
$G_{Ep}/G_{Mp}$ for $Q^2 > 1$ (GeV/$c$)$^2$, but the agreement with
recent recoil-polarization data is only qualitative.  The LFCBM
calculation for $G_{Mp}/G_D$ also decreases too rapidly at large
$Q^2$.  Calculations using this model agree relatively well with the
$G_{En}$ data for $Q^2 \gtrsim 1$ (GeV/$c$)$^2$ but are too small at
lower $Q^2$.  Interestingly, this model predicts much stronger values
for $G_{En}/G_D$ at large $Q^2$ than the Lomon parametrization.
However, the LFCBM calculations for three of the four form factors
show complicated and rather implausible shapes for $Q^2 < 1$
(GeV/$c$)$^2$ that disagree strongly with data.

Chiral effective field theory \cite{fuchs04} provides a more
systematic procedure that includes intermediate excitation and can be
extended to two pion loops \cite{schindler04}.  Alternatively,
two-loop contributions can be evaluated in dispersion theory
\cite{hammer04}.  Recently it has become possible also to include both
pion loops and vector meson diagrams in a consistent manner
\cite{schindler05}; however, we do not show curves here because this
approach remains limited to $Q^2 \lesssim 0.4$ (GeV/$c$)$^2$.

Friedrich and Walcher \cite{friedrich03} performed a phenomenological
analysis of the nucleon electromagnetic form factors using a
parametrization motivated by pion cloud models.  The core form factor
is represented by two dipole form factors with different ranges while
the pion cloud contribution, represented as a ``bump'' at low $Q^2$,
is described by two Gaussians.  These fits, with 5 free parameters for
$G_{En}$ and 6 for each of the other form factors, are also compared
with data in Fig.\ \ref{fig:pion}.  The quality of these fits is
generally satisfactory, but it is not clear that the postulated
oscillation in $G_{En}/G_D$ is warranted by the available data;
considerably better experimental precision at $Q^2 \sim 0.3$
(GeV/$c$)$^2$ would be needed to justify such a structure.

\begin{figure}
\includegraphics[scale=0.47]{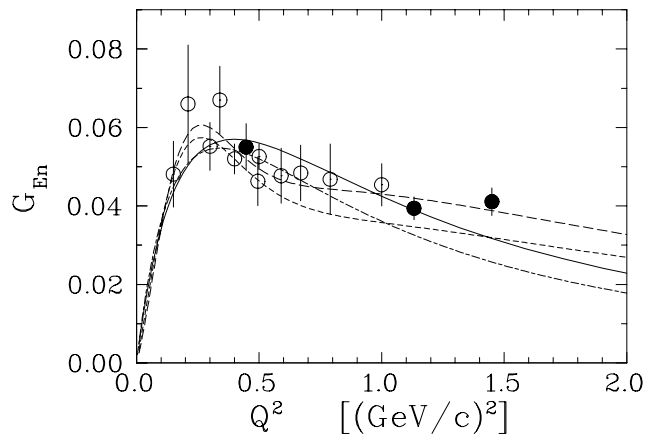}
\caption{Closer look at comparison of representative pion cloud models
with data on $G_{En}$ (data from
\cite{passchier99,herberg99,becker99,zhu01,bermuth03,warren04,glazier05}).
Solid curve: a fit based upon the pion cloud model of Kaskulov and
Grabmayr \cite{kaskulov04}.  Short-dashed curve: parametrization of
Friedrich and Walcher \cite{friedrich03}.  Long-dashed curve:
re-analysis by Glazier \textit{et al}.\ \cite{glazier05} using the
Friedrich and Walcher model.  The dash-dotted curve is the original
Galster parametrization \cite{galster71}.}
\label{fig:kaskulov}
\end{figure}

A closer look at the $G_{En}$ data is given in Fig.\
\ref{fig:kaskulov}.  The original Friedrich and Walcher fit
(short-dashed curve) used a very preliminary version of the data from
the present experiment and falls systematically below the final data
for this and other more recent experiments for $Q^2 > 0.5$
(GeV/$c$)$^2$.  A re-analysis using final data for this experiment
plus new data \cite{bermuth03,warren04,glazier05} was made by Glazier
\textit{et al}.\ \cite{glazier05} and is shown as the long-dashed
curve featuring a bump for $Q^2 \sim 0.3$ (GeV/$c$)$^2$ superimposed
upon a much flatter core form factor.  With 5 parameters it is
obviously possible to fit the data very well, perhaps too well --- the
simple two-parameter fit of Kelly \cite{kelly04} based upon the
Galster parametrization already provides $\chi^2_\nu = 0.8$ without
distinguishing between soft and hard structures.  The data presently
available do not require this complication.  Data at higher $Q^2$
should test whether such a hard core is needed but significantly more
precise data for low $Q^2$ would be needed to establish the soft pion
cloud contribution to $G_{En}$.

\begin{figure*}
\includegraphics[scale=0.71,angle=90]{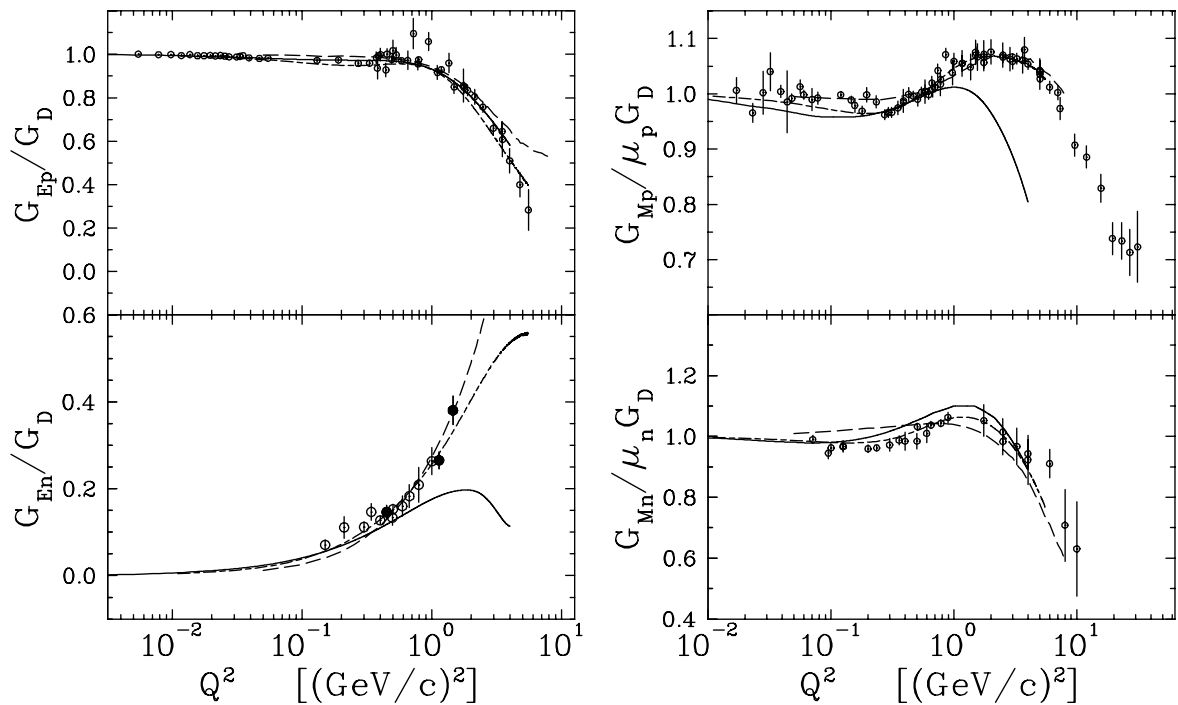}
\caption{Comparison of representative quark models with nucleon form
factor data ($G_{Ep}$ from \cite{simon80,price71}; $G_{Ep}/G_{Mp}$
from \cite{jones00,gayou02,pospischil01,milbrath99}; $G_{Mp}$ from
\cite{hohler76,brash02}; $G_{En}$ from
\cite{passchier99,herberg99,becker99,zhu01,bermuth03,warren04,glazier05};
$G_{Mn}$ from \cite{anklin94,anklin98,lung93,kubon02,xu03}).  Solid
curve: PFSA using pointlike constituents \cite{wagenbrunn01}.
Long-dashed curve: light-front using OGE interaction and
constituent-quark form factors \cite{simula01}.  Dash-dotted curve:
hCQM with constituent-quark form factors \cite{desanctis05}.}
\label{fig:qm}
\end{figure*}

Finally, Kaskulov and Grabmayr \cite{kaskulov04} used a chiral quark
model ($\chi$QM) to derive a relationship
\begin{equation}
G_{En} = \bar{S} (1-F_\pi)G_C,
\end{equation}
between $G_{En}$, the pion form factor $F_\pi$, and the core form
factor $G_C$ for the 3-quark component of the nucleon.  The
coefficient $\bar{S}$ is a weighted average over spectroscopic factors
for $N$ and $\Delta$ intermediate states in the one-pion loop
contribution to the self-energy but is treated as an adjustable
parameter.  If one stipulates a monopole for $F_\pi =
(1+Q^2/\Lambda_\pi^2)^{-1}$ and a dipole for $G_C =
(1+Q^2/\Lambda_C^2)^{-2}$, the neutron electric form factor
\begin{equation}
G_{En} = \bar{S} \frac{b \tau}{1+b\tau} G_C,
\end{equation}
with $b=4m_N^2/\Lambda_\pi^2$ reduces to a Galster-like form with up
to 3 free parameters ($\bar{S}$, $\Lambda_\pi$, $\Lambda_C$);
however, $\bar{S}$ is largely determined by the neutron radius
\begin{equation}
\langle r^2 \rangle_n = -6 \left( \frac{\partial G_{En}}{\partial Q^2} \right)
_{Q^2 \rightarrow 0} = -\frac{3 \bar{S} b}{2 m_N^2}.
\end{equation}
If we further assume that $\Lambda_\pi$ within a loop is the same as
that for pion electroproduction, only $\Lambda_C$ remains to be fit to
data for $G_{En}$.  Thus, using fixed parameters $\bar{S} = 0.26$ and
$b = 6.65$ suggested by Kaskulov and Grabmayr, we fit
$\Lambda_C^2=1.00 \pm 0.03$ (GeV/$c$)$^2$ to the current $G_{En}$
data.  The value given in \cite{kaskulov04} for $\Lambda_C$ is
slightly smaller because they used the same preliminary data as
\cite{friedrich03} that are smaller than the final results.  Our fit
is shown in Fig.\ \ref{fig:kaskulov} and is practically
indistinguishable from the two-parameter Galster fit given in
\cite{kelly04}.  The Kaskulov and Grabmayr model has the same physical
basis as that of Friedrich and Walcher, but is much more constrained;
nevertheless, it fits the $G_{En}$ data quite well.  This result
suggests that the radius of the $3q$ nucleon core is
\begin{equation}
\langle r^2 \rangle_{3q}^{1/2} = {\frac{\sqrt{12}}{\Lambda_C}} 
= (0.68 \pm 0.01)~\mathrm{fm}.
\end{equation}

\subsubsection{Quark models}
\label{sec:final-results-data-theory-quarkmodels}

The predictions of several recent relativistic constituent quark
models are compared with the data in Fig.\ \ref{fig:qm}.  All employ a
linear confining potential.  The solid curves show calculations of the
Pavia-Graz collaboration \cite{wagenbrunn01} that used the point-form
spectator approximation (PFSA) for pointlike constituent quarks and a
Goldstone boson exchange interaction fitted to spectroscopic data.  No
additional parameters were adjusted to fit the form factors.  The data
for $G_{Ep}/G_{D}$ are reproduced very well and the data for magnetic
form factors are also described relatively well for $Q^2 \lesssim 1$
(GeV/$c$)$^2$, but the calculated value of $G_{Mp}/\mu_{p}G_{D}$
decreases too rapidly for larger $Q^2$.  The prediction for
$G_{En}/G_{D}$ lies well below the data for $Q^2 > 1$ (GeV/$c$)$^2$.
The long-dashed curves show calculations of Simula \cite{simula01},
based upon the model of Cardarelli \textit{et al}.\
\cite{cardarelli95}, that used the light-front approach and the
one-gluon exchange (OGE) interaction.  Here, constituent-quark form
factors were fitted to data for $Q^2 < 1$ (GeV/$c$)$^2$ and the
calculations were extrapolated to larger $Q^2$.  This approach
provides good fits up to about 4 (GeV/$c$)$^2$.  Finally, the
dash-dotted curves show the results for a semirelativistic
hypercentral constituent quark model (hCQM) \cite{desanctis05} where
the constituent-quark form factors, chosen as linear combinations of
monopole and dipole forms, were also fitted to recent data.  Of the
selected quark model calculations, their results clearly achieve the
best overall agreement with the data.

Finally, the most recent lattice QCD calculations of nucleon form
factors were reported by the QCDSF collaboration \cite{gockeler05}
using nonperturbatively improved Wilson fermions in the quenched
approximation.  Unfortunately, straightforward chiral extrapolation
\cite{ashley04} does not provide adequate agreement with data for $Q^2
< 1.5$ (GeV/$c$)$^2$.  Matevosyan \textit{et al}.\ \cite{matevosyan05}
proposed a model-dependent extrapolation procedure based upon the
LFCBM.  This extrapolation is quite severe because the lattice
calculations remain limited to quark masses that correspond to $m_\pi
\geq 0.5$ GeV, lattice spacings with $a \geq 0.05$ fm, and volumes
that might not fully contain the pion cloud; therefore, comparison
with data is probably premature.

\section{Summary and conclusions}
\label{sec:summary-conclusions}

We reported values for the neutron electric to magnetic form factor
ratio, $G_{En}/G_{Mn}$, deduced from measurements of the neutron's
recoil polarization in quasielastic $^{2}$H$(\vec{e},e'\vec{n})^{1}$H
kinematics at three acceptance-averaged $Q^{2}$ values of 0.45, 1.13,
and 1.45 (GeV/$c$)$^{2}$.  In the one-photon exchange approximation
for elastic scattering from a free neutron, the polarization vector of
the recoil neutron is confined to the scattering plane and consists of
a longitudinal component, $P^{(h)}_{\ell} \propto G_{Mn}^{2}$, and a
transverse component, $P^{(h)}_{t} \propto G_{En}G_{Mn}$.  The use of
a deuteron target to access the neutron form factor ratio via the
quasielastic $^{2}$H$(\vec{e},e'\vec{n})^{1}$H reaction has the
advantage, as established by Arenh\"{o}vel \textit{et al}.\
\cite{arenhovel87,arenhovel88}, that both $P^{(h)}_{t}$ and
$P^{(h)}_{\ell}$ are relatively insensitive to final-state
interactions (FSI), meson-exchange currents (MEC), isobar
configurations (IC), and to theoretical models of deuteron structure.

A high-luminosity neutron polarimeter designed specifically for our
experiment, Jefferson Laboratory E93-038, was used to measure neutron
polarization-dependent scattering asymmetries proportional to the
projection of the polarization vector on the transverse axis.  A
dipole magnet located upstream of the polarimeter was used to precess
the neutron polarization vector in the transverse-longitudinal plane,
thereby permitting access to the ratio $P^{(h)}_{t}/P^{(h)}_{\ell}
\propto G_{En}/G_{Mn}$.  Values for the scattering asymmetries were
extracted from neutron time-of-flight measurements in our polarimeter
via the cross ratio technique.  The merit of the cross ratio technique
is that the scattering asymmetries are independent of the luminosities
for the two electron beam helicity states, and independent of the
efficiencies and acceptances of the top and bottom halves of the
polarimeter.  Systematic uncertainties in our results are minimal as
the analyzing power of the polarimeter and the polarization of the
electron beam cancel in the form factor ratio.  Further, other sources
of uncertainty, such as radiative corrections and neutron
depolarization by lead shielding, are small as they nearly cancel in
the ratio.

To account for the finite experimental acceptance and nuclear physics
effects (i.e., FSI, MEC, and IC), we used two independent simulation
programs to average theoretical $^{2}$H$(\vec{e},e'\vec{n})^{1}$H
recoil polarization calculations computed according to the model of
Arenh\"{o}vel \textit{et al}.\ \cite{arenhovel87,arenhovel88} over the
acceptance.  The results from these two simulation programs agreed to
better than 1\% at our two lower $Q^{2}$ points and 2\% at our highest
$Q^{2}$ point.  Further, by averaging two different sets of
theoretical calculations assuming different parametrizations for
$G_{En}$, our acceptance-averaged and nuclear physics corrected values
for $G_{En}$ were found to be insensitive to the choice of the $Q^{2}$
dependence of $G_{En}$.

Our results for $G_{En}$ and data on the nucleon form factors were
compared with selected theoretical model calculations.  All of the
model calculations based upon vector meson dominance and those
emphasizing the pion cloud presented here provide qualitative
agreement with some of the four nucleon form factors, but no model
achieves simultaneous agreement with all four form factors.  The
predictions of several recent relativistic quark models also achieve
qualitative agreement with the data, with the most successful models
utilizing form factors for the constituents; the results from a chosen
model assuming pointlike constituents are not as successful.  Although
a comparison between data and the results of lattice QCD calculations
is probably premature, the recent precise data obtained from
experiments employing polarization degrees of freedom will no doubt
serve as a future challenging test of QCD as formulated on the
lattice.

In conclusion, our results at $Q^{2} = 1.13$ and 1.45 (GeV/$c$)$^{2}$
are the first direct measurements of $G_{En}$ using polarization
degrees of freedom in the $Q^{2} > 1$ (GeV/$c$)$^{2}$ region and are
the most precise determinations of $G_{En}$ over all ranges of
$Q^{2}$.  The achievement of relative statistical uncertainties in the
form factor ratio $G_{En}/G_{Mn}$ of 8.4\% and 9.5\%, respectively, at
these two $Q^{2}$ points, together with relative systematic
uncertainties on the level of 2\%, was a triumph for our high
figure-of-merit and high luminosity neutron polarimeter.

\begin{acknowledgments}

This work was supported in part by grants from the U.S.\ National
Science Foundation, the U.S.\ Department of Energy, and the Deutsche
Forschungsgemeinschaft.  Two of us (R.M.\ and T.E.) acknowledge
support during the planning phase of E93-038 from the Nuclear and High
Energy Physics Center at Hampton University.  The Southeastern
Universities Research Association (SURA) operates the Thomas Jefferson
National Accelerator Facility under the U.S.\ Department of Energy
Contract No.\ DE-AC05-84ER40150.  We thank the Jefferson Laboratory
Hall C engineering and technical staff for their skillful assistance
during our experiment.

\end{acknowledgments}

\begin{appendix}

\section{Formalism for the quasielastic
         $\bm{^{2}}$H$\bm{(}\vec{\bm{e}}\bm{,e'}\vec{\bm{n}}\bm{)^{1}}$H
         reaction}
\label{sec:appendix-a}

Our notation for the kinematics and nucleon recoil polarization for
the quasielastic $^{2}$H$(\vec{e},e'\vec{n})^{1}$H reaction follows
that of Arenh\"{o}vel \textit{et al}.\ (e.g., \cite{arenhovel04}).
For ease of notation, all kinematic quantities in the center-of-mass
(c.m.)  frame of the recoiling neutron-proton ($n$-$p$) system will
carry a superscript $^{\mathrm{c.m.}}$; however, kinematic quantities
referred to the laboratory frame will not be adorned with a
superscript $^{\mathrm{lab}}$.

\begin{figure*}
\includegraphics[scale=0.55,angle=270]
{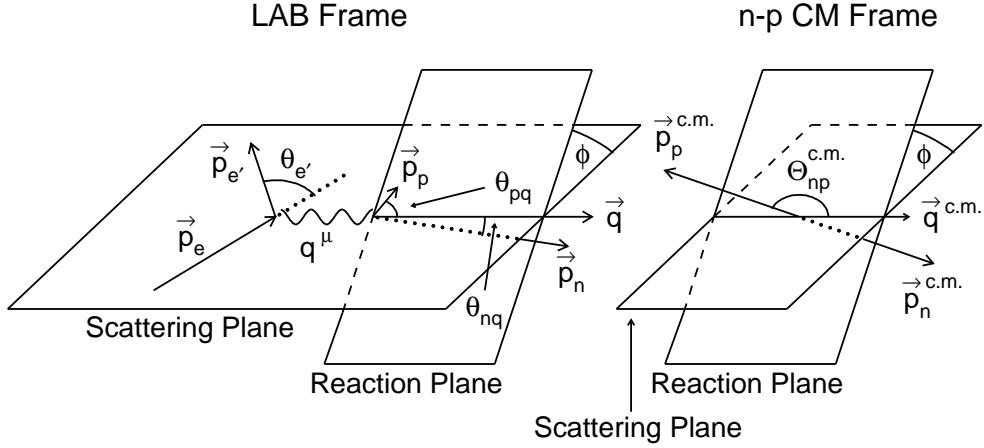}
\caption{Schematic diagram of the kinematics for the
electrodisintegration of the deuteron in the one-photon exchange
approximation as viewed from the laboratory frame and the recoiling
$n$-$p$ c.m.\ frame.}
\label{fig:deuteron-electrodisintegration-schematic}
\end{figure*}

\subsection{Kinematic notation}
\label{sec:appendix-a-kinematics}

A schematic diagram of the kinematics for the electrodisintegration of
the deuteron in the one-photon exchange approximation is shown in
Fig.\ \ref{fig:deuteron-electrodisintegration-schematic}.  Our
notation for the electron kinematics is as usual, and we assume the
electron scatters from a deuteron with initial four-momentum
$(m_{d},{\bf{0}})$.  Following the break-up of the deuteron, the
proton and neutron exit with three-momenta ${\bf{p}}_{p}$ and
${\bf{p}}_{n}$, respectively.  As is customary, we use ${\theta}_{pq}$
(${\theta}_{nq}$) to denote the polar angle between ${\bf{p}}_{p}$
(${\bf{p}}_{n}$) and ${\bf{q}}$ in the laboratory frame, and a
reaction plane is defined by any two of ${\bf{q}}$, ${\bf{p}}_{p}$,
and ${\bf{p}}_{n}$.  As is shown in Fig.\
\ref{fig:deuteron-electrodisintegration-schematic}, the reaction plane
is tilted at a dihedral angle ${\phi}$ with respect to the scattering
plane.  It should be noted that in the $n$-$p$ c.m.\ frame, this
dihedral angle, ${\phi}^{\mathrm{c.m.}}_{np}$, is, obviously, just
equal to ${\phi}$.

The $n$-$p$ c.m.\ frame is reached via a boost along ${\bf{q}}$.  In
the laboratory frame, the $n$-$p$ final state has an invariant mass,
$W_{np}$, of $W_{np} = \sqrt{E_{np}^{2} - {\bf{q}}^{2}}$, where the
relative $n$-$p$ energy in the laboratory frame, $E_{np}$, is $E_{np}
= \omega + m_{d}$.  With these definitions, it is clear that the
Lorentz factor for the boost from the laboratory frame to the $n$-$p$
c.m.\ frame is
\begin{equation}
\gamma = \frac{E_{np}}{W_{np}} = \frac{\omega+m_{d}}
{\sqrt{(\omega+m_{d})^{2}-{\bf{q}}^{2}}}.
\label{eq:gamma-lab-npcm}
\end{equation}
We denote the polar angle between the relative $n$-$p$ motion in the
c.m.\ frame, ${\bf{p}}^{\mathrm{c.m.}}_{np} =
\frac{1}{2}({\bf{p}}^{\mathrm{c.m.}}_{p} -
{\bf{p}}^{\mathrm{c.m.}}_{n}) = {\bf{p}}^{\mathrm{c.m.}}_{p}$
(assuming equal nucleon masses), and ${\bf{q}}^{\mathrm{c.m.}}$ as
$\Theta^{\mathrm{c.m.}}_{np}$.  As can be shown easily, this angle can
be written solely in terms of the laboratory frame observables $E_{n}
= \sqrt{{\bf{p}}_{n}^{2}+m_{n}^{2}}$, $|{\bf{p}}_{n}|$, $\theta_{nq}$,
and $\omega$ as
\begin{equation}
\cos\Theta^{\mathrm{c.m.}}_{np} =
-\frac{|{\bf{p}}_{n}|\cos\theta_{nq} -
|{\bf{q}}|E_{n}/E_{np}}
{\sqrt{A + B}},
\label{eq:thetacmnp-definition}
\end{equation}
where
\begin{subequations}
\begin{eqnarray}
A &=& {\bf{p}}_{n}^{2}\sin^{2}\theta_{nq}
\left(1 - \displaystyle{\frac{{\bf{q}}^{2}}{E_{np}^{2}}}\right), \\
B &=& \left(|{\bf{p}}_{n}|\cos\theta_{nq} -
\displaystyle{\frac{|{\bf{q}}|E_{n}}{{E_{np}}}}\right)^{2}.
\end{eqnarray}
\end{subequations}

Clearly, ${\Theta}^{\mathrm{c.m.}}_{np} = 0^{\circ}$ ($180^{\circ}$)
corresponds to perfect quasifree emission of the proton (neutron);
however, it should be noted that there is vanishing phase space for
perfect quasifree emission.

\subsection{Recoil polarization}
\label{sec:appendix-a-recoilpolarization}

The five-fold differential coincidence cross section for the
electrodisintegration of the deuteron in polarized-electron,
unpolarized-deuteron scattering is of the simple form
\cite{arenhovel88}
\begin{equation}
\sigma(h,0,0) \equiv
\frac{\mathrm{d}^{5}\sigma}{\mathrm{d}E_{e'}\mathrm{d}\Omega_{e'}
\mathrm{d}\Omega^{\mathrm{c.m.}}_{np}} =
\sigma_{0}\left(1 + hP_{e}A_{e}\right),
\end{equation}
as the electron asymmetry, $A_{e}$, is the only polarized contribution
to the cross section.  As usual, $\sigma_{0}$ denotes the unpolarized
cross section.  The above expression for the cross section can also be
written in terms of structure functions as \cite{arenhovel88}
\begin{eqnarray}
\sigma(h,0,0) &=& C \big(\rho_{L}f_{L} + \rho_{T}f_{T} +
\rho_{LT}f_{LT}\cos\phi^{\mathrm{c.m.}}_{np} \nonumber \\
&& ~~~+ \rho_{TT}f_{TT}\cos2\phi^{\mathrm{c.m.}}_{np} \nonumber \\
&& ~~~+ hP_{e}\rho'_{LT}f'_{LT}\sin\phi^{\mathrm{c.m.}}_{np}\big),
\end{eqnarray}
where the $f_{i}$ structure functions are evaluated in the $n$-$p$
c.m.\ frame, the $\rho_{i}$ are elements of the virtual photon density
matrix and functions of kinematics, and $C$ is a function of
kinematics.  It should be noted that the above expression for the
cross section is differential in $E_{e'}$, $\Omega_{e'}$, and
$\Omega^{\mathrm{c.m.}}_{np}$.  The Jacobian, $\mathcal{J} =
\partial\Omega^{\mathrm{c.m.}}_{np}/\partial\Omega_{n}$, which
transforms $\Omega^{\mathrm{c.m.}}_{np} \rightarrow \Omega_{n}$ is
given by \cite{arenhovel04}
\begin{eqnarray}
\mathcal{J} &=& \frac{1}{\gamma}\left(\frac{\beta_{n}\gamma_{n}}
{\beta_{n}^{\mathrm{c.m.}}\gamma_{n}^{\mathrm{c.m.}}}\right)^{3}
\left(1+\frac{\beta}{\beta_{n}^{\mathrm{c.m.}}}
\cos(\pi-\Theta^{\mathrm{c.m.}}_{np})\right)^{-1}. \nonumber \\
\end{eqnarray}
Here, $\gamma$ is as given in Eq.\ (\ref{eq:gamma-lab-npcm}),
$\gamma_{n}^{\mathrm{c.m.}}$ is the Lorentz factor for the boost which
takes the neutron from its rest frame to the $n$-$p$ CM frame,
\begin{equation}
\gamma_{n}^{\mathrm{c.m.}} = \frac{W_{np}}{2m_{n}},
\label{eq:gamma-n-cm}
\end{equation}
and $\gamma_{n}$ is the Lorentz factor for the boost which takes the
neutron from its rest frame to the laboratory frame,
\begin{equation}
\gamma_{n} = \gamma\gamma^{\mathrm{c.m.}}_{n}\left[1+\beta
\beta_{n}^{\mathrm{c.m.}}
\cos(\pi-\Theta_{np}^{\mathrm{c.m.}})\right].
\label{eq:gamma-n-lab}
\end{equation}
$\beta$, $\beta^{\mathrm{c.m.}}_{n}$, and $\beta_{n}$ are the
velocities associated with $\gamma$, $\gamma^{\mathrm{c.m.}}_{n}$, and
$\gamma_{n}$, respectively.

The nucleon recoil polarization in the $n$-$p$ c.m.\ frame,
${\bf{P}}^{\mathrm{c.m.}}$, is of the form \cite{arenhovel88}
\begin{equation}
\frac{\mathrm{d}^{5}\sigma}{\mathrm{d}E_{e'}\mathrm{d}\Omega_{e'}
\mathrm{d}\Omega^{\mathrm{c.m.}}_{np}}{\bf{P}}^{\mathrm{c.m.}} =
\sigma_{0}\big[\big({\bf{P}}^{(0)}\big)^{\mathrm{c.m.}} +
hP_{e}\big({\bf{P}}^{(h)}\big)^{\mathrm{c.m.}}\big],
\end{equation}
where ${\bf{P}}^{(0)}$ and ${\bf{P}}^{(h)}$ denote, respectively, the
helicity-independent and helicity-dependent recoil polarization.
Written in terms of $g^{t,n,\ell}_{i}$ structure functions, the
helicity-independent polarization components are
\begin{subequations}
\begin{eqnarray}
\big(P^{(0)}_{t}\big)^{\mathrm{c.m.}} &=& \frac{C}{\sigma_{0}}
\left(\rho_{LT}g^{t}_{LT}\sin\phi^{\mathrm{c.m.}}_{np} +
\rho_{TT}g^{t}_{TT}\sin \phi^{\mathrm{c.m.}}_{np}\right), \nonumber \\
&& \\
\big(P^{(0)}_{n}\big)^{\mathrm{c.m.}} &=& \frac{C}{\sigma_{0}}
\big(\rho_{L}g^{n}_{L} + \rho_{T}g^{n}_{T} +
\rho_{LT}g^{n}_{LT}\cos\phi^{\mathrm{c.m.}}_{np} \nonumber \\
&&~~~~~+ \rho_{TT}g^{n}_{TT}\cos2\phi^{\mathrm{c.m.}}_{np}\big), \\
\big(P^{(0)}_{\ell}\big)^{\mathrm{c.m.}} &=& \frac{C}{\sigma_{0}}
\left(\rho_{LT}g^{\ell}_{LT}\sin\phi^{\mathrm{c.m.}}_{np} +
\rho_{TT}g^{\ell}_{TT}\sin2\phi^{\mathrm{c.m.}}_{np}\right), \nonumber \\
\end{eqnarray}
\end{subequations}
and the helicity-dependent polarization components are
\begin{subequations}
\begin{eqnarray}
\big(P^{(h)}_{t}\big)^{\mathrm{c.m.}} &=& \frac{C}{\sigma_{0}}
\left(\rho'_{LT}g'^{t}_{LT}\cos\phi^{\mathrm{c.m.}}_{np} +
\rho'_{T}g'^{t}_{T}\right), \nonumber \\
&& \\
\big(P^{(h)}_{n}\big)^{\mathrm{c.m.}} &=& \frac{C}{\sigma_{0}}
\rho'_{LT}g'^{n}_{LT}\sin\phi^{\mathrm{c.m.}}_{np}, \\
\big(P^{(h)}_{\ell}\big)^{\mathrm{c.m.}} &=& \frac{C}{\sigma_{0}}
\left(\rho'_{LT}g'^{\ell}_{LT}\cos\phi^{\mathrm{c.m.}}_{np} +
\rho'_{T}g'^{\ell}_{T}\right). \nonumber \\
\end{eqnarray}
\end{subequations}

The boost from the laboratory frame to the $n$-$p$ c.m.\ frame is
along ${\bf{q}}$, which is not, in general, parallel to either
nucleon's momentum vector; therefore, the recoil polarization
components in the laboratory frame are related to the recoil
polarization components in the $n$-$p$ c.m.\ frame via a relativistic
Wigner spin rotation.  As the nucleons' momenta span the
$\hat{t}$-$\hat{\ell}$ plane, the $\hat{n}$-component is unchanged,
while the $\hat{t}$- and $\hat{\ell}$-components mix according to
\begin{equation}
P_{i} = \mathcal{R}_{ij}\left(\theta^{W}_{n}\right)P^{\mathrm{c.m.}}_{j},
\end{equation}
where $i,j~\in~\{t,n,\ell\}$, $\mathcal{R}_{ij}(\theta^{W}_{n})$
denotes a matrix element of the Wigner rotation matrix,
\begin{equation}
\mathcal{R}\left(\theta^{W}_{n}\right) =
\left( \begin{array}{ccc}
\cos\theta^{W}_{n}& 0& \sin\theta^{W}_{n} \\
0& 1& 0 \\
-\sin\theta^{W}_{n}& 0& \cos\theta^{W}_{n}
\end{array} \right),
\end{equation}
and $\theta^{W}_{n}$, the Wigner rotation angle for the neutron,
is expressed in terms of kinematics as
\cite{giebink85,arenhovel04}
\begin{equation}
\theta^{W}_{n} = \sin^{-1}\left[\frac{1+\gamma}
{\gamma^{\mathrm{c.m.}}_{n}+\gamma_{n}}\sin\left(\theta^{\mathrm{c.m.}}_{n} -
\theta_{n}\right)\right].
\end{equation}
Here, $\theta^{\mathrm{c.m.}}_{n} (= \pi -
\Theta^{\mathrm{c.m.}}_{np})$ and $\theta_{n}$ denote, respectively,
the polar angle of the neutron's momentum vector relative to
${\bf{q}}$ in the $n$-$p$ c.m.\ frame and the laboratory frame.  For
non-relativistic boosts (i.e., $\gamma$, $\gamma^{\mathrm{c.m.}}_{n}$,
and $\gamma_{n}$ all $\sim1$), it is clear that we recover the
non-relativistic result, $\theta^{W}_{n} \rightarrow
\theta^{\mathrm{c.m.}}_{n} - \theta_{n}$.  Also, it is obvious that
for perfect quasifree emission (i.e., $\Theta^{\mathrm{c.m.}}_{np}=0$
or $\pi$), the recoil polarization components in the $n$-$p$ c.m.\
frame are identical to those in the laboratory frame.

\section{Sensitivity to nuclear physics effects and deuteron structure}
\label{sec:appendix-b}

\begin{figure}
\includegraphics[scale=0.47]{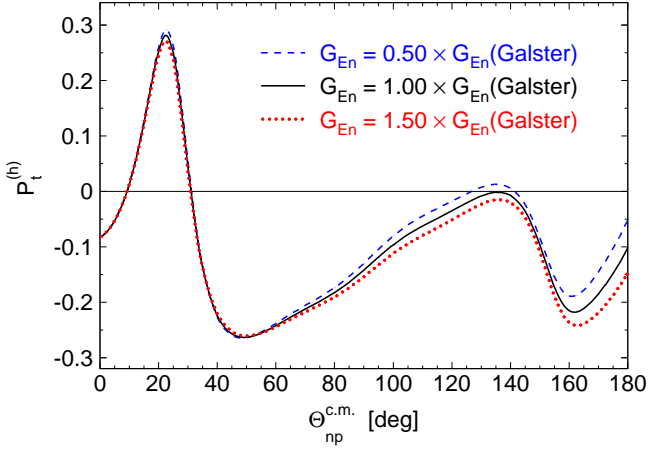}
\caption{(Color online) Sensitivity of FSI+MEC+IC+RC calculations of
$P^{(h)}_{t}$ to the value of $G_{En}$ for the central kinematics of
our $Q^{2} = 1.136$ (GeV/$c$)$^{2}$ point.  The results shown are for
$\phi^{\mathrm{c.m.}}_{np} = 0^{\circ}$ and the Bonn potential.}
\label{fig:phx-bonn-fsi-mec-ic-compare-galster}
\end{figure}

\begin{figure}
\includegraphics[scale=0.47]{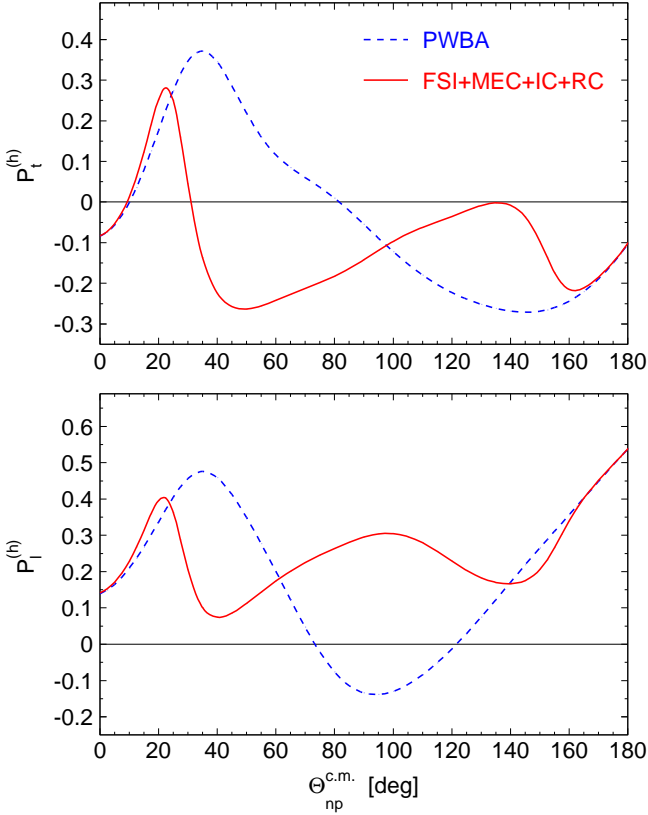}
\caption{(Color online) Comparison of PWBA and FSI+MEC+IC+RC calculations
of $P^{(h)}_{t}$ (top panel) and $P^{(h)}_{\ell}$ (bottom panel) for
the central kinematics of our $Q^{2} = 1.136$ (GeV/$c$)$^{2}$ point.
The results shown are for $\phi^{\mathrm{c.m.}}_{np} = 0^{\circ}$ and
the Bonn potential.}
\label{fig:phxphz-bonn-compare-pwba-fsi-mec-ic}
\end{figure}

\begin{figure}
\includegraphics[scale=0.47]{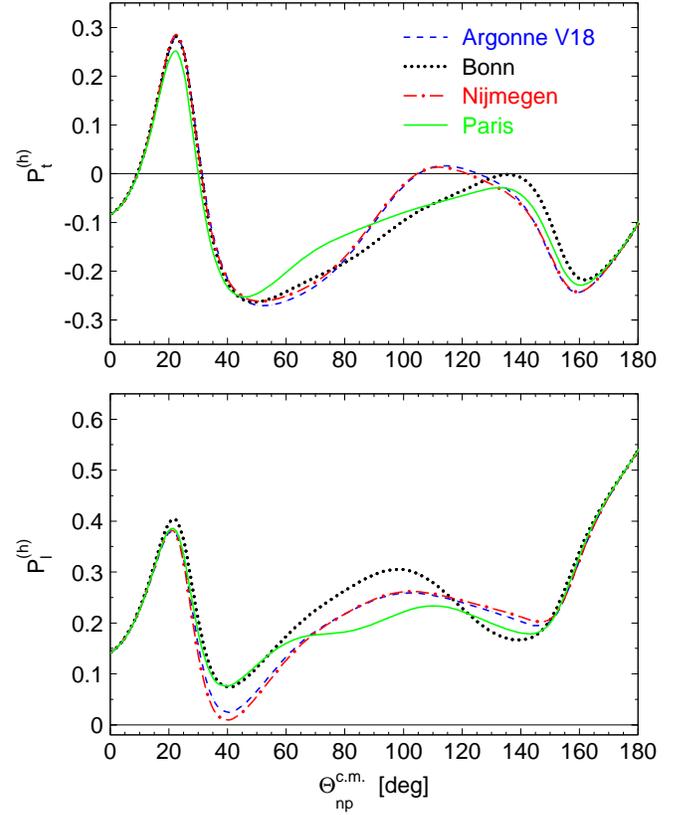}
\caption{(Color online) Comparison of FSI+MEC+IC+RC calculations of
$P^{(h)}_{t}$ (top panel) and $P^{(h)}_{\ell}$ (bottom panel) for the
Argonne V18, Bonn, Nijmegen, and Paris potentials.  The results shown
are for the central kinematics of our $Q^{2} = 1.136$ (GeV/$c$)$^{2}$
point and $\phi^{\mathrm{c.m.}}_{np} = 0^{\circ}$.}
\label{fig:phxphz-fsi-mec-ic-compare-potential}
\end{figure}

To demonstrate the sensitivity of $P^{(h)}_{t}$ to the value of
$G_{En}$ and the insensitivity of $P^{(h)}_{t}$ and $P^{(h)}_{\ell}$
to FSI, MEC, IC, and the choice of the $NN$ potential, we present
several examples of $^{2}$H$(\vec{e},e'\vec{n})^{1}$H recoil
polarization calculations performed within the PWBA and FSI+MEC+IC+RC
models of Arenh\"{o}vel \textit{et al}.\
\cite{arenhovel87,arenhovel88,arenhovel-private} in Figs.\
\ref{fig:phx-bonn-fsi-mec-ic-compare-galster},
\ref{fig:phxphz-bonn-compare-pwba-fsi-mec-ic}, and
\ref{fig:phxphz-fsi-mec-ic-compare-potential}.  We have (arbitrarily)
chosen to show examples of these calculations for the central
kinematics of our $Q^{2} = 1.136$ (GeV/$c$)$^{2}$ point (i.e., $E_{e}
= 2.326$ GeV, $E_{e'} = 1.718$ GeV, $\theta_{e'} = 30.93^{\circ}$).

First, FSI+MEC+IC+RC calculations of $P^{(h)}_{t}$ are shown in Fig.\
\ref{fig:phx-bonn-fsi-mec-ic-compare-galster} as a function of
$\Theta^{\mathrm{c.m.}}_{np}$ for three values of $G_{En}$ scaled by
the Galster parametrization: 0.5, 1.0, and 1.5.  A strong (nearly
linear) sensitivity of $P^{(h)}_{t}$ to the value of $G_{En}$ is seen
at and near quasifree emission.  Second, the insensitivity of
$P^{(h)}_{t}$ and $P^{(h)}_{\ell}$ to FSI, MEC, and IC for quasifree
emission is shown in Fig.\
\ref{fig:phxphz-bonn-compare-pwba-fsi-mec-ic} where little difference
between the PWBA and FSI+MEC+IC+RC calculations is observed at and near
quasifree emission.  Finally, we compare FSI+MEC+IC+RC calculations of
$P^{(h)}_{t}$ and $P^{(h)}_{\ell}$ for the Argonne V18
\cite{wiringa95}, Bonn \cite{machleidt87}, Nijmegen \cite{nagels78},
and Paris \cite{lacombe81} $NN$ potentials in Fig.\
\ref{fig:phxphz-fsi-mec-ic-compare-potential}.  Again, at and near
quasifree emission, there is little model dependence.

\clearpage

\end{appendix}


\end{document}